# Mechanistic Insights into Active Sites for Electrochemical $CO_2$ and CO Reduction over the Strain-Engineered Dealloyed Cu

*Yuxiang Zhou, Ayman A. El-Zoka*, Oliver R. Waszkiewicz, Benjamin Bowers, Rose P. Oates, James Murawski, Anna Winiwarter, Guangmeimei Yang, Oleg Konovalov, Maciej Jankowski, Caiwu Liang, Paul-Benjamin Amar, Andreas G. Kafizas, Ludmilla Steier, Ifan E. L. Stephens* and Mary P. Ryan **

Yuxiang Zhou. Author 1, Department of Materials, Imperial College London, SW7 2AZ, UK

Ayman A. El-Zoka. Author 2 and Corresponding Author, Univ Rouen Normandie, CNRS, INSA Rouen Normandie, Groupe de Physique des Matériaux,
Rouen, France E-mail: ayman.el-zoka@univ-rouen.fr

Oliver R. Waszkiewicz. Author 3, Benjamin Bowers. Author 4, Rose P. Oates, Author 5, James Murawski. Author 6, Anna Winiwarter. Author 7, Guangmeimei Yang. Author 8, Caiwu Liang. Author 11, Paul-Benjamin Amar. Author 12, Department of Materials, Imperial College London, SW7 2AZ, UK

Andreas G. Kafizas. Author 13, Department of Chemistry, Imperial College London, Molecular Science Research Hub, W12 0BZ, UK

Oleg Konovalov. Author 9, Maciej Jankowski. Author 10, The European Synchrotron - ESRF, 71 Av. des Martyrs, 38000, Grenoble, France

Ludmilla Steier. Author 14, Department of Chemistry, Oxford University, 12 Mansfield Road, Oxford, OX1 3TA, UK

Ifan E. L. Stephens. Corresponding Author, Department of Materials, Imperial College London, SW7 2AZ, UK
E-mail: i.stephens@imperial.ac.uk

Mary P. Ryan. Corresponding Author, Department of Materials, Imperial College London, SW7 2AZ, UK

E-mail: m.p.ryan@imperial.ac.uk

Keywords: dealloying, strain, cryo-APT, in situ synchrotron XRD, electrochemical $CO_2$ reduction, electrochemical CO reduction

**Abstract**

Nanoporous Cu produced by chemical dealloying is a promising catalyst for electrochemical $CO_{(2)}$ reduction owing to its tunable chemistry, morphology, and surface defect sites. However, how dealloying controls the atomic-scale structure of Cu ligaments and how these features govern catalytic behavior remain unclear, particularly in nanostructured catalysts under realistic operating conditions. Here, we synthesize nanoporous Cu by dealloying $Cu_{20}Zn_{80}$ in $H_3PO_4$ at different temperatures, enabling control over ligament sizes from the nanoscale to the microscale. Nanoporous Cu outperforms polycrystalline Cu for CO reduction, with the sample dealloyed at 15 °C reaching 60% Faradaic efficiency at -0.65 V *vs.* RHE. Using *in situ* synchrotron X-ray diffraction and cryogenic atom probe tomography, we trace the structural and chemical evolution during dealloying, revealing, for the first time, the sequential phase transitions from ε-brass to γ-brass to Cu and chemical segregation of Cu and Zn within nano-ligaments. We further establish a quantifiable strain metric linking surface defect density to ligament surface strain, quantified from the asymmetry of synchrotron XRD peaks. This approach reveals a direct correlation between catalytic activity and ligament surface strain, identifying surface strain as a practical descriptor for designing nanostructured Cu catalysts for $CO_{(2)}$ reduction under realistic operating conditions.

# 1. Introduction

The electrochemical $CO_2$ reduction reaction ($CO_2$RR) offers a promising route to convert renewable electricity into fuels and commodity chemicals, providing a potential pathway for carbon utilization and storage of intermittent renewable energy.[1-3] Among metal catalysts, Cu remains unique in its ability to produce appreciable amounts of multi-carbon products,[2-4] owing to its intermediate binding strength toward *H and *CO key intermediates.[5] However, $CO_2$ reduction on Cu still suffers from limited selectivity,[6] substantial overpotentials,[1] and poor energy and carbon efficiency under practically relevant operating conditions.[7-12]

Nanostructured Cu catalysts are especially attractive because their high surface-area-to-volume ratio and abundance of surface defect sites can strongly influence catalytic behavior.[13-14] Numerous studies have shown that steps, kinks, low-coordination atoms, and other surface defect sites can alter the adsorption energies of key intermediates and thereby steer both activity and selectivity.[13-19] Hori *et al.* investigated the $CO_2$ reduction properties on different Cu single crystals. They found that by introducing step atoms to Cu(111) basal terrace, selectivity towards $C_{2+}$ molecules can be significantly promoted.[15-16, 20] Recently, Scholten *et al* [21-22] also showed that pristine Cu(111) and (100) facets are actually inert sites, and defects are necessary for $CO_2$RR. The above results are consistent with many recent works.[17-18, 23-32]

Devising a scalable, reproducible synthesis method for making nano-structured Cu with surface defect sites is challenging. Dealloying offers a facile, yet controllable approach to synthesize nano-structured porous catalysts with surface defects.[33-34] Nanoporous metals made by dealloying have been widely used in catalysis,[34-35] chemical sensors,[36] fuel cells,[37] lithium-ion batteries,[38] *etc*. In the dealloying process, the selective dissolution of a less noble metal from an alloy of two or three metal(s) yields the formation of an open-pore, bi-continuous, sponge-like nanoporous structure that is chemically rich in the more noble metal(s). Significant attention has been paid during the last two decades to the dealloying of Ag-Au system, which yields the formation of nanoporous gold. Nanoporous gold has shown promising catalytic performance for the electrochemical reduction of $CO_2$,[39-43] with highly tunable morphology and chemistry through the control of dealloying conditions,[44] precursor composition,[44-45] and post-dealloying thermal treatments.[46-48] Yet the prospects of using nanoporous gold for $CO_2$ reduction are limited due to the lack of ability to further reduce CO to more valuable $C_{2+}$ molecules.

Although solution-phase dealloying is a versatile route to nanoporous metals, it also has notable limitations. The process typically requires corrosive acidic or alkaline etchants, generating waste streams containing dissolved heavy metal ions from the less noble alloy component.[49] It can also be less suitable for fabricating nanoporous structures from relatively noble metals. Alternative approaches, including liquid-metal,[38] and vapor-phase dealloying,[50] have been developed; however, these methods generally produce substrates with larger pores (and hence lower surface areas and strain) because these processes are carried out at high temperatures.[38, 50] In addition, dealloyed nanoporous materials are often mechanically fragile, particularly in free-standing architectures such as gas diffusion electrodes for flow cells and membrane electrode assemblies.[51] Other synthetic routes, such as sol-gel and bottom-up wet-chemical methods, also face challenges including sacrificial phases, limited scalability, and particle agglomeration during operation.[13, 52-59] Nevertheless, chemical dealloying was employed here to prepare nanoporous Cu because of its simplicity and straightforward implementation.

A wide range of conditions has been explored for producing nanoporous Cu by dealloying (Table S1).[60-72] Zhong *et al* reported dealloyed Cu-Al as a highly active $CO_2$RR catalyst, achieving 80% Faradaic Efficiency (FE) for ethylene at a total current density of 400 mA $cm^{-2}$ in 1 M KOH using a flow cell.[35] However, the origin of this high activity and its relationship to nanoporous Cu structure and chemistry remain unclear. Similar to nanoporous Au, ligament size in nanoporous Cu can also be well tuned from the nano- to microscale,[60-72] but there is less consensus on suitable etchants for Cu-Zn systems. Previous studies have shown that $H_3PO_4$ yields more uniform chemically dealloyed structures from Zn-rich brass (*e. g.* ε-brass) than other acids, especially HCl.[68-69] These studies mainly focused on electrolyte effects and post-dealloying thermal treatment. Building on this work, we hypothesize that additional control over nanoporous Cu structure and reactivity could be achieved by tuning the dealloying temperature.

Additionally, several key questions remain regarding the dealloying of ε-brass in $H_3PO_4$ and its impact on nanoporous Cu structure and chemistry, limiting the rational optimization of nanoporous Cu for $CO_2$RR. Because the Cu-Zn phase diagram predicts crystallographic changes as Zn is progressively dissolved,[73] high-resolution characterization is needed to track the structural and chemical evolution of nano-ligaments during dealloying. In the Ag-Au system, nanoporous Au retains the crystallographic structure of the parent alloy while developing Ag-rich cores and Au-rich shells,[33] but whether dealloyed nanoporous Cu follows similar trends remains unknown. The atomic-scale chemistry of Cu-Zn nano-ligaments during dealloying is likewise unresolved, largely because these air-sensitive porous structures are prone to oxidation and difficult to analyze using conventional transmission electron microscopy (TEM) and atom probe tomography (APT). Moreover, porosity infiltration strategies like electrodeposition used to enable APT analysis of nanoporous Au are expected to perturb the chemistry of Cu-Zn ligaments.[45, 74] To the best of our knowledge, the most detailed study of Zn-rich brass dealloying was reported by Pickering, who identified γ-brass as an intermediate phase and observed a compositional gradient along the Cu-enrichment direction, suggesting Cu-Zn interdiffusion during dealloying.[75] However, these results were based on *ex situ* diffraction, and *in situ* studies have not yet been reported. In addition, Pickering's work focused on γ- and η-brass under electrochemical dealloying and different chemical conditions. A refined X-ray Diffraction (XRD), together with other compositional analysis approaches that preserve the native chemistry of dealloyed ligaments, is therefore needed to follow both crystallographic and composition evolution of chemically dealloyed ε-brass through complete Zn removal.

Another important aspect of nanoporous Cu structure relevant to catalysis that remains unknown is the effective measurement of surface defect sites on the nanoporous Cu, and whether these surface defect atoms at ligament surfaces are also directly dependent on ligament size in nanoporous Cu. More analysis is needed to quantify the amount of surface defect sites on nano-ligaments during the dealloying process, which could be challenging using TEM. A high-resolution TEM study from Fujita *et al* showed that the amount of under-coordinated atoms (surface defect sites) on nanoporous Au surfaces with a ligament size of around 30 nm is comparable to that of Au nanoparticles with the diameter of 3-5 nm,[33] highlighting the potential catalytic significance of these features in nanoporous metals. A more practical method to quantify surface defect sites in nanoporous Cu is therefore needed to better understand the promising $CO_2$RR activity previously reported for dealloyed Cu catalysts by Zhong *et al*.[35] One possible route is to use microstrain measured by XRD, because lattice defects generate local strain and are closely associated with surface defect sites.[34]

More broadly, most mechanistic insight into strain effects has been obtained from model systems, including epitaxial thin films [76] and idealized planar electrodes,[77] where surface structure can be tightly controlled. However, such systems do not fully capture the complexity of practical Cu electrocatalysts, which are typically nanostructured, porous, defect-rich, and dynamically evolving under reaction conditions. Although strain effects have been discussed for model Cu surfaces and selected Cu-based nanomaterials, direct structure-strain-reactivity correlations under operando $CO_2$RR and CO reduction (CORR) conditions remain limited for realistic nanostructured Cu catalysts. Establishing such correlations is important not only for catalyst design, but also for determining whether strain can serve as a meaningful descriptor beyond idealized model electrodes. Additionally, dedicated CORR measurements are also important because they isolate the downstream reduction of *CO from the initial $CO_2$-to-CO step and reduce complications associated with pH gradients in neutral $CO_2$RR electrolytes.[78-81]

Herein, we overcome the challenges in understanding how dealloying shapes nanoporous Cu structure and chemistry through a systematic investigation of the evolution of nanoporous Cu during the dealloying of a Zn-rich ε-brass ($Cu_{20}Zn_{80}$, atomic ratio) in $H_3PO_4$. The single-phased ε-brass (Figure S1) was chosen as the precursor for dealloying due to its high Zn concentration, allowing us to optimize the morphology and chemistry for enhanced catalytic behaviors. A single-phased brass precursor would also lead to more uniform dealloying behavior, and a more homogeneous structure. We examined its crystallographic restructuring *via* both *ex situ* and *in situ* synchrotron XRD. Inspired by the works from Schofield *et al* [82] and Dotzler *et al*,[83] where they used the asymmetry of the diffraction peaks, using two peaks (strain component + bulk component) to fit each Au(200) peaks, to study the strain evolution during the dealloying process of Ag-Au system, we proposed a new method for measuring and calculating the ligament surface strain of such dealloyed nanoporous Cu that is also based on the asymmetry of the diffraction peak. The magnitude of strain was evaluated as a practical descriptor for surface defect sites. Furthermore, we used cryogenic atom probe tomography (cryo-APT) to atomically map nano-ligaments in frozen water- nanoporous Cu samples dealloyed at different times in 3D, enabling observations on the chemical composition of nano-ligaments without altering the as-dealloyed chemistry. The impact of dealloying medium, and temperature on the electrolyte-assisted coarsening process and ligament size and morphology was studied systematically. The catalytic performance of all the dealloyed nanoporous Cu samples was also tested for $CO_2$ reduction to examine whether ligament surface strain correlates with $CO_2$RR activity. Finally, the CO reduction properties of these nanoporous Cu materials were also investigated to assess the influence of ligament surface strain on further CO reduction. Overall, this investigation aims to establish how dealloying conditions control nanoporous Cu structure, chemistry, and ligament surface strain, and how these structural descriptors relate to $CO_2$RR and CORR catalysis.

## 2. Results and Discussion

### 2.1. Structural and Chemical Evolution during ε-Brass Dealloying

*In situ* XRD results in **Figure 1** summarize the phase changes in ε-$Cu_{20}Zn_{80}$ (**Figure S1**) during chemical dealloying in 5 M $H_3PO_4$ at 15 $^oC$. Details of the *in situ* synchrotron diffraction setup can be found in Supplementary Information and **Figure S2**. Prior to dealloying, only ε-phase peaks were observed, which remained the dominant phase during the first 10 minutes of dealloying. As the dealloying progresses, more Zn is removed from the Cu-Zn alloy, which was expected to induce phase transformations according to the Cu-Zn phase diagram.[75] After 10 min, as shown **in Figure 1b** and **S3a**, peaks attributed to γ-$Cu_5Zn_8$ (222) at 2.47 $Å^{-1}$, and γ (420) at 3.18 $Å^{-1}$ began to appear between 25-445 min, and 25-490 min of dealloying respectively. Interestingly, Zn diffraction peaks at 2.52 $Å^{-1}$ and 3.73 $Å^{-1}$ were also detected over a similar time range, appearing at around 400 min, and disappearing at 480 min, which can be assigned to Zn (002) and (102), respectively. This indicates the transient presence of a separated Zn phase during dealloying. Finally, after around 490 min, only Cu (α-brass) diffraction peaks remained, indicating the completion of the chemical dealloying process. The electrolyte composition during the 15 $^oC$ dealloying process was monitored by ICP-OES, showing the Zn dissolution ended at around 400 min (**Figure S4**), in good agreement with the *in situ* synchrotron XRD results. The same transitions in crystallographic structure were observed when performing the *in situ* synchrotron XRD at a higher dealloying temperature (40 $^oC$) but occurred more rapidly, ending at around 260 min (**Figure S5**).

Our in situ results are consistent with the *ex situ* XRD and electron microscopy observations reported by Pickering,[75] where γ-brass ($Cu_5Zn_8$) was identified as an intermediate phase. This intermediate phase was later rationalized by the surface diffusion-based model in theoretical and experimental studies of nanoporous metals.[74, 84-85] Our findings further corroborate the important role of Cu surface diffusion during dealloying. As mentioned above, the phase transitions at 40 $^oC$ occurred almost twice as fast as those at 15 $^oC$ dealloying, suggesting that faster Cu surface diffusion acts to facilitate both primary and secondary dealloying by exposing Zn atoms, leading to faster dissolution.

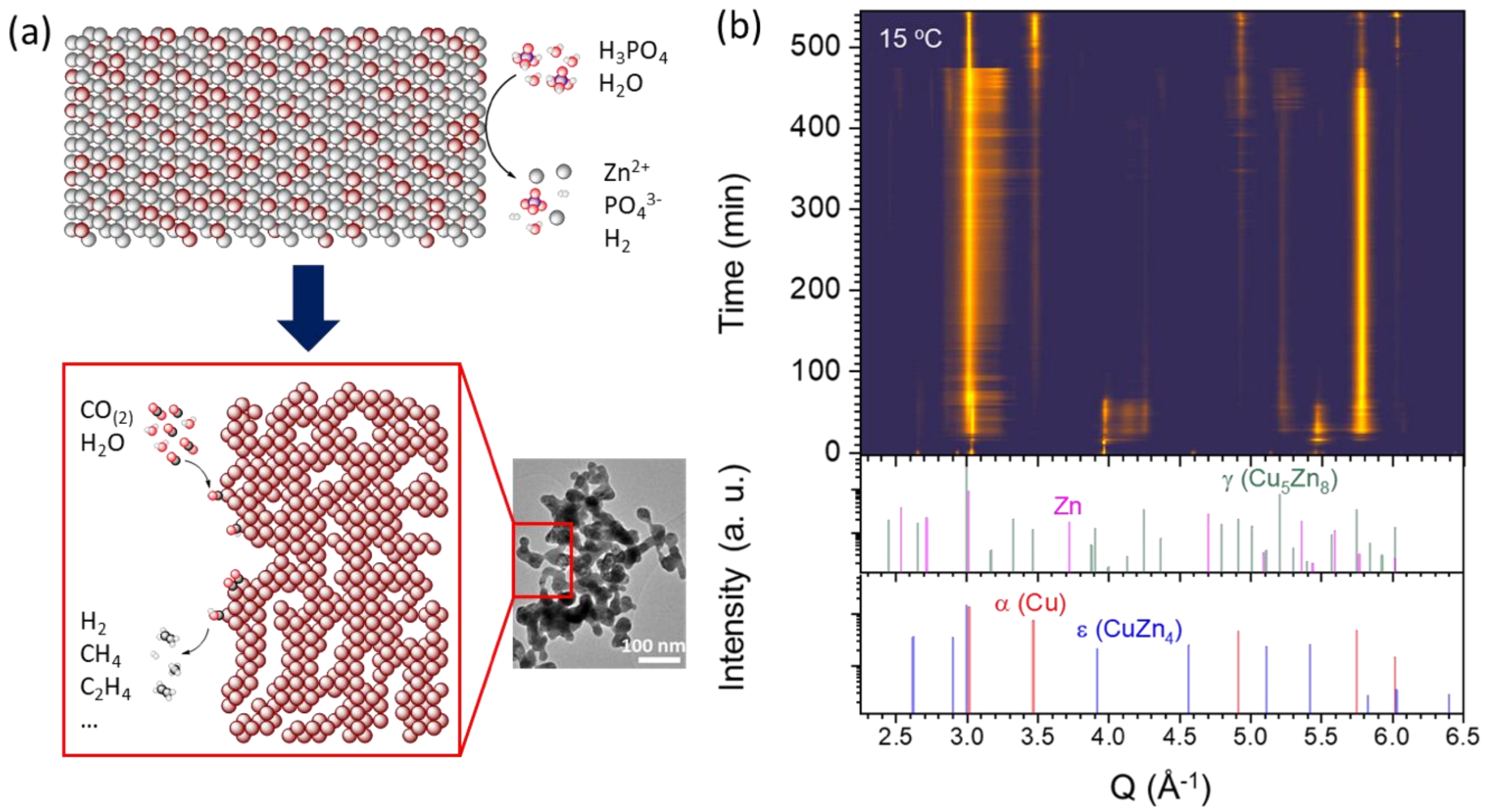

**Figure 1.** (a) Schematic illustration of the dealloying and $CO_{(2)}$ reduction work done in this paper, with an insert showing a TEM image of the 5 °C dealloyed nanoporous Cu. (b) *In situ* synchrotron XRD patterns for the chemical dealloying of $Cu_{20}Zn_{80}$ in 5 M $H_3PO_4$ at 15 °C.

To further probe the chemical evolution of nano-ligaments during dealloying, cryo-APT studies were carried out on samples dealloyed for different durations. An electrophoresis deposition method was developed which allows the cryo-APT test on these nano-structured samples, based on the previous works from El-Zoka *et al* [86], Stender *et al* [87], and Vorlaufer *et al* [88]. Further details of the cryo-APT sample preparation can be found in Supplementary Information, and **Figure S6**, and **S7**. The comparison between the chemical composition of ligaments after 3 hours, 6 hours, and 24 hours of dealloying at 5 °C (**Figure 2**, **3**, and **S8**) showed a progressive decrease in Zn content relative to the non-dealloyed $Cu_{20}Zn_{80}$ precursor measured under different field conditions in **Figure S9**. After 3 h of dealloying (**Figure 2**) ligaments showed surface enrichment in Cu. The composition of the ligament core after 3 h was already much lower in Zn than the nominal precursor composition. Zn depletion at the ligament surface supports the surface diffusion assisted mechanism of dealloying, as previously observed for other nanoporous metals in earlier APT investigations.[45, 74] Zinc-rich regions were also found within the nano-ligaments. As shown in **Figure 2e**, a single ligament exhibited a Zn-rich interface, where the Zn concentration rose to 45 atomic % (at. %) in the center of the ligament. This chemical heterogeneity is consistent with our *in situ* XRD results, which showed that multiple phases, including γ-brass and transient Zn, were present during intermediate stages of dealloying. The Zn-rich regions detected by cryo-APT therefore provide local chemical evidence supporting the coexistence of different phases within individual nano-ligaments during dealloying.

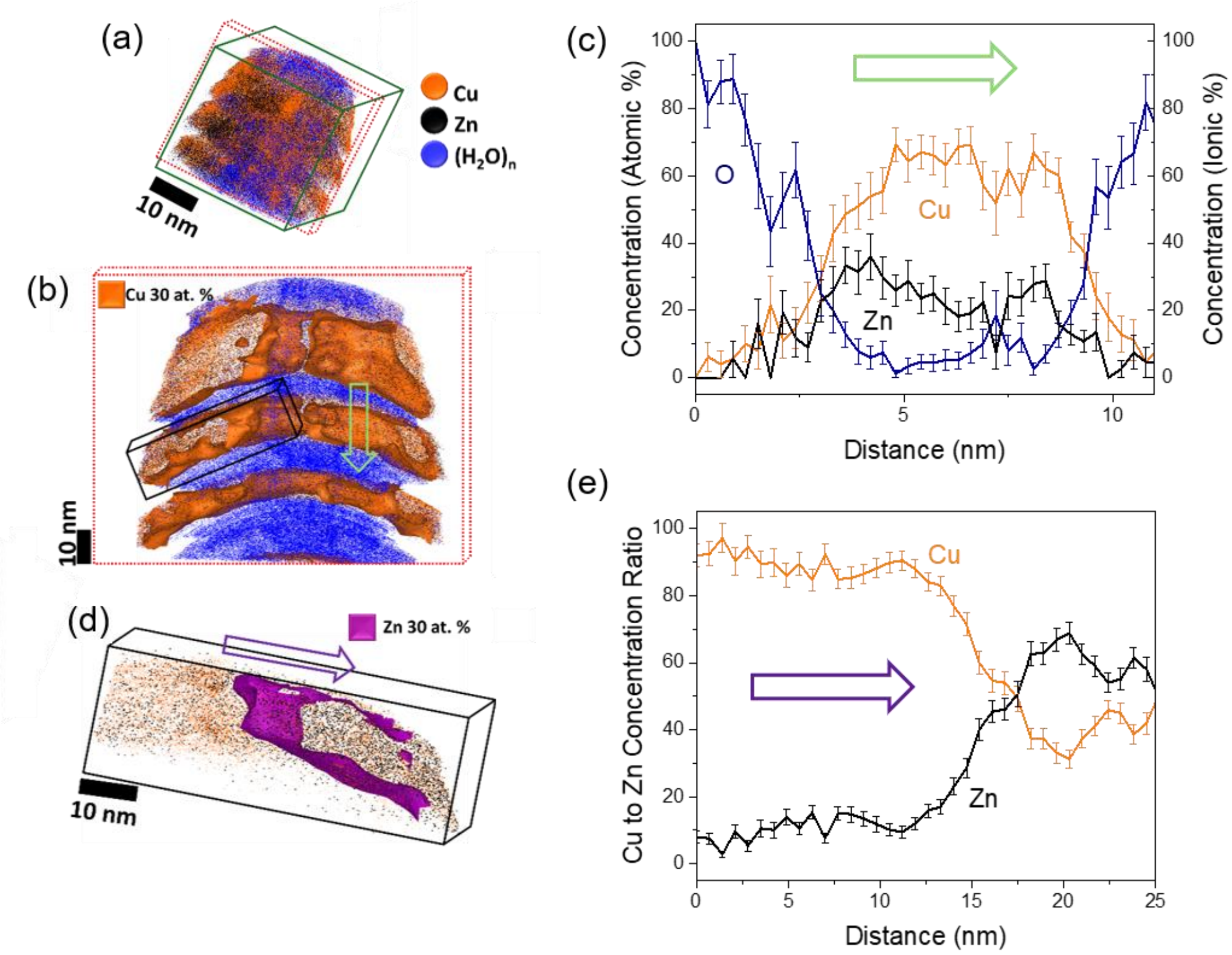

**Figure 2.** (a) Cryo-APT analysis of brass dealloyed for 3 hours. (b) 5 nm longitudinal cross-section showing the different nano-ligaments, intertwined with ice. (c) Chemical analysis across a single nano-ligament in (b). (d) A sub-volume of the reconstruction in (b) showing the interface between areas with different levels of Zn concentrations inside a single nano-ligament. (e) Chemical analysis within a single nano-ligament shows the transition from Zn-lean areas to areas with predominant Zn composition.

The cryo-APT analysis for nanoporous Cu after 6 h of dealloying is shown in **Figure 3**, where a further increase in the overall Cu to Zn ratio was observed, due to the progression of dissolution of Zn. The chemical compositional profile across a single nano-ligament in **Figure 3c**, showed that the highest concentration of Zn in the ligament is 20 at. %, with the Zn concentration peaking near the ligament surface. This result suggests that the core-shell structure of a nano-ligament may break down when Zn concentration in the ligament becomes sufficiently low. Similar to the case with the 3 h sample, a Zn rich region was still observed inside the ligament, although its Zn concentration was lower than that in the 3 h sample.

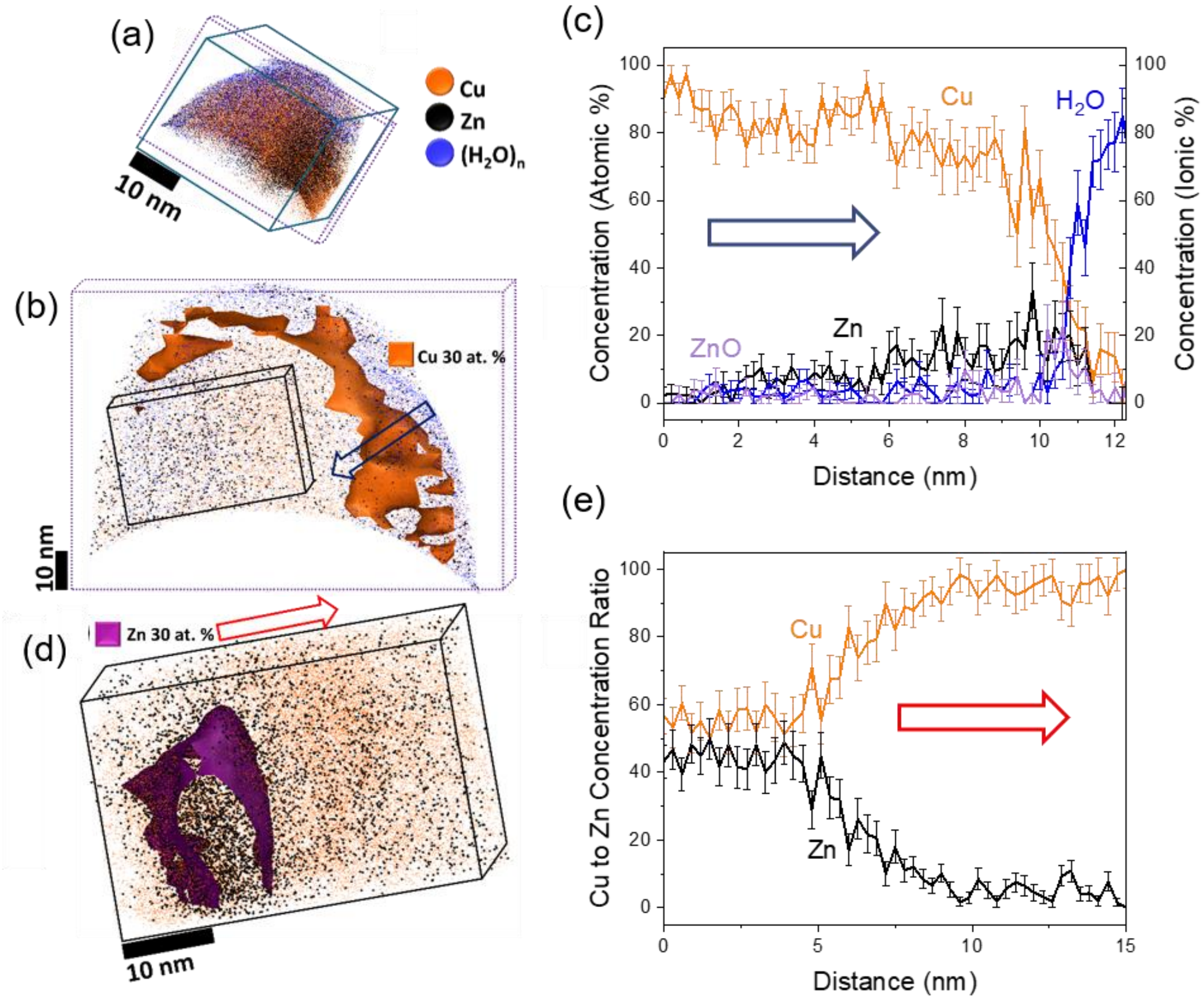

**Figure 3.** (a) Cryo-APT analysis of brass dealloyed for 6 hours. (b) 5 nm longitudinal cross-section showing the different nano-ligaments, intertwined with ice. (c) Chemical analysis across a single nano-ligament in (b). (d) A sub-volume of the reconstruction in (b) showing the interface between areas with different levels of Zn concentrations inside a single nano-ligament. (e) Chemical analysis within a single nano-ligament showing the transition from Zn-lean areas to areas with predominant Zn.

## 2.2. Temperature-Dependent Morphology and Ligament Surface Strain

The impact of temperature on phase transition during dealloying prompted us to further examine the microstructures of nanoporous Cu formed at different temperatures. As shown in **Figure 4**, the typical sponge-like bi-continuous structure was observed in all nanoporous Cu materials dealloyed in $H_3PO_4$. We noticed that room temperature dealloying of CuZn yields the formation of large ligament sizes (> 200 nm), relative to those formed in the Ag-Au system [33, 84] during dealloying at room temperature, due to the higher mobility of Cu, compared to Au.[89] In order to further refine pore sizes and obtain a higher surface area-to-volume ratio, the mobility of Cu during the dealloying process had to be controlled. Therefore, we conducted the chemical dealloying at temperatures as low as 5 $^{o}$C. **Figure 4a-g** shows the temperature effect on the ligament sizes of nanoporous Cu for a fixed dealloying time (24 h) in 5 M $H_3PO_4$. Ligament sizes varied markedly with temperature and could be tuned from 32 ± 7 nm (for 5 $^{o}$C dealloying) to 1045 ± 329 nm (for 60 $^{o}$C dealloying) (**Figure 4g** and **S10**), due to electrolyte assisted coarsening during the dealloying process. The nanoporous Cu dealloyed at 5 $^{o}$C, 10 $^{o}$C, and 15 $^{o}$C displayed the common bi-continuous structures, where the ligaments were more interconnected. However, ligaments became larger and more elongated in nanoporous Cu samples dealloyed at higher temperatures (**Figure 4d-f**, and **S11**). With increasing temperature, the aspect ratio of ligaments also increased (*e. g.* 5.23 for 25 $^{o}$C, and 6.81 for 40 $^{o}$C), and the structure became more analogous to stacked needles. In the meantime, the pore sizes of the nanoporous Cu also increased, but quantifying the porosity here is difficult due to limitations associated with the 2D nature of SEM imaging.

Another interesting feature of nanoporous Cu dealloyed at temperatures higher than 10 $^{o}$C is the appearance of small nanoscale spherical ligaments, especially in samples at 15 $^{o}$C and 25 $^{o}$C, as highlighted in **Figure S12**. With further increases in dealloying temperatures, the number of spherical features decreased, while the ligaments became increasingly elongated and developed locally thickened, approximately spherical regions. At lower temperatures, the ligament network remained relatively homogeneous, whereas at elevated temperatures the morphology became more irregular. This evolution can be attributed to the enhanced mobility and surface diffusion of Cu atoms at elevated temperatures, which promotes mass redistribution along the ligament network. Similar temperature-dependent coarsening and surface reconstruction have been reported for nanoporous Cu, where increasing dealloying temperature leads to larger ligament dimensions and a transition from relatively smooth ligament surfaces toward more irregular and semispherical morphologies.[90-91] The formation of elongated ligaments with locally thickened regions is also consistent with the surface-diffusion-controlled morphological instability described by Erlebacher.[92] In his kinetic Monte Carlo simulations, ligament pinch-off and the formation of thicker regions were associated with a solid-state Rayleigh instability driven by surface diffusion.[92] Thus, increasing the dealloying temperature does not merely accelerate the increase in characteristic ligament size, but can also modify the morphology and evolution pathway of the nanoporous network.

Two models can be used to describe the coarsening process (**Figure S13**).[46, 84, 93-94] Ostwald ripening, also known as surface-diffusion-controlled coarsening, usually happens when there is a significant difference in ligament sizes throughout the material. The larger ligaments further grow at the expense of smaller ones, to reduce the surface energy of the whole system.[95-96] On the other hand, coalescence-controlled coarsening, which refers to the merging of two separated ligaments, usually caused by lattice diffusion and volume diffusion.[97-98] The increased aspect ratio with dealloying temperature (**Figure S11**), together with the emergence of spheroidal regions at elevated temperatures (**Figure S12**) suggests that surface-diffusion-driven coarsening plays a dominant role here, although further analysis of ligament geometry and its evolution with temperature is required to establish the detailed mechanism conclusively.

The surface diffusion coefficient ($D_s$) can be calculated by the following equation:[99-101]

$$D_s = \frac{L^4 kT}{2\gamma t a^4} \quad (1)$$

where $L$ is the ligament size at dealloying time ($t$); $k$ is Boltzmann constant ($1.38 \times 10^{-23}$ J K$^{-1}$); $T$ is the dealloying temperature in K; $\gamma$ is the surface energy of Cu (1.79 J m$^{-2}$);[102] and $a$ is the lattice parameter of Cu ($3.6199 \times 10^{-10}$ m).[103] According to Arrhenius equation,

$$D_s = D_0 \exp\left(-\frac{E_a}{RT}\right) \quad (2)$$

where $D_0$ is pre-exponential factor; $E_a$ is the activation energy for Cu atom surface diffusion in J mol$^{-1}$; and R is the gas constant (8.314 J mol$^{-1}$ K$^{-1}$), the activation energy of the Cu surface diffusion in 5 M $H_3PO_4$ can be estimated (208 kJ mol$^{-1}$, as shown by the slope in **Figure S14**), by coupling Equation 1 and 2. Compared with the literature values (**Table S2**),[89, 104-107] a slightly higher Cu surface diffusion activation energy and Cu surface diffusion coefficients were obtained herein, where we postulate that this increase may be related to the surface passivation effect caused by the adsorbed phosphate.

The traditional HCl etchant was also attempted for dealloying CuZn, albeit leading to much rougher ligament surfaces, and less homogeneously dealloyed structures (**Figure S15**). This might be explained by the complexing effect caused by the $Cl^-$ anions. Trace amounts of $O_2$ in the acid would oxidize and dissolve Cu, and the existence of $Cl^-$ could accelerate this process by forming $[CuCl_4]^{2-}$ complex. The oxidation and dissolution of Zn would further reduce $Cu^{2+}$ back to Cu, causing re-deposition. According to such complexing effect, rougher ligament surfaces and less unstable under-coordinated Cu atoms (surface defect sites) are formed in HCl, relative to the smoother ligament surfaces formed in $H_3PO_4$.[68] During the dealloying process, Cu atoms also tend to restructure *via* surface diffusion, to reduce the surface energy, minimizing the amount of under-coordinated atoms. The existence of $Cl^-$ anions will accelerate both surface diffusion and the dissolution and re-deposition of Cu, as discussed as above, resulting in not only a more stable stepped surface, but also a less controlled dealloying process, and hence, less homogeneous structures (**Figure S15**).[101, 108-112] On the other hand, in $H_3PO_4$, the adsorption of bulk phosphate anions would suppress the surface diffusion of Cu atoms,[113] leading to a smoother ligament surface (**Figure 4a-f**). Remarkably, no obvious differences in morphology were observed when $H_3PO_4$ of different concentrations were used (**Figure S16**). Similar phenomena were also observed by Egle, and Cheng *et al*.[68-69]

Since the morphology of nanoporous Cu is directly controlled by dealloying temperature, we anticipated a similar effect on strain in dealloyed nanoporous Cu layers. Two types of strain usually exist in materials, namely microstrain and macrostrain. Macrostrain is the global strain throughout the whole material, induced by the lattice distortion, causing the XRD peak position to shift. On the other hand, microstrain refers to the localized strain that exists within just a few lattices. Microstrain is usually caused by defects, including grain boundaries, dislocations, stacking faults, leading to the broadening of diffraction peaks, as shown in **Figure 4h**.[34] **Figure 4h** shows the synchrotron XRD patterns of *ex situ* nanoporous Cu samples dealloyed at different temperatures. Diffraction peak broadening instead of peak shifting was observed, where the peak width decreased with increases in the dealloying temperature, indicating the existence of microstrain, especially for those dealloyed at lower temperatures.

XRD peak broadening can be induced by 3 factors, including instrumental error, crystallite size, and microstrain.[34] Instrumental error effect can be excluded by measuring the peak width using a standard sample (*e. g.* $LaB_6$).[114] In this work, the peak broadening caused by any instrumental error is negligible compared to the full width at half maximum (FWHM) of dealloyed nanoporous Cu samples (**Figure S17**). One of the most common methods for calculating the microstrain, which separates out any crystallite size effect is to use the Williamson-Hall equation (See Experimental and SI for more details).[115] By plotting $\beta cos\theta$ against $sin\theta$ of the XRD peaks of dealloyed nanoporous Cu samples, poor linearity was obtained (**Figure S18**). When using the Williamson-Hall equation to calculate the microstrain, an isotropic assumption of the microstrain is typically applied, which assumes that microstrain is uniform in all directions within the material's crystallites.[115] However, this is not the case for these dealloyed nanoporous Cu samples. As shown in **Figure 4i**, the peak cannot be fit by a single pseudo-Voigt function, and the asymmetry of the Cu(200) diffraction peak indicates anisotropy of the microstrain. Additionally, the term, microstrain, is usually a simplified scalar quantity, only used to represent the number of atoms that deviate from their idea lattice positions.[34] Therefore, a new method for describing the strain of these dealloyed nanoporous materials is required.

A few studies have shown that, for these dealloyed nanoporous materials, the strain mainly comes from the massive surface curvature on the ligaments,[33, 82-83] which were also observed on our nanoporous Cu samples (**Figure 4a-f**). Inspired by the work from Schofield *et al*,[82] and Dotzler *et al*,[83], where the strain values of dealloyed nanoporous Au materials were calculated according to the asymmetry of the diffraction peak, we propose the following new method for quantifying the strain. Ligaments can be divided into three groups, including the bulk (grey) region corresponding to the atoms inside the ligaments, and concave (blue) and convex (green) regions leading to the compressive and tensile strains, respectively (**Figure S19**). Thus, each diffraction peak can be fitted by three pseudo-Voigt functions. **Figure 4i** shows an example of the fitting for the Cu(200) peak of the nanoporous Cu dealloyed at 5 $^{o}$C, which was chosen for simplicity, because the strain resides along the lattice cell in the [100] direction. For tensile and compressive peak components, the amount of Q shifted from their center bulk peak position can be used to respectively calculate the amplitude of tensile and compressive strains *via* Bragg's Law. As mentioned above, the instrumental error has only negligible influence on the peak asymmetry and width (i.e. < 0.0001$^{o}$, **Figure S17**). Therefore, the FWHM of each 3 fitting components mainly represents the crystallite size. It can also be caused by the deviations of the strain measured. Due to the complex and irregular structures of these dealloyed nanoporous Cu materials, each surface atom might deviate from its idea lattice positions differently. Tracking the positions of each single atom on such a large scale is impossible. This method only provides a way to estimate the amplitude of averaged tensile and compressive strains. It is also worth mentioning that atoms that contribute to either tensile or compressive strain can be treated as surface defects, and might show different electronic properties compared with the pristine terrace surface. To quantify the amount of surface defects and differentiate from the traditional concept of 'microstrain', we introduce here a more specific term, Ligament Surface Strain, which is equal to the sum of the absolute values of the tensile and compressive strain values obtained from the fitting method described above.

Results from the fittings are shown in **Figure 4j** and **Table S3**. Ligament surface strain indicates the density of surface defects on Cu ligaments. During dealloying, Cu atoms tend to restructure to reduce these high-surface-energy surface defect sites *via* the surface diffusion.[69, 105, 116] According to the Arrhenius equation, Equation 2, the surface diffusion coefficient ($D_s$) increases exponentially with temperature (T). Therefore, the overall ligament surface strain decreases with the increase in dealloying temperatures, due to strain relaxation at higher temperatures (**Figure 4j**). Remarkably, no obvious tensile strain component was found for the nanoporous Cu dealloyed at 40 $^{o}$C. Only two pseudo-Voigt functions, corresponding to bulk and compressive components, can fit its (200) diffraction peak very well (**Figure S20**). Similarly, for the 60 $^{o}$C dealloyed nanoporous Cu, no obvious asymmetry of the (200) diffraction peak can be observed, of which the FWHM is similar to that of the $LaB_6$ standard diffraction peak at a similar Q, indicating the slight peak broadening is caused mainly by the instrumental error.

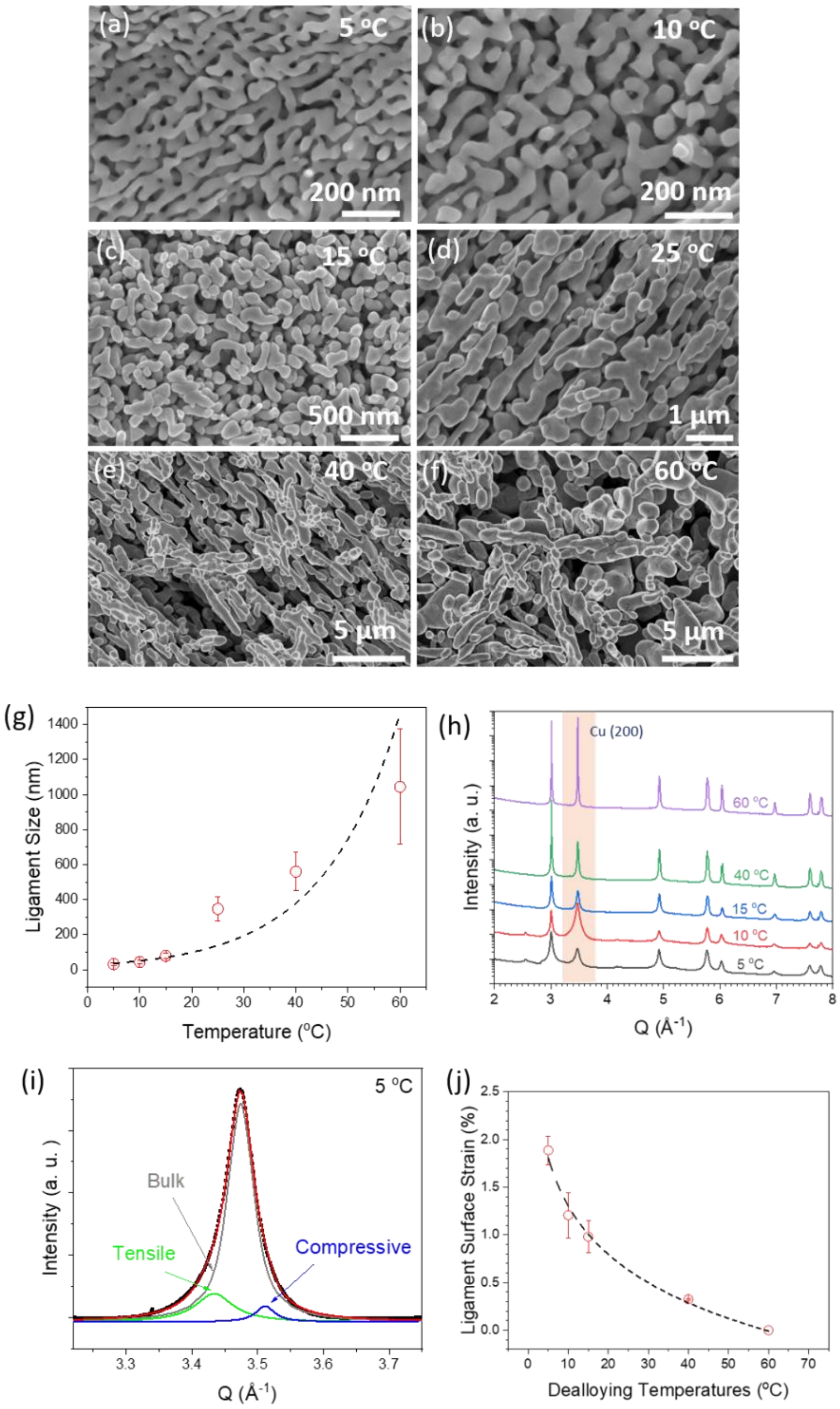

**Figure 4.** (a) – (f) SEM images of the nanoporous Cu dealloyed in 5 M $H_3PO_4$ for 24 hours under different temperatures. (g) Plot of ligament sizes of NPC against dealloying temperature. (h) Synchrotron XRD patterns of nanoporous Cu dealloyed in 5 M $H_3PO_4$ for 24 hours under different temperatures. (i) Close up and fitting of the (200) diffraction peak of the nanoporous Cu dealloyed at 5 $^oC$. Black dotted line is the diffraction pattern after the baseline subtraction by polynomial equation. Grey, green, and blue solid lines are the pseudo-Voigt fittings to the data, and red solid line is the summary of all 3 fitting components, representing the quality of fittings. (j) Plot of ligament surface strain values on Cu(200) facet against the dealloying temperature. Error bar was obtained by repeating the fitting for three times.

### 2.3. $CO_2$ Reduction on Dealloyed Cu

The capability of precisely tuning the nanostructure and surface strain of Cu-based materials *via* temperature controlled dealloying provides new strategies for electrochemical $CO_{(2)}$ reduction catalyst design. Electrochemical $CO_2$RR performance of these dealloyed nanoporous Cu materials were first studied using the conventional H-cell setup. The same 'sandwich' H-cell as Kuhl *et al* was used in our studies (**Figure S21**).[117] Our H-cell setup was calibrated and benchmarked with the electropolished polycrystalline Cu (**Figure S22-24**). Details of the cell configuration and analysis procedures are provided in the Supporting Information.

The $CO_2$RR properties on each dealloyed nanoporous Cu samples were tested at -1.17 V and -1.35 V *vs.* RHE for 25 min. Compared with the more conventional 1 h chronoamperometry test,[117-119] the shorter potential hold time used in this study leads to a much better-preserved nanoporous structures (**Figure S25**). However, we acknowledge that there is still some extent of restructuring happening. Numerous papers have looked at the cathodic corrosion of Cu under the $CO_2$ reduction conditions.[120-127] Raaijman *et al* discovered that pure Cu itself should be stable under the cathodic $CO_2$RR conditions (at -0.75 V *vs.* RHE, in 0.5 M $KHCO_3$).[120] It is the initial brief exposure to the $O_2$ or oxidizing conditions, even for only 1 min, would cause Cu restructuring during $CO_2$RR, due to the stripping away of O atoms.[120] However, recently, Amirbeigiarab and co-workers also proposed a CO-induced Cu cluster formation restructuring mechanism.[121] There are still a lot of debates on the mechanisms behind this,[121-127] and practically speaking, to completely blocking $O_2$ exposure of synthesized Cu catalysts is difficult. Therefore, solving this Cu-based catalysts restructuring under $CO_2$RR conditions issue still remains challenging, and is beyond the scope of the present work.

Compared with electropolished polycrystalline Cu, the dealloyed nanoporous Cu samples showed lower overall $CO_2$RR activity and selectivity (**Figure S23-S24, S26- S27**). This is likely due to the deliberately low catalyst loading used here (150 µg $cm^{-2}$), which was chosen to minimize $CO_2$ mass transport limitations. As a result, the nanoporous Cu coverage on the graphite foil was only about 20% (**Figure S28**), which limited the total catalytic current. In addition, the dealloyed nanoporous Cu samples displayed similar double-layer capacitance values (**Figure S29-S30**), indicating comparable electrochemical active surface area (ECSA) under the present conditions. We therefore use geometric area normalized partial current density as a practical measure of intrinsic activity for comparing the different nanoporous Cu samples.

**Figure 5a** summarizes the Pearson correlations [128] between the structural characteristics of nanoporous Cu and the partial current densities of different $CO_2$RR products. No clear trends were observed with Zn content, ECSA, or binder content. The weak dependence on ECSA is consistent with the similar double-layer capacitance values measured for all dealloyed nanoporous Cu samples at this low mass loading. Binder volumetric percentage factor (= Sustainion Solution Volume / (Ligament Size)$^3$) was used to estimate the influence of Sustainion ionomer added to the nanoporous Cu catalyst ink here. However, the effects of ionomer are likely more complex than simple surface blocking, and may also influence the adsorption energies of reaction intermediates on the Cu surface.[129] Further studies on the effect of Sustainion ionomer are currently in progress.

Interestingly, our statistical analysis (**Figure 5a**) also showed a strong correlation between CO partial current density and descriptors linked to dealloying temperature, including ligament size and ligament surface strain. Because increasing dealloying temperature accelerates Cu surface diffusion, leading to larger ligaments and relaxation of ligament surface strain, these variables are not independent. Among them, ligament surface strain is the most direct descriptor of the local structural state of the ligament surface and therefore provides the clearest physical basis for interpreting the activity trend.

Based on the work from Fujita *et al* on dealloyed AgAu,[33] and the ligament surface strain that we measure, we conjecture that a pronounced surface curvature will be introduced to the material during the dealloying process (**Figure 4**), resulting in a substantial presence of surface defect sites. In essence, ligament surface strain can be viewed as a structural descriptor for the abundance of such surface defect sites. **Figure 5b** shows a clear linear relationship between the geometric surface area normalized partial current density toward CO and the ligament surface strain of the nanoporous Cu samples. This result indicates that the intrinsic activity toward CO production increases with increasing density of surface defect sites. Because *CO is the key intermediate in Cu-catalyzed $CO_2$ reduction,[5, 130] the observed trend suggests that these strained, defect-rich ligament surfaces are especially important for the initial *CO forming step. This interpretation is also consistent with many previous studies showing that Cu steps, kinks, and other non-terrace surface sites are more active than flat terraces for $CO_2$ reduction.[13, 22-23, 26, 57, 131-132] However, for the more reduced $CH_4$ and other $C_{2+}$ products, no similarly robust trend was observed, presumably because these further reduction pathways involve additional elementary steps,[130, 133-135] and are more sensitive to factors such as local CO coverage, local pH, electrolyte effects, and mass transport.[79, 136-137]

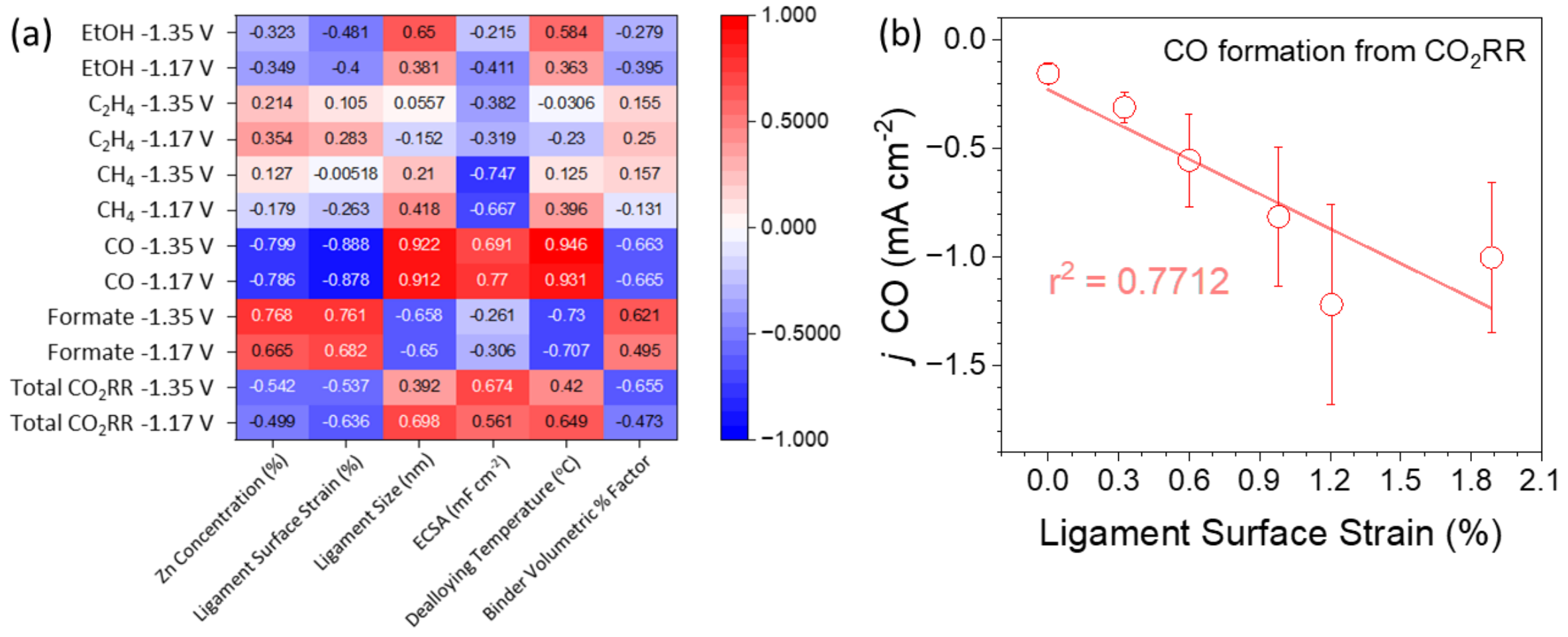

**Figure 5.** (a) Heatmap of Pearson correlation between the physical characteristic analysis of nanoporous Cu samples (x-axis) and their geometric surface area normalized current densities towards different $CO_2$RR products at -1.17 V and -1.35 V *vs.* RHE (y-axis). Sustainion Solution Volume / (Ligament Size)$^3$ was used to calculate the Binder Volumetric % Factor. (b) The plot of geometric surface area normalized partial current density of CO formation from $CO_2$RR test at -1.17 V *vs.* RHE, against the ligament surface strain values of nanoporous Cu samples. $r^2$ value also represents the quality of the linear relationship obtained. All tests were conducted in 0.1 M $KHCO_3$, H-cell setup. The error bar was obtained from the repeating experiments, conducted on differently prepared and drop-casted working electrodes.

## 2.4. CO Reduction on Dealloyed Cu

To more directly probe the influence of ligament surface strain on $C_{2+}$ formation, CO reduction measurements were conducted using on-chip electrochemistry mass spectrometry (EC-MS). In this setup (**Figure 6a**), the working electrode was positioned 100 μm above a hydrophobic porous chip and separated from it by a thin layer of 0.1 M KOH electrolyte. CO was purged through microchannels inside the chip, and volatile products formed on the working electrode diffused through the thin electrolyte layer within around 100 ms, after which they were captured by the vacuum beneath the chip and detected by the mass spectrometer. More detailed descriptions of this setup have been reported by Scott *et al*,[138] Qiao *et al*,[139] and Hochfilzer *et al*.[140] We calibrated and benchmarked this CORR setup using electropolished polycrystalline Cu (**Figure S31**). Details for the EC-MS calibration procedure can be found in Supplementary Information, and **Figure S32-S34**. As shown in **Figure S31c**, the overall CORR partial current densities obtained in our setup agree well with bulk electrolysis results reported by Hori *et al* [141] and Bertheussen *et al* [142] using conventional H-cell with gas chromatography to measure the products, as well as Qiao *et al* [139] on a similar EC-MS setup. This agreement supports the reliability of the present CORR measurements. Dedicated CORR measurements are also mechanistically informative, because Gao *et al* [143] have shown that the $CO_2$-to-CO and CO-to-$C_{2+}$ steps on Cu can involve distinct classes of active sites, making it useful to isolate the downstream reduction of *CO from the more complex overall $CO_2$RR pathway.

Unlike the $CO_2$RR test, post-mortem SEM suggested only negligible restructuring on nanoporous Cu samples after CORR (**Figure S35**). This is presumably because of the shorter measurement duration. Additionally, our extended-term stability test (**Figure 6d**) shows that the partial current densities towards CORR only starts to decrease after 2 hours of electrolysis, which is much longer than the ~20 min measurement time used for each nanoporous Cu sample. These observations could suggest that the ligament surface strain of the nanoporous Cu catalysts remained essentially unchanged during the CORR experiments, allowing more direct comparison between catalytic behavior and the pre-characterized structural descriptors. This interpretation is also consistent with the recent study from Wang *et al* [19], which reported that Cu nanocrystal catalysts did not show evident morphological reconstruction under CORR conditions, in contrast to the more severe restructuring often observed during $CO_2$RR.

**Figure 6b-c** compared the selectivity towards $C_2H_4$ (green) and total CORR products (black) for electropolished polycrystalline Cu and nanoporous Cu materials dealloyed at 5, 10, 15, 25, and 40 $^oC$. A volcano-like relationship, with ligament surface strain was observed. The nanoporous Cu dealloyed at 15 $^oC$, with the ligament surface strain of 0.9%, showed both the highest CORR and $C_2H_4$ faradic efficiency (FE) of 60%, and 21%, respectively, at -0.65 V *vs.* RHE in 0.1 M KOH. Under the same conditions, polycrystalline Cu showed total CORR and $C_2H_4$ FE of only around 45%, and 8%, respectively. All nanoporous Cu samples displayed substantially higher CORR selectivity than polycrystalline Cu (**Figure 6c, S36,** and **Table S4**), compared with the polycrystalline Cu, under the same reaction conditions. Moreover, despite the extremely low catalyst loading used for nanoporous Cu (4.3 μg $cm^{-2}$) (**Figure S35**), the nanoporous Cu dealloyed at 15 $^oC$ still showed a $C_2H_4$ partial current density comparable to that of polycrystalline Cu at -0.65 V *vs.* RHE, and 25 $^oC$ dealloyed nanoporous Cu showed an even higher $CH_4$ partial current density at 0.7 V *vs.* RHE (**Table S4**, and **Figure S37**). Comparison with selected literature reports further indicates that the intrinsic CORR activity of the dealloyed nanoporous Cu catalysts is comparable, albeit slightly lower than the highest performing Cu-based catalysts in the literature (**Figure S38**).

Pearson correlation analysis [128] was also conducted for CORR (**Figure S39**), where the dataset did not show any strong linear correlation between selectivity and the structural descriptors considered. This is consistent with the volcano-like behavior in **Figure 6b-c**, indicating that CORR selectivity in this nanoporous Cu series is not governed by a simple monotonic dependence on any single parameters. The volcano-like relationships further suggest that an intermediate level of ligament surface strain is most favorable for CORR selectivity in this dealloyed nanoporous Cu system. If the strain is too low, as in the more coarsened ligaments obtained at higher dealloying temperatures, the density of surface defect sites may be insufficient to promote efficient CO reduction beyond that of polycrystalline Cu. Conversely, if the strain is too high, as in the finest ligaments obtained at the lowest dealloying temperatures, the binding of key intermediates or the local arrangement of active sites may become less optimal for selective $C_{2+}$ formation. Our results are consistent with the recent work from Wang *et al* [19] on strained Cu nanocrystal catalysts for electrochemical CO reduction, which showed that strain can be served as a physically meaningful descriptor for $CO_{(2)}$RR selectivity on Cu catalysts.

However, we acknowledge that errors in both partial current densities for $CO_2$RR (**Figure 5b**) and FE for CORR (**Figure 6b-c**) are relatively large. The standard deviations from at least three independent experiments were used as the error bars in **Figure 5b**, and **6b-c**. Such discrepancy might be caused by factors including the relatively inhomogeneous structures of nanoporous Cu samples due to the dealloying nature, local pH effect, and bubble formation during the electrocatalysis. The measurements were repeated on separately drop-casted nanoporous Cu working electrodes, and relatively high currents were also obtained on these nanoporous structures.[144] Despite this variability, the overall trend remains clear: dealloyed nanoporous Cu catalysts outperform polycrystalline Cu in CORR selectivity under the present conditions; there is also a strong correlation between the $CO_2$RR intrinsic activity towards CO and ligament surface strain. Overall, these results show that dealloyed nanoporous Cu is a promising Cu catalyst platform for CO reduction and that ligament surface strain provides a useful descriptor for rationalizing both $CO_2$RR and CORR properties trends of Cu-based catalysts. More broadly, most recent efforts to introduce strain or defect-rich structures into Cu electrocatalysts have relied on wet-chemically synthesized nanoparticles or epitaxial thin films.[13, 19, 145-151] By contrast, the present results demonstrate that chemical dealloying offers a simple and effective route to generate strain-engineered, defect-rich realistic Cu architectures that are highly active for CORR. The combined structural analysis and catalytic measurements presented here therefore provide direct experimental support for the importance of surface defect sites and strain in Cu-based $CO_{(2)}$ reduction catalysis.

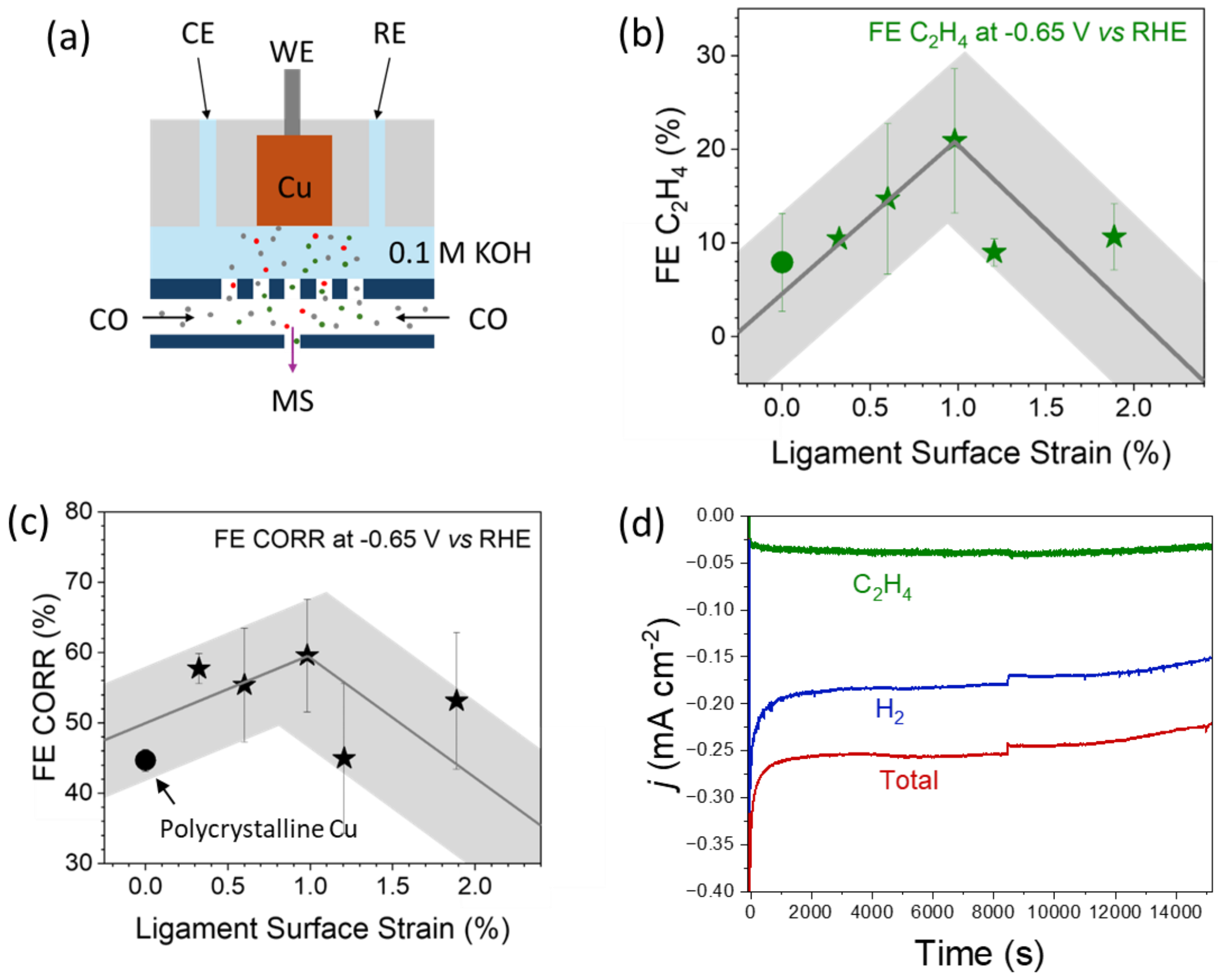

**Figure 6.** (a) Schematic illustration on the on-chip EC-MS setup used in this work. (b) FE of $C_2H_4$ (green) on electropolished polycrystalline Cu (circle) and dealloyed Cu (star). (c) FE of CORR (black) on electropolished polycrystalline Cu (circle) and dealloyed Cu (star). (d) Extended-term CORR stability test of the nanoporous Cu sample dealloyed at 15 °C. All measurements were conduction in 0.1 M KOH, EC-MS setup.

## 3. Conclusion

In this study, nanoporous Cu materials with homogeneous structures and tuneable ligament sizes ranging between tens of nanometres to a few micrometres were synthesized through chemical dealloying of a Zn-rich ε-brass alloy ($Cu_{20}Zn_{80}$, atomic ratio) in 5 M $H_3PO_4$. The intermediate γ-phase was observed, and phase changes from ε- to γ-, and finally to α-brass (mainly Cu) was confirmed by monitoring the dealloying process *via in situ* synchrotron XRD. Cryo-APT studies on the partially dealloyed ε-brass samples further revealed a complex dealloying process, where the mixed crystal structures can exist within a single nano-ligament. The morphology studies with SEM suggested Ostwald ripening to be the dominant coarsening mechanism. Furthermore, an analytical method was developed to account for the asymmetry of the synchrotron XRD peaks and to estimate the ligament surface strain. This parameter can be used as a practical structural descriptor for the density of surface defect sites in dealloyed nanoporous Cu. The ligament surface strain was found to decrease with increasing of dealloying temperatures, caused by the relaxation of surface energy.  A linear relationship between the intrinsic activity toward CO formation during $CO_2$ reduction and ligament surface strain was observed, indicating that Cu surface defect sites play a key role in the CO forming step. CO reduction measurements using EC-MS further showed that these nanoporous Cu samples are promising CO reduction catalysts. A volcano-like relationship and an optimal ligament surface strain (0.981% in the nanoporous Cu sample dealloyed at 15 °C) was found to promote CO reduction, providing additional insight into the factors that govern $CO_{(2)}$RR and offering an alternative strategy for engineering Cu-based catalysts. This work also demonstrates how the combined use of synchrotron X-ray diffraction and EC-MS can facilitate more efficient and accurate optimization of Cu-based catalysts for electrochemical $CO_{(2)}$ reduction.

## 4. Experimental Section

*Dealloying:* 85% $H_3PO_4$ (ACS reagent, Merck), and 37% HCl (ACS reagent, Merck) were used to prepare the chemical dealloying solution mentioned in the paper. The 1 mm thick brass precursor plate was ultrasonicated in acetone (≥99%, TECHNICAL, VWR), ethanol (≥99.8%, AnalaR NORMAPUR® ACS, Reag. Ph. Eur. analytical reagent, VWR), and deionized water (DI water, Millipore, 18.2 MΩ cm) for 10 min respectively, followed by drying at 70 °C for 15 min. Further details of the alloy precursor fabrication can be found in SI. The dealloying electrolyte was cooled or heated in a jacket beaker with a water bath to the desired temperature. The cleaned brass plate was then immersed into the electrolyte for 24 h. Apart from the 2.5 M HCl + 2.5 M $H_3PO_4$, no obvious $H_2$ bubbles can be observed after the 24 h for all other etchants, indicating the finish of the dealloying process. However, 24 h is not enough for the enough for the dealloying process to complete for the 2.5 M HCl + 2.5 M $H_3PO_4$ etchant. The electrolyte was also taken during the dealloying process, and measured by the Inductively Coupled Plasma

Optical Emission Spectroscopy (ICP-OES, iCAP 6000 series, Thermo Scientific) to track the composition change of the brass plate. After dealloying, the nanoporous Cu was rinsed with DI water, and ethanol for at least 10 times to clean the residual acids, and dried in the $N_2$ flow. The nanoporous Cu samples were stored in the $N_2$ filled vial and kept in the freezer (-20 $^oC$) to avoid further oxidation and coarsening.

*Morphology and Composition Characterization:* nanoporous Cu samples were dissolved in 10% $HNO_3$ (diluted from the 68% $HNO_3$, TECHNICAL, VWR), and measured by the Inductively Coupled Plasma Mass Spectrometry (ICP-MS, Agilent 8900 triple quadrupole), for the determination of the residual Zn amount after the dealloying. The machine was calibrated by the Cu (TraceCERT®, 10 g $L^{-1}$ Cu in nitric acid (nominal concentration)) and Zn standards (TraceCERT®, 10 g $L^{-1}$ Zn in nitric acid (nominal concentration)) at 0, 0.1, 1, 10, 100, 200, 1000, 2000 ppb. The Scanning Electron Microscopy (SEM) images of nanoporous Cu samples were taken on the Zeiss Auriga machine, with the accelerating voltage of 5 kV. To better visualise the 5$^o$C dealloyed nanoporous Cu samples, which has the ligament size less than 50 nm, TEM images were taken on a JEOL STEM/TEM 2100Plus machine. The acceleration energy of electrons was set at 200 kV.

*Synchrotron X-Ray Diffraction (XRD):* The synchrotron XRD measurement was conducted at ID10 beamline, European Synchrotron Radiation Facility (ESRF). 22.5 keV x-ray with the vertical and horizontal beam size of 8.8 µm and 19.6 µm, respectively, was used as the incident beam. Eiger 4M detector, with the pixel size of 75 µm, was used for all measurements. The as-collected two-dimensional diffraction patterns were converted to one- dimensional plots for the further analysis, using pyFAI python package.[152] For the *in situ* measurements to detect the phase change during the dealloying process, the brass precursor plates were further polished down to 20 µm thick, allowing the penetration of the x-ray. The brass precursor film was mounted onto a piece of Kapton tape, acting as the one side of the *in situ* cell. The other side of the x-ray window of the *in situ* cell was another piece of bare Kapton tape. Transmission mode diffraction was conducted on all in situ samples. The dealloying electrolyte (5 M $H_3PO_4$) was cooled or heated in a separated water bath and circulated into the cell via a peristaltic pump, with the flow rate of 10 mL $min^{-1}$. The temperature gradient due to the long electrolyte tube was also measured. The *in situ* cell was designed to have a thin electrolyte layer (1 mm) between the x-ray windows, minimizing both the attenuation and scattering of the x-ray caused by the aqueous electrolyte. The schematic illustration of the *in situ* cell can be found in **Figure S2**. For the *ex situ* measurements to detect the ligament surface strain, nanoporous Cu powder samples were mounted between two pieces of Kapton tapes. Transmission mode diffraction was also performed. The diffraction patterns of $LaB_6$ standard were also collected under the same conditions as both the *in situ* and *ex situ* powder samples, to calibrate the errors caused by the energy, detector and sample positions *etc*.

*Williamson-Hall Plot:* Using Williamson-Hall equation to plot $\beta cos\theta$ against $sin\theta$ of the XRD peaks is one of the most common methods for calculating the microstrain, excluding the crystallite size effect. Williamson-Hall equation can be written as:

$$(\frac{\beta cos\theta}{\lambda})^D = \varepsilon^D \cdot (\frac{4sin\theta}{\lambda})^D + (\frac{K}{L})^D \quad (3)$$

where $D$ is the transposed convolution coefficient. $D = 2$, when the diffraction peak shape is more close to Gaussian function, and $D = 1$ when the diffraction peak is more close to a Lorentzian function; $\beta$ is FWHM of the diffraction peak in radians; $\theta$ is the diffraction angle; $\lambda$ is the wavelength of the incident X-ray; $K$ is Scherrer constant, which can be treated approximately as unity here; $L$ is the crystallite size; and $\varepsilon$ is the microstrain.[115] The results were plotted in **Figure S18**, where poor linearity was obtained.

*Cryo-Atom Probe Tomography:* Atom probe tomography is very useful in obtaining chemical analysis at the near atomic scale. However, conventional Ultra-high vacuum-cryogenic sample transfer is required to keep the sample frozen in its native state and to limit any frosting from residual water vapor. This transfer mechanism is detailed in previous cryo-APT studies, pioneered in the Laplace project from the Max-Planck-Institute for Sustainable Materials, in Dusseldorf.[86] A transfer suitcase maintained at ultra-high vacuum (10-11 mbar) and cryogenic temperatures ($\approx$ 90 K) from Ferrovac was used to transport the frozen samples between an inert nitrogen glovebox, a Thermo Fisher Hydra/Helios dual beam PFIB (Xe plasma source) fitted with a cryo-stage and Cameca LEAP 5000 XR (Cameca Instruments). The dealloyed samples immersed in the DI water were rapidly frozen inside the glovebox by plunging into liquid nitrogen. The frozen samples were then transferred from the glovebox into the transfer suitcase *via* a load lock which reached a vacuum of $10^{-6}$ mBar. APT tips of water-nanoporous Cu were created in the Thermo Fisher Hydra PFIB (Plasma Focussed Ion Beam) following the same procedure as reported by El-Zoka *et al* [86] and Stender *et al*.[87] Ion beam currents ranged from 1.3 µA to 0.1 nA with a voltage of 30 kV. APT analysis was carried out in a Cameca LEAP 5000 XR (Cameca Instruments) using Laser pulsing mode with a pulse of 60 pJ and voltage mode pulse rate of 50 kHz. The detection rate was varied between 0.1-1% (0.001-0.01 ions per pulse). The stage temperature was kept constant at 50 K for all experiments. Reconstructions were carried out in AP Suite 6.1, IVAS 6.1.3.42.

*Working Electrode Preparation:* For the benchmarking experiments, polycrystalline Cu (0.5 mm thick, Alfa Aesar Supratronic®, purity 99.9999%) and polycrystalline Cu rotating disk electrode (RDE) insert (5.0 mm OD x 4.0 mm thick, mirror polished, PINE) were used as the working electrode for $CO_2RR$ and CORR, respectively. Both polycrystalline Cu electrodes were electropolished at +2.1 V for 5 min in 85% $H_3PO_4$, followed by rinsing by the DI water, prior to the test. Afterwards, they were installed into the cells and immersed into the electrolyte immediately, to minimize the time exposed to air, oxidizing the surface. For nanoporous Cu samples, 8.5 mg of the specimen, and 280 μL of Sustainion (XA-9 Alkaline Ionomer 5% in ethanol, Dioxide Materials) were ultrasonicated in 720 μL of isopropanol (≥99.7%, AnalaR NORMAPUR® ACS, Reag. Ph. Eur. analytical reagent) for 1 h to form the catalyst ink. Ice was added to the ultrasonication bath to avoid the temperature increasing and the coarsening of the nanoporous Cu catalysts. For $CO_2RR$ measurement, 200 μL of the catalyst ink was drop-casted onto the 2.7 cm × 4.2 cm graphite foil (0.5 mm thick, 99.8%, metals basis, Alfa Aesar), giving the mass loading of 150 μg $cm^{-2}$. For CORR test, the above-mentioned ink was diluted with the isopropanol and DI water mixture (with volume ratio of 4 : 1). Certain amount of the diluted catalyst ink, depending on the concentration, was drop-casted onto the glassy carbon disk (SIGRADUR G, 5.0 mm OD x 4.0 mm thick, 0.196 $cm^2$), to reach the mass loading of the nanoporous Cu 4.3 μg $cm^{-2}$. For both $CO_2RR$ and CORR test, the as drop-casted electrodes were dried in air for at least 1 h before using.

*$CO_2$ and CO Reduction:* The Pt mesh and Ag/AgCl electrode were used as the counter electrode and reference electrode, respectively, for both $CO_2$ and CO reduction. The Ag/AgCl reference electrode was calibrated with the shift from the RHE scale as:

$$V_{RHE} = V_{Ag/AgCl} + 0.241 + 0.059 \times pH \qquad (4)$$

0.1 M KOH (pH = 13), prepared from potassium hydroxide hydrate (99.995%, Merck), was used as the electrolyte for CO reduction test. For $CO_2RR$, 0.1 M $KHCO_3$ (pH = 6.8) was used as electrolyte, prepared by purging the same KOH electrolyte with $CO_2$ gas (99.999%, Research Grade, BOC) until saturation. $CO_2$ and CO reduction measurements were conducted on the H-cell gas chromatography (GC) and EC-MS setup, respectively. Further details can be found in Supplementary Information.

**Acknowledgements**

The authors acknowledge the European Synchrotron Facility for provision of synchrotron radiation facilities and ID 10 diffractometer.

The authors would like to thank Prof Christopher Gourlay, Dr Liuqing Peng, and Dr Yi Cui from the Engineering Alloy Group, Department of Materials, Imperial College London for their help on the brass precursor fabrication.

The authors would like to thank Peter Haycock from the Department of Chemistry, Imperial College London for helping us run the aqueous NMR measurements.

The authors also would like to thank Prof Brain Seger, and Dr Yu Qiao from SURFCAT, Department of Physics, Technical University of Denmark, for their help and suggestions on the CO reduction benchmarking on EC-MS.

We also acknowledge the support from Dr Paula Sebastian Pascual for advice on the electropolishing method for polycrystalline Cu.

Authors would like to thank Prof Finn Giuliani, Head Imperial Cryomicroscopy Centre, funded by EPSRC Grant EP/V007661/1.

A.A. El-Zoka acknowledges support from the Department of Materials, at Imperial College London.

Benjamin Bowers thanks UKRI EPSRC (EP/2445673) for the funding.

Guangmeimei Yang thanks Imperial College London for the President’s scholarship.

**Data Availability Statement**

All raw data that support the findings of this study is available from the corresponding author upon reasonable request. The data that support the findings of this study are available in the SI of this article.

**Supporting Information**

Supporting Information is available from the Wiley Online Library or from the author.

# Supporting Information

Mechanistic Insights into Active Sites for Electrochemical $CO_2$ and CO Reduction over the Strain-Engineered Dealloyed Cu

Yuxiang Zhou, Ayman A. El-Zoka*, Oliver R. Waszkiewicz, Benjamin Bowers, Rose P. Oates, James Murawski, Anna Winiwarter, Guangmeimei Yang, Oleg Konovalov, Maciej Jankowski, Caiwu Liang, Paul-Benjamin Amar, Andreas G. Kafizas, Ludmilla Steier, Ifan E. L. Stephens* and Mary P. Ryan *

Yuxiang Zhou. Author 1, Department of Materials, Imperial College London, SW7 2AZ, UK

Ayman A. El-Zoka. Author 2 and Corresponding Author, Univ Rouen Normandie, CNRS, INSA Rouen Normandie, Groupe de Physique des Matériaux,
Rouen, France E-mail: ayman.el-zoka@univ-rouen.fr

Oliver R. Waszkiewicz. Author 3, Benjamin Bowers. Author 4, Rose P. Oates, Author 5, James Murawski. Author 6, Anna Winiwarter. Author 7, Guangmeimei Yang. Author 8, Caiwu Liang. Author 11, Paul-Benjamin Amar. Author 12, Department of Materials, Imperial College London, SW7 2AZ, UK

Andreas G. Kafizas. Author 13, Department of Chemistry, Imperial College London, Molecular Science Research Hub, W12 0BZ, UK

Oleg Konovalov. Author 9, Maciej Jankowski. Author 10, The European Synchrotron - ESRF, 71 Av. des Martyrs, 38000, Grenoble, France

Ludmilla Steier. Author 14, Department of Chemistry, Oxford University, 12 Mansfield Road, Oxford, OX1 3TA, UK

Ifan E. L. Stephens. Corresponding Author, Department of Materials, Imperial College London, SW7 2AZ, UK
E-mail: i.stephens@imperial.ac.uk

Mary P. Ryan. Corresponding Author, Department of Materials, Imperial College London, SW7 2AZ, UK
E-mail: m.p.ryan@imperial.ac.uk

***1. Support Experimental Details***

***Alloy Precursor Fabrication and Characterization***

*Pure Cu (purity, 99.997%, Advent) and Zn (purity, 99.99%, Advent) metal with the mass ratio of 1 : 4.12 were placed and encapsulated into an Ar filled quartz tube. The metals were melted at 850 °C for 3 days to fabricate the $Cu_{20}Zn_{80}$ (atomic ratio) alloy precursor. The brass ingot was then cut to 1 mm thick alloy plates by the low-speed diamond saw (IsoMet 1000 precision saw, Buehler). The composition and phase of the brass precursor was then verified using a PANalytical X-ray diffractometer with Cu Kα radiation (λ = 1.5406 Å). 2θ angle range, step size, and time per step were set to be 10° to 100°, 0.00418°, and 24.13 s, respectively.*

***Cryo-APT Sample Preparation***

*Dealloyed nanoporous Cu (NPC) samples were dispersed into isopropanol (IPA)* via *ultrasonication, followed by loading onto the pre-sharpened W needle* via *the electrophoresis deposition (**Figure S6a**). By placing the nanoparticles into a dielectric medium with an electric field, dipole moment would be generated on those nanoparticles. The unevenly distributed charges and the electric field could cause the motion of those nanoparticles, as shown by **Figure S6d**. Therefore, nanoparticles can be deposited and physiosorbed onto the electrode. **Figure S6e** shows the transient current curve for the electrophoresis deposition of NPC at -8 V, where a Pt coil was used as the counter electrode, holding around 100 μL of the NPC ink, due to the surface tension. The W needle was pre-sharpened by FIB, and placed inside the Pt coil, serving as the working electrode (**Figure S6a**). Chronoamperometry measurements at -8 V were conducted for around 400 s in total, with 3 gaps to allow the diffusion of NPC particles. The NPC dispersion ink was also refreshed in the middle. The current was fluctuated at around -3.2 μA. **Figure S6a** shows the SEM images of the W needle after the electrophoresis deposition. A massive amount of NPC particles was loaded and physiosorbed onto the electrode via the Van der Waals forces. The as prepared samples can then be submerged into DI water and placed under the vacuum to infiltrate the nano-pores, followed by dipped into liquid nitrogen ($LN_2$). As shown by **Figure S6b**, the nanoporous structures preserved from both the electrophoresis deposition and quenching in $LN_2$. NPC was then embedded into and mounted together by the ice. The cryo-needle would need to be sharpened again by FIB for the final APT measurement (**Figure S6c**).*

***$CO_2$ Reduction Test***

*All $CO_2$RR tests in this work were conducted in a PTFE sandwich H-cell, similar to the setup initially designed by Kuhl* et al *(**Figure S21**).[1] Working and counter electrodes were placed between the electrolyte chamber and cell wall. Al foil was used as the current collector. The Ag/AgCl reference electrode was connected to the catholyte chamber using a plastic capillary. The electrolyte volume for both anolyte and catholyte was 8 mL. During the measurement, anolyte was ventilated to air to prevent the $O_2$ accumulation and pressuring the cell, while catholyte was purged with either Ar (99.9999%, BOC) or $CO_2$ (99.999%, Research Grade, BOC) without exposed to the air. Ar with the flow rate of 20 mL $min^{-1}$ was purged into the electrolyte for 15 min first for deaeration, then switched to $CO_2$. Bio-Logic potentiostat (SP300) equipped with a 1A/48V booster was used for all the electrochemical tests. After 15 min of the electrolyte deaeration, a few CV scans between -1 V and -1.6 V* vs. *Ag/AgCl with the scan rate of 50 mV $s^{-1}$ were run in Ar until the current stabilized, to reduce the Cu surface oxidised layers. After the electrolyte was saturated with $CO_2$ again, impedance was performed at 0 V* vs. *RHE between 30 kHz and 300 Hz to determine the ohmic drop of the cell. For $CO_2$RR test, chronoamperometry measurement was performed on each working electrode for only 25 min, and each drop-casted working electrode was only tested for one* *potential to minimise the NPC catalysts restructuring. The Ohmic drop for each $CO_2$RR chronoamperometry measurement was compensated* in situ *with 85% on the Bio-Logic potentiostat. The rest 15% was corrected after experiments during the data analysis process, using the following equation:*

$$V_{actual} = V_{measured} - 15\% \cdot I \cdot R$$

*where $I$ is the averaged current measured, and $R$ is the ohmic drop measured by impedance. $CO_2$ gas was purged into the cell from the bottom with the flow rate of 20 mL min$^{-1}$. A mass flow controller (MFC) (Bronkhorst) was used to control the inlet flow rate. Another flow metre (Bronkhorst) was also connected to the cell outlet to monitor the actual flow rate changing during the electrochemical process. A one-way valve (water bubbler) was also installed as the exhaust of the whole setup to prevent the catholyte exposed to air.*

*All gas phased products were analysed with an on-line connected SRI8610C Gas Analyzer MG#5 gas chromatography (GC), equipped with a 0.5 metre Haysep D precolumn, a 2-metre molecular sieve 5A column, and a 2-metre Haysep D column. $H_2$ was quantified with the thermal conductivity detector (TCD), while all other carbon-containing gas molecules were measured by the flame ionisation detector (FID).* ***Figure S40*** *shows an example of the GC spectrum. Prior to the measurements, GC was calibrated, where standard gas mixtures (containing 500 ppm $N_2$ + 500 ppm $O_2$ + 1000 ppm $H_2$ + 1000 ppm CO + 1000 ppm $CH_4$ + 1000 ppm $C_2H_4$ in $CO_2$ as balancing gas, purchased from BOC) was mixed with another pure $CO_2$ gas. 2 MFCs were used to control the flow rate of the 2 gases separately, so that the concentration of the target gas in the final gas mixture can be changed, allowing the multi-points calibration (**Figure S41**). The partial current and FE of gas product* X *can be calculated by the following two equations:*

$$I_X = \frac{Area_X}{slope_X} \times \frac{z_X \cdot F}{V_m} \times Flow\ Rate$$

$$FE_X = \frac{I_X}{I_{Total}}$$

*where $Area_X$ is the GC peak area for gas* X *measured, $V_m$ is molar volume (22.4 L mol$^{-1}$), $z_X$ is the number of electron transferred for the generation of gas* X, F *is the Faraday constant (96485.3321 s A mol$^{-1}$), $I_{Total}$ is the total current measured, and $slope_X$ is from the calibration curve of gas* X *(**Figure S41**).*

*All liquid products were analysed by Bruker Av500 MHz nuclear magnetic resonance (NMR) spectrometer. The pre-saturation pulse sequence for water suppression technique was used for collecting the aqueous $^1$H NMR spectrum. Delay time and saturation delay was set to be 0.5 s and 5 s, respectively. 52 scans were run for each sample. Each NMR sample contains 250 µL of the electrolyte after 25 min of chronoamperometry test, 350 µL of $D_2O$, and 12.3 µL of the internal standards (10 mM phenol + 50 mM DMSO dissolved in $D_2O$).* ***Figure S42*** *shows an example of the NMR spectrum. Similar to the gas products, standard solutions with known concentrations of the possible $CO_2$RR products were prepared to obtain the calibration curves (Figure S36). The partial current of liquid product* X *can be calculated by:*

$$I_X = \frac{Area_X}{slope_X} \times \frac{z_X \cdot F}{t} \times V_{cell}$$

*where $Area_X$ is the NMR peak area for liquid* X *measured, $V_{cell}$ is electrolyte volume of the cell (8 mL for both anolyte and catholyte chamber), $z_X$ is the number of electron transferred for the generation of liquid* X, t *is the reaction time (25 min),* F *is the Faraday constant (96485.3321 s A mol$^{-1}$), and $slope_X$ is from the calibration curve of liquid* X *(**Figure S43**).*

***CO Reduction Test***

*All CO reduction measurements were conducted on the EC-MS. Details of the cell geometry can be found in the paper. After assembling the cell and injecting the electrolyte, He was purged for deaeration with the flow rate of 1 mL min$^{-1}$. Meanwhile, CV between -0.1 V and -0.6 V* vs. *RHE with the scan rate of 50 mV s$^{-1}$ was conducted until stabilization, to remove the Cu surface oxidation layer, which usually takes around 15 cycles. Afterwards, gas was switched to CO with the same flow rate (1 mL min$^{-1}$), and another 15 cycles of CV were taken in CO. The current detected for the CVs in CO was lower than that in He, because of the Cu surface poisoning by CO. Finally, the CO reduction measurement was performed in the chronoamperometry mode. Potentials between -0.3 V and -0.8 V* vs. *RHE were applied for 3 min each. The working electrode was rested at -0.05 V* vs. *RHE for 3 to 5 min, depending on the chip conditions, between each applied potential, to allow the mass spectrometry signal to reduce down to the background value. $H_2$, $CH_4$, and $C_2H_4$ were detected and quantified* via *the mass spectrometer (MS), through mass 2, 15, and 26, respectively. Taking $H_2$ as an example, its partial current and FE can be calculated by:*

$$I_{H_2} = \frac{n_{H_2} \cdot z_{H_2} \cdot F}{t}$$

$$FE_{H_2} = \frac{I_{H_2}}{I_{total}} \times 100\%$$

*where $n_{H_2}$ is the number of moles of $H_2$ generated in the time span of t, $z_{H_2}$ is the electron transferring number for $H_2$, $I_{total}$ is the overall current that measured, and* F *is the faraday constant. $n_{H_2}$ can be calculated by:*

$$n_{H_2} = \frac{MS_{H_2} \cdot t}{SF_{H_2}}$$

*where $MS_{H_2}$ is the MS signal intensity of $H_2$,* t *is the time span, and $SF_{H_2}$ is the sensitivity factor of $H_2$ (**Figure S32**). The sensitivity factor for each gas is the slope of the calibration curve (**Figure S32**). Prior to the measurements, MS was calibrated. For the $H_2$ calibration, electrochemical calibration was performed in the same setup, except the pristine glassy carbon was used as the working electrode. Different currents were applied to the working electrode, to allow the multi-points calibration. Both $CH_4$ and $C_2H_4$ were calibrated by the standard gas mixtures. 1000 ppm $CH_4$ in Ar was used for $CH_4$ calibration, while $C_2H_4$ was calibrated in the gas mixture containing 1000 ppm $C_2H_4$ + 1000 ppm $CO_2$ + 1000 ppm CO + 1000 ppm $H_2$ + 1000 ppm $O_2$ with Ar as the balancing gas. Only single-point calibration was done for $CH_4$ and $C_2H_4$, due to the fixed concentration of the standard gas mixtures. **Figure S32** shows the calibration curves for all the three gases. The EC-MS CO reduction results were analysed* via *the open-source NumPy [2] and IXDAT [3] python package (available from https://github.com/ixdat).*

***ECSA Measurement***

*Electrochemical active surface area (ECSA) for NPC was estimated by the double layer capacitance measurements. CVs at around 0 V* vs. *RHE were taken in Ar saturated 0.1 M $HClO_4$ (diluted from 70% $HClO_4$, 99.999% trace metals basis, Merck) at various scan rates. The same H-cell as $CO_2$RR test, and EC-MS setup as CORR test were used. Further details can be found in **Figure S29-S30**.*

## *2. Support Tables*

**Table S1**. State-of-art examples of the brass dealloying to produce NPC.

| Alloy Precursor | Dealloying Mode | Conditions | Morphology | Ref |
|---|---|---|---|---|
| Electrodeposited Cu-rich brass | Chemical Dealloying | 2 wt% NaOH for 15 h then 4 M HCl for 20 s | Pore size: 200 – 800 nm | [4] |
| $Cu_{20}Zn_{80}$ | Chemical Dealloying | 0.05 M HCl for 4 h at 50 °C | Ligament size and pore size: around 80 nm | [5] |
| $Cu_{63}Zn_{37}$ | Electrochemical Dealloying | Melted CsCl, KCl, and LiCl mixture | Pore size: 250 – 500 nm Porosity: 25 – 30% | [6] |
| α-CuZn | Chemical Dealloying | 20 wt% HCl | Pore size: 0.8 – 1 μm Ligament size: 1 μm | [7] |
| Electrodeposited $Cu_{64}Zn_{36}$, and $Cu_{37}Zn_{63}$ | Chemical Dealloying | 1 M HCl | No obvious sponge-like structures | [8] |
| Electrodeposited γ-CuZn | Chemical Dealloying | 0.25 M HCl for 6 h | No obvious sponge-like structures | [9] |
| | | 0.1 M NaOH for 17 h and 3 M HCl for 60 s | Pore and ligament size: around 100 nm | |
| | | 0.1 M NaOH for 17 h and 0.25 M HCl for 6 h | Pore size: 200 nm Ligament size: 500 nm | |
| $Cu_{70}Zn_{30}$ | Chemical Dealloying | 1 M NaOH for 96 h | Pore size: 100 nm Ligament size: 50 nm | [10] |
| $Cu_{30}Zn_{70}$ | Chemical Dealloying | 1 M $NH_4Cl$ + HCl | Pore size: 100 nm Ligament size: 30 nm | [11] |
| $Cu_{20}Zn_{80}$ | Chemical Dealloying | 15 M $H_3PO_4$ | Pore size: 10 nm Ligament size: 46 ± 9 nm | [12] |
| $Cu_{35}Zn_{65}$ | Electrochemical Dealloying | 5 wt% HCl | Pore size: 500 nm Ligament size: 200 nm | [13] |
| $Cu_{20}Zn_{80}$ | Chemical Dealloying | 85 wt% $H_3PO_4$ | Pore size: 50 nm Ligament size: 50 nm | |
| Electrodeposited Zn-rich brass | Electrochemical Dealloying | Zinc chloride-1-ethyl-3-methyl imidazolium chloride ionic liquid | Hierarchical porous structures Ligament size: 50 nm | [14] |
| Electrodeposited Zn-rich brass | Chemical Dealloying | NaOH | Pore size: 100 nm Ligament size: 50 nm | [15] |
| ε-brass | Chemical Dealloying | 1.3 M NaOH | Hierarchical porous structures | [16] |
| | | 0.1 M HCl | Pore size: 20 nm Ligament size: 20 nm | |

**Table S2**. Cu surface diffusion activation energy comparison.

| Conditions | T (K) | $E_a$ (kJ mol$^{-1}$) | $D_s$ (m$^2$ s$^{-1}$) | Ref |
|---|---|---|---|---|
| Inner layers of CuAl with 35 wt% Cu dealloyed in 10 wt% NaOH | 333 | 47.7 | $1.63\times10^{-18}$ | [17] |
| | 348 | | $3.54\times10^{-18}$ | |
| | 363 | | $6.83\times10^{-18}$ | |
| Out layers of CuAl with 35 wt% Cu dealloyed in 10 wt% NaOH | 333 | 145.3 | $5.41\times10^{-17}$ | |
| | 348 | | $4.36\times10^{-16}$ | |
| | 363 | | $4.27\times10^{-15}$ | |
| $Mn_{72}Cu_{28}$ dealloyed in 0.1 M HCl | 298.15 | N/A | $2.03\times10^{-18}$ | [18] |
| $Mn_{72}Cu_{27}Ni_1$ dealloyed in 0.1 M HCl | | | $5.83\times10^{-19}$ | |
| $Mn_{72}Cu_{25}Ni_3$ dealloyed in 0.1 M HCl | | | $1.91\times10^{-19}$ | |
| CuMn with 34.6wt% Mn chemically dealloyed in 0.1 M HCl | 293.15 | 36.6 | $3.84\times10^{-15}$ | [19] |
| | 363.15 | | $5.41\times10^{-14}$ | |
| CuMn with 34.6wt% Mn electrochemically dealloyed in 5% $HBF_4$ at 5.2 mA cm$^{-2}$ | 293.15 | 28.5 | $1\times10^{-9}$ | |
| CuMn with 34.6wt% Mn electrochemically dealloyed in 5% $HBF_4$ at 5.2 A cm$^{-2}$ | 293.15 | 17.1 | $1\times10^{-7}$ | |
| Cu free diffusion at Cu(100) facet | 298.15 | 26.78 | $1\times10^{-9}$ | [20] |
| $Cu_{20}Zn_{80}$ chemically dealloyed in $H_3PO_4$ | 278.15 | 208 | $7.62\times10^{-19}$ | This work |
| | 283.15 | | $2.77\times10^{-18}$ | |
| | 288.15 | | $2.79\times10^{-17}$ | |
| | 298.15 | | $1.13\times10^{-14}$ | |
| | 313.15 | | $8.05\times10^{-14}$ | |
| | 333.15 | | $1.04\times10^{-12}$ | |

**Table S3**. The ligament surface strain states along the lattice cell axis direction on dealloyed NPC samples.

| Dealloying Temperature ($^{o}$C) | Compressive Strain (%) | Tensile Strain (%) | Ligament Surface Strain (%) |
|---|---|---|---|
| 5 | -0.8 ± 0.2 | 1.1 ± 0.2 | 1.9 ± 0.3 |
| 10 | -0.6 ± 0.2 | 0.6 ± 0.2 | 1.2 ± 0.5 |
| 15 | -0.6 ± 0.3 | 0.3 ± 0.01 | 0.9 ± 0.3 |
| 40 | -0.32 ± 0.01 | 0 | 0.32 ± 0.01 |
| 60 | 0 | 0 | 0 |

**Table S4**. FE and partial current densities normalized by geometric surface area of $H_2$, $CH_4$, $C_2H_4$ and overall CORR for electropolished polycrystalline Cu, and dealloyed NPC materials at -0.60, -0.65 and -0.70 V *vs* RHE.

| Potential (V *vs* RHE) | | | Polycrystalline Cu | NPC – Dealloying Temperatures (°C) | | | | |
|---|---|---|---|---|---|---|---|---|
| | | | | 5 | 10 | 15 | 25 | 40 |
| -0.60 | FE (%) | $H_2$ | 51.63 | 47.25 | 48.41 | 43.58 | 46.04 | 42.61 |
| | | $CH_4$ | 0.80 | 0 | 0.09 | 0 | 0 | 0 |
| | | $C_2H_4$ | 9.85 | 5.57 | 4.28 | 12.39 | 9.53 | 5.15 |
| | | CORR | 48.37 | 52.75 | 51.59 | 56.42 | 53.96 | 57.39 |
| | -j (mA $cm^{-2}$) | $H_2$ | 0.519 | 0.034 | 0.048 | 0.052 | 0.027 | 0.041 |
| | | $CH_4$ | 0.004 | 0 | $2.7\times10^{-5}$ | 0 | 0 | 0 |
| | | $C_2H_4$ | 0.101 | 0.004 | 0.004 | 0.014 | 0.005 | 0.005 |
| | | CORR | 0.487 | 0.037 | 0.051 | 0.065 | 0.031 | 0.055 |
| -0.65 | FE (%) | $H_2$ | 55.31 | 46.88 | 55.05 | 40.42 | 44.61 | 42.29 |
| | | $CH_4$ | 1.89 | 0.52 | 0.41 | 1.33 | 0.64 | 0.95 |
| | | $C_2H_4$ | 7.95 | 10.63 | 9.00 | 20.90 | 14.69 | 10.42 |
| | | CORR | 44.69 | 53.12 | 44.95 | 59.58 | 55.39 | 57.71 |
| | -j (mA $cm^{-2}$) | $H_2$ | 0.780 | 0.069 | 0.109 | 0.105 | 0.054 | 0.085 |
| | | $CH_4$ | 0.016 | 0.001 | 0.001 | 0.004 | 0.001 | 0.002 |
| | | $C_2H_4$ | 0.122 | 0.016 | 0.018 | 0.049 | 0.017 | 0.021 |
| | | CORR | 0.624 | 0.074 | 0.090 | 0.144 | 0.067 | 0.116 |
| -0.70 | FE (%) | $H_2$ | 55.04 | 45.66 | 60.46 | 40.85 | 40.41 | 39.45 |
| | | $CH_4$ | 2.00 | 2.90 | 3.16 | 7.36 | 3.50 | 5.77 |
| | | $C_2H_4$ | 9.92 | 14.88 | 15.33 | 22.15 | 21.37 | 18.62 |
| | | CORR | 44.96 | 54.34 | 39.54 | 59.15 | 59.59 | 60.55 |
| | -j (mA $cm^{-2}$) | $H_2$ | 1.236 | 0.249 | 0.384 | 0.302 | 0.128 | 0.250 |
| | | $CH_4$ | 0.022 | 0.016 | 0.019 | 0.056 | 0.011 | 0.036 |
| | | $C_2H_4$ | 0.223 | 0.080 | 0.096 | 0.146 | 0.067 | 0.117 |
| | | CORR | 1.010 | 0.280 | 0.255 | 0.404 | 0.186 | 0.379 |

**Table S4**. Residual Zn mass percent of the NPC samples. Results were obtained by dissolving the NPC in 10% $HNO_3$ followed by dilution and measuring the metal concentrations *via* ICP-MS.

| Dealloying Temperature (ºC) | Mass % of the Residual Zn |
|---|---|
| 5 | 6.8% |
| 10 | 3.3% |
| 15 | 2.0% |
| 25 | 1.3% |

### *3. Supporting Figures*

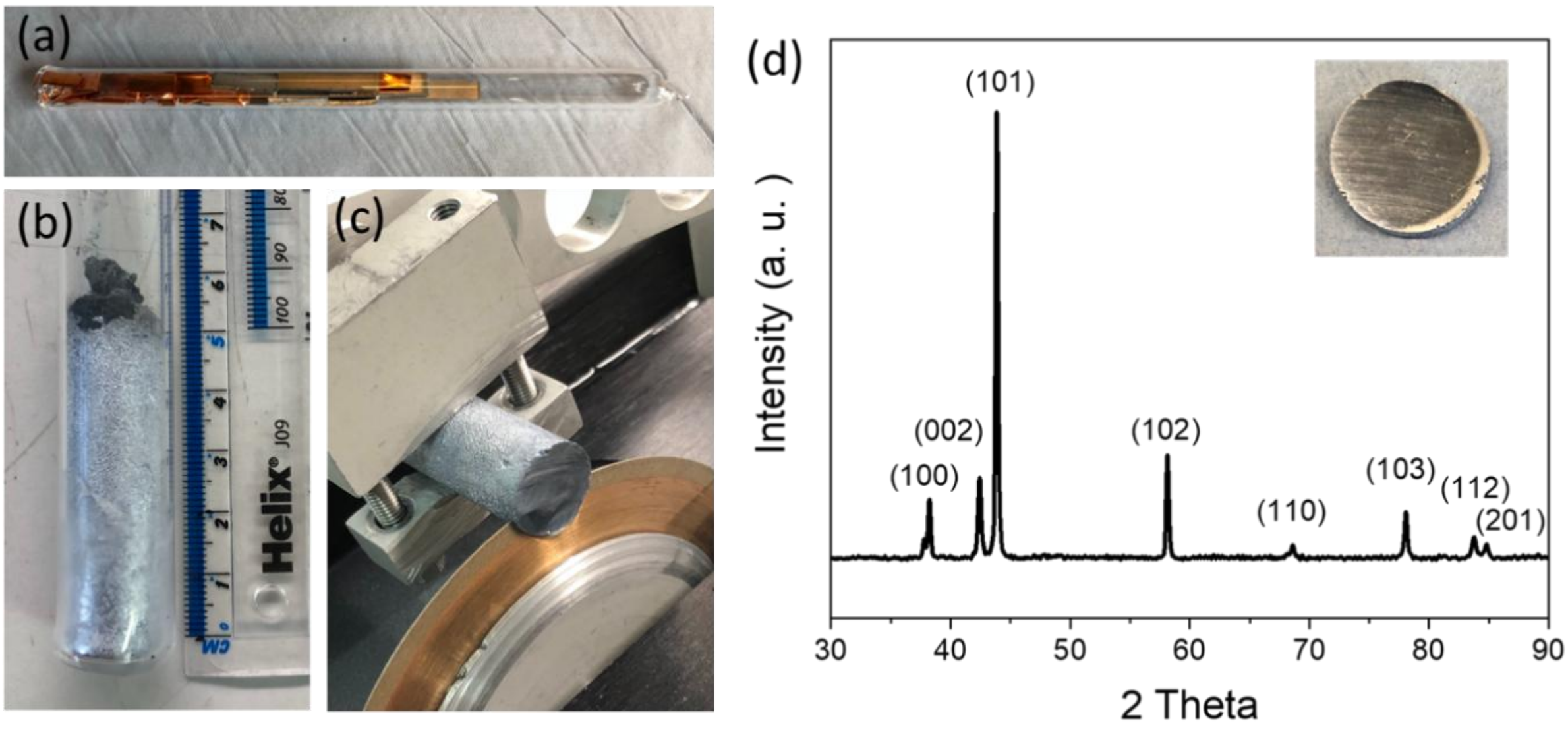

**Figure S1.** The pictures of (a) encapsulated pure Cu and Zn metal, (b) as prepared $Cu_{20}Zn_{80}$ brass precursor, and (c) cuttings of the brass precursor. (d) XRD patterns for the $Cu_{20}Zn_{80}$ brass precursor, showing it is a single ε-phased brass.

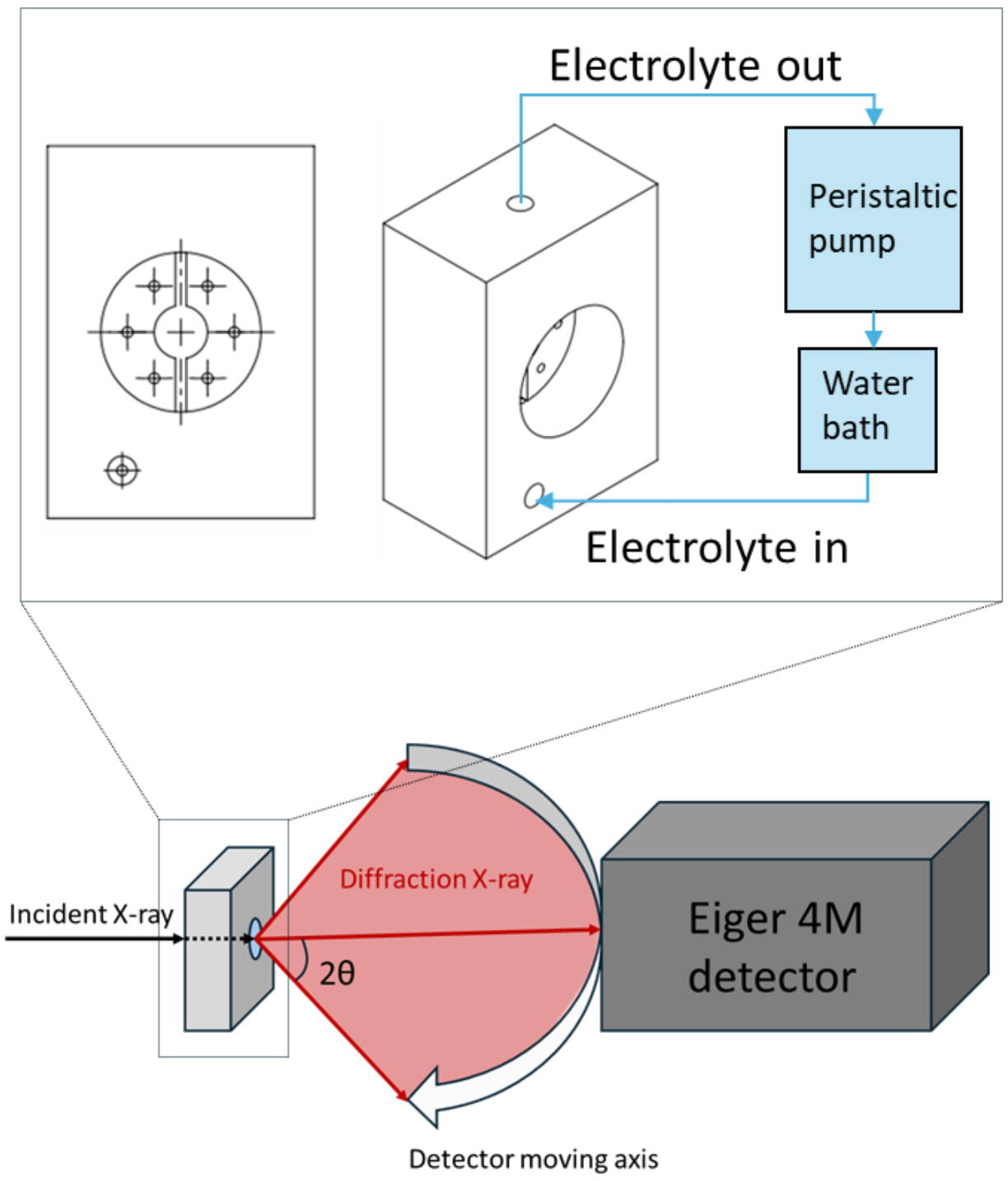

**Figure S2.** Schematic illustration of the *in situ* synchrotron X-ray diffraction setup.

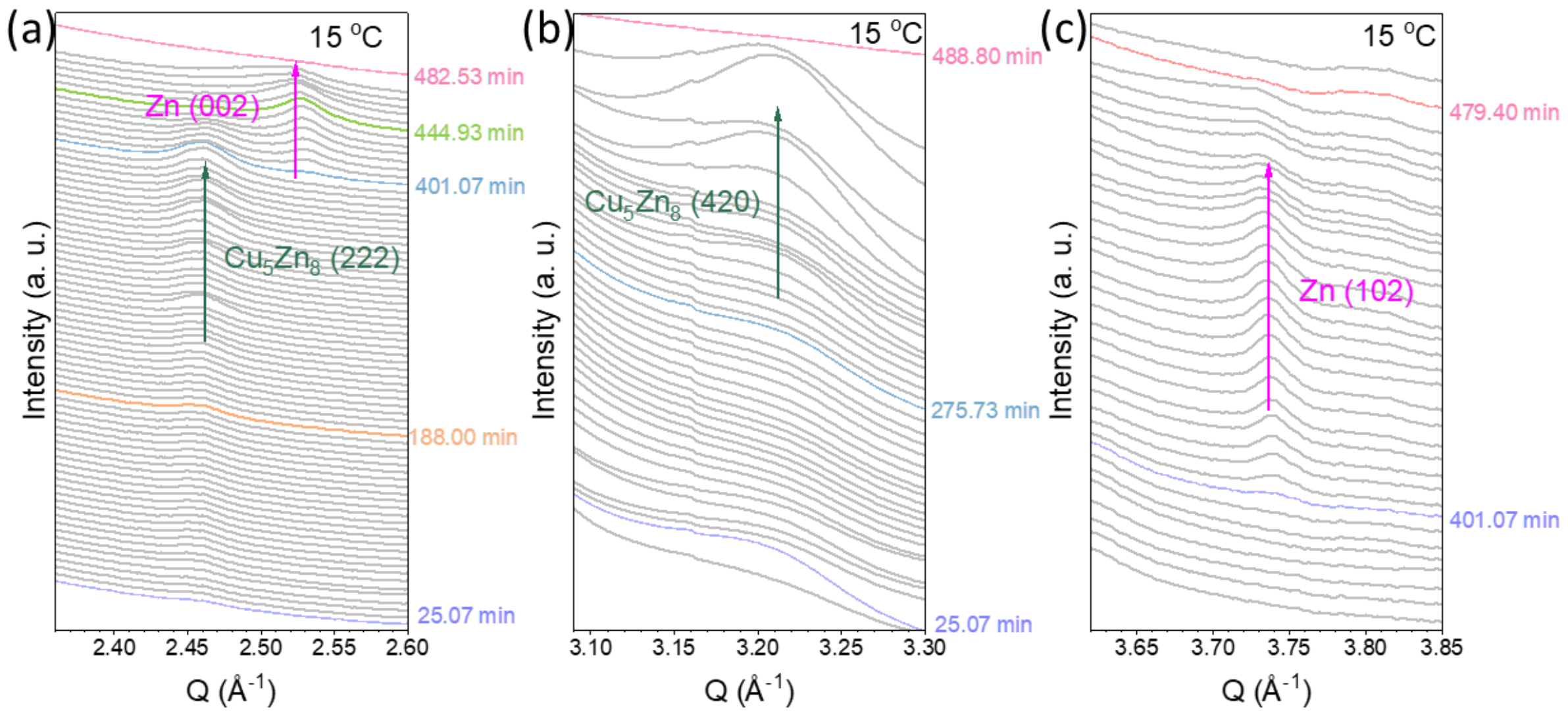

**Figure S3.** Close ups of Figure 1b showing the phase changes that occur during the chemical dealloying process of $Cu_{20}Zn_{80}$ in 5 M $H_3PO_4$ at 15 °C.

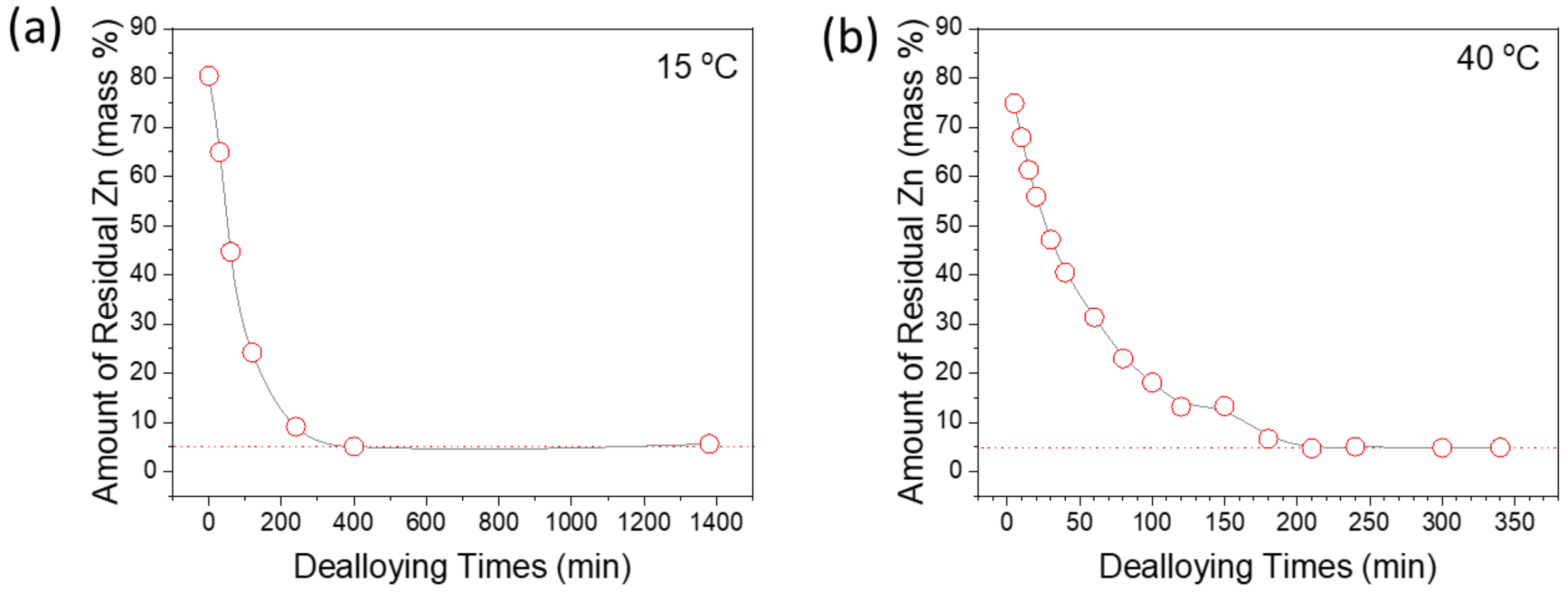

**Figure S4.** Plots of the mass % of residual Zn against the dealloying time in 5 M $H_3PO_4$ at (a) 15 °C, and (b) 40 °C, measured by the ICP-OES of the electrolyte composition.

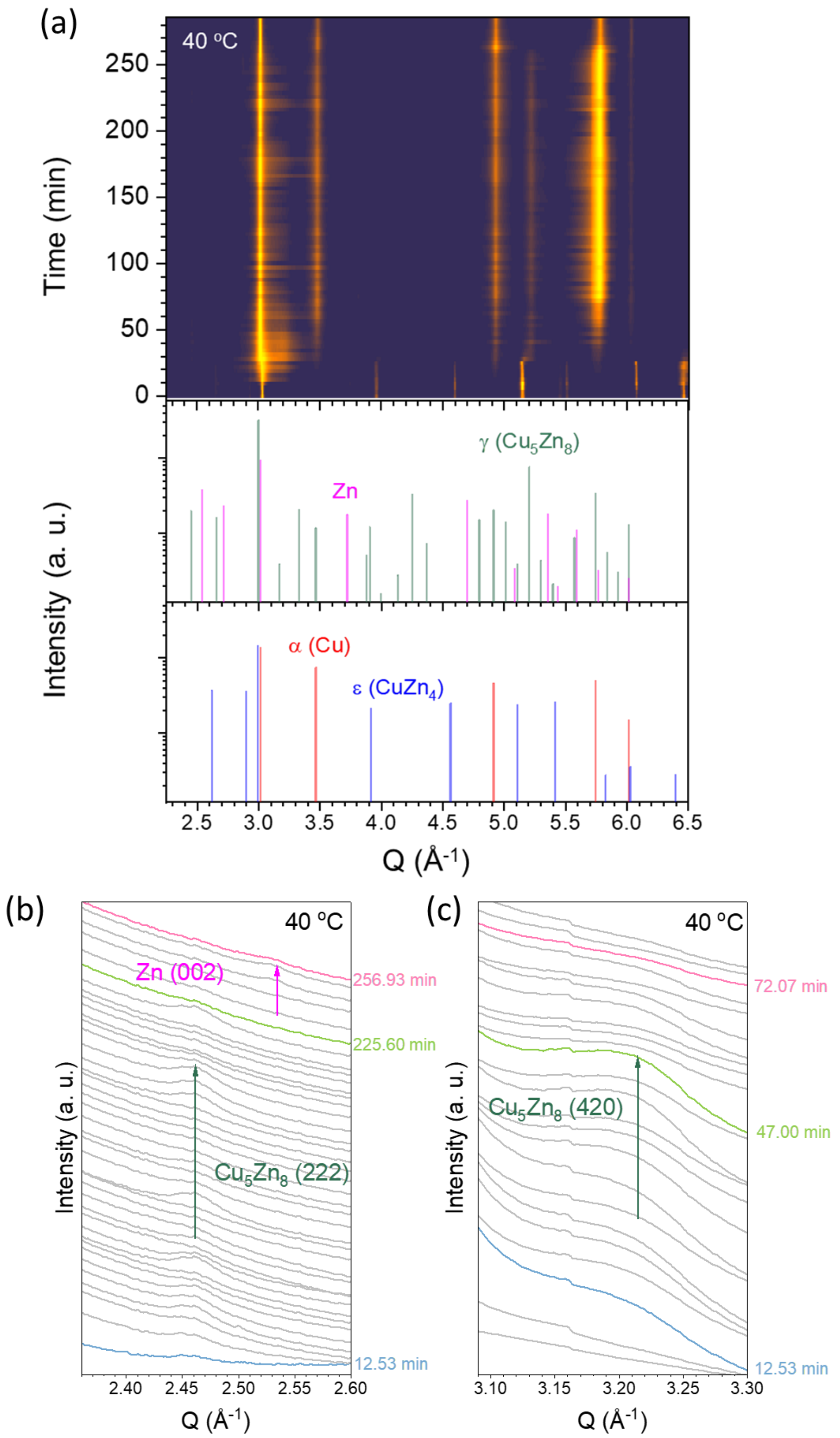

**Figure S5.** (a) *In situ* synchrotron XRD patterns for the dealloying of $Cu_{20}Zn_{80}$ in 5 M $H_3PO_4$ at 40 °C. (b) and (c) Close ups of (a) showing the phase changes seen during the dealloying process.

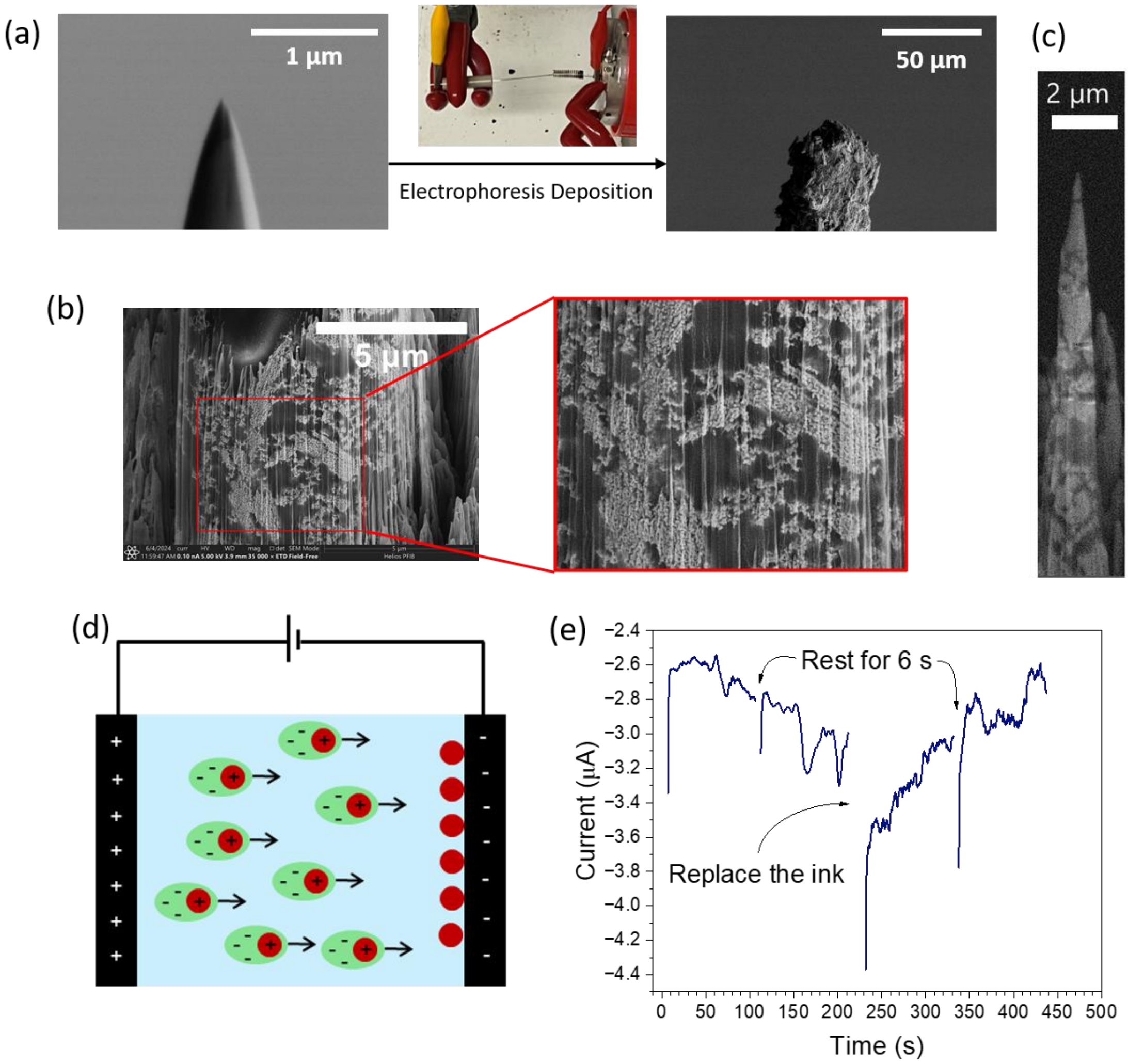

**Figure S6.** (a) Schematic illustration for the process of the NPC samples loading onto the pre-sharpened W needle *via* the electrophoresis deposition. (b) The SEM images of the 5 $^{o}$C dealloyed NPC embedded in ice. (c) The SEM image of the FIB-sharpened needle ready for the APT measurements. (d) Schematic illustration of the electrophoresis deposition. (e) Transient current plots for the electrophoresis deposition of NPC.

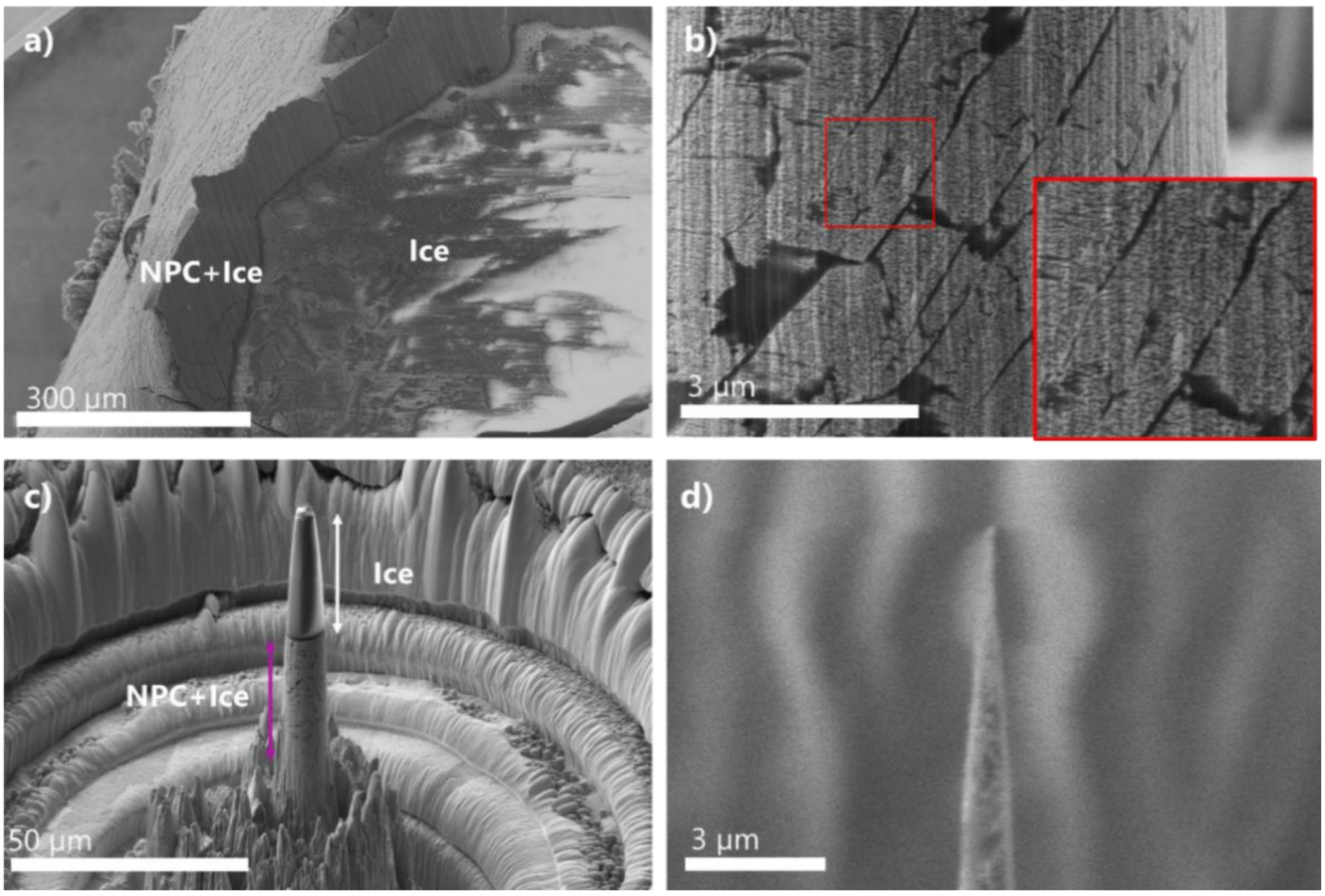

**Figure S7.** (a-b) The SEM images of the 5 ºC dealloyed NPC embedded in ice. (c)The SEM images showing the sharpening process under the focused ion beam (FIB). The top ice layer would protect the NPC underneath. (d) The SEM image of the prepared APT needle. The NPC samples is still mounted together by ice without collapsing, and the bi-continuous nano-porous morphology is preserved.

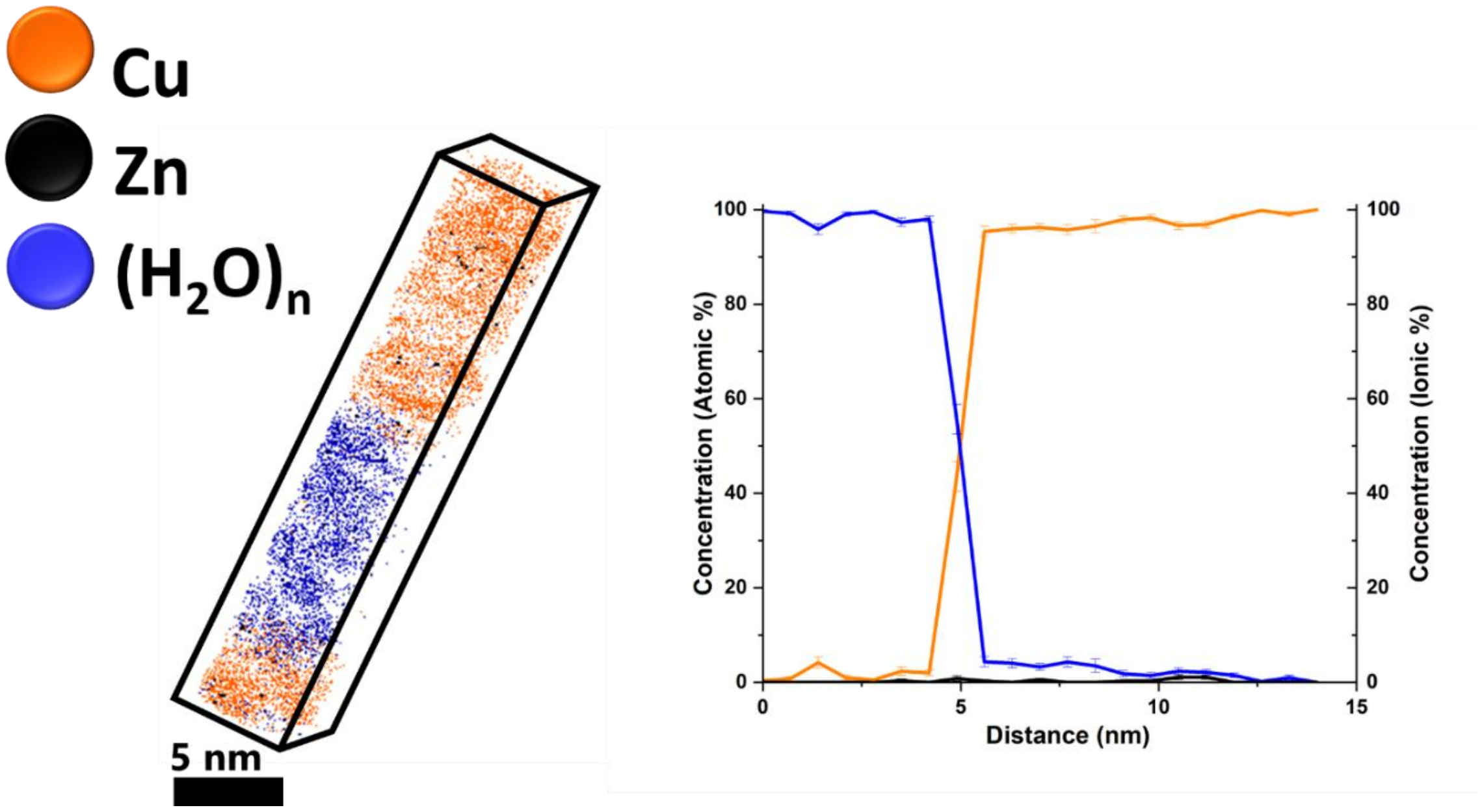

**Figure S8.** APT results for the fully dealloyed NPC sample (5 ºC for 24 h) showing almost pure copper compositions in nano-ligaments with trace amounts of Zn (<1 at. %).

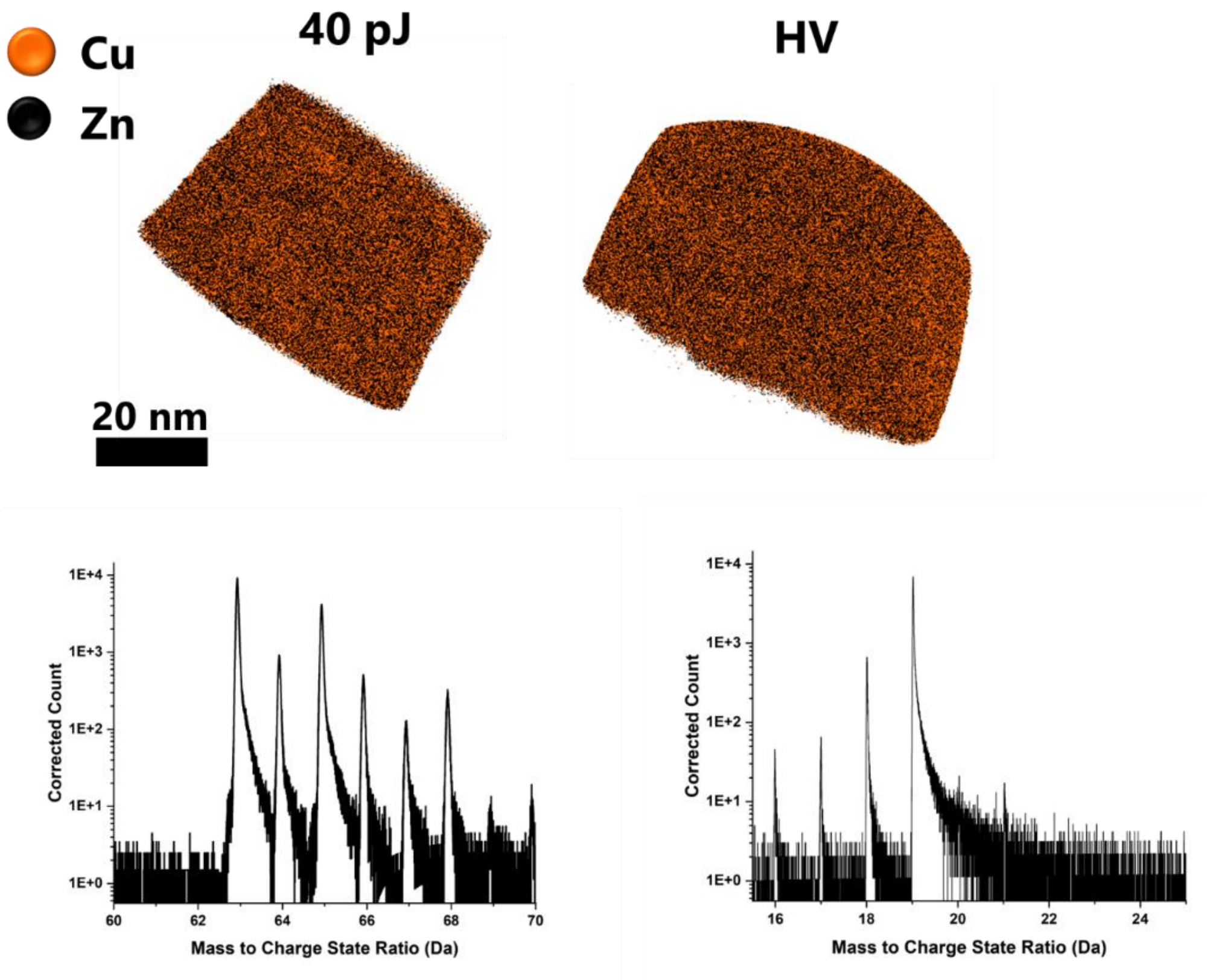

**Figure S9.** Atom maps of $Cu_{20}Zn_{80}$ precursor at different field evaporation parameters, showing similar Cu-Zn ratios, and model spectra showing the Cu signals, Zn signals and signals alluding to water.

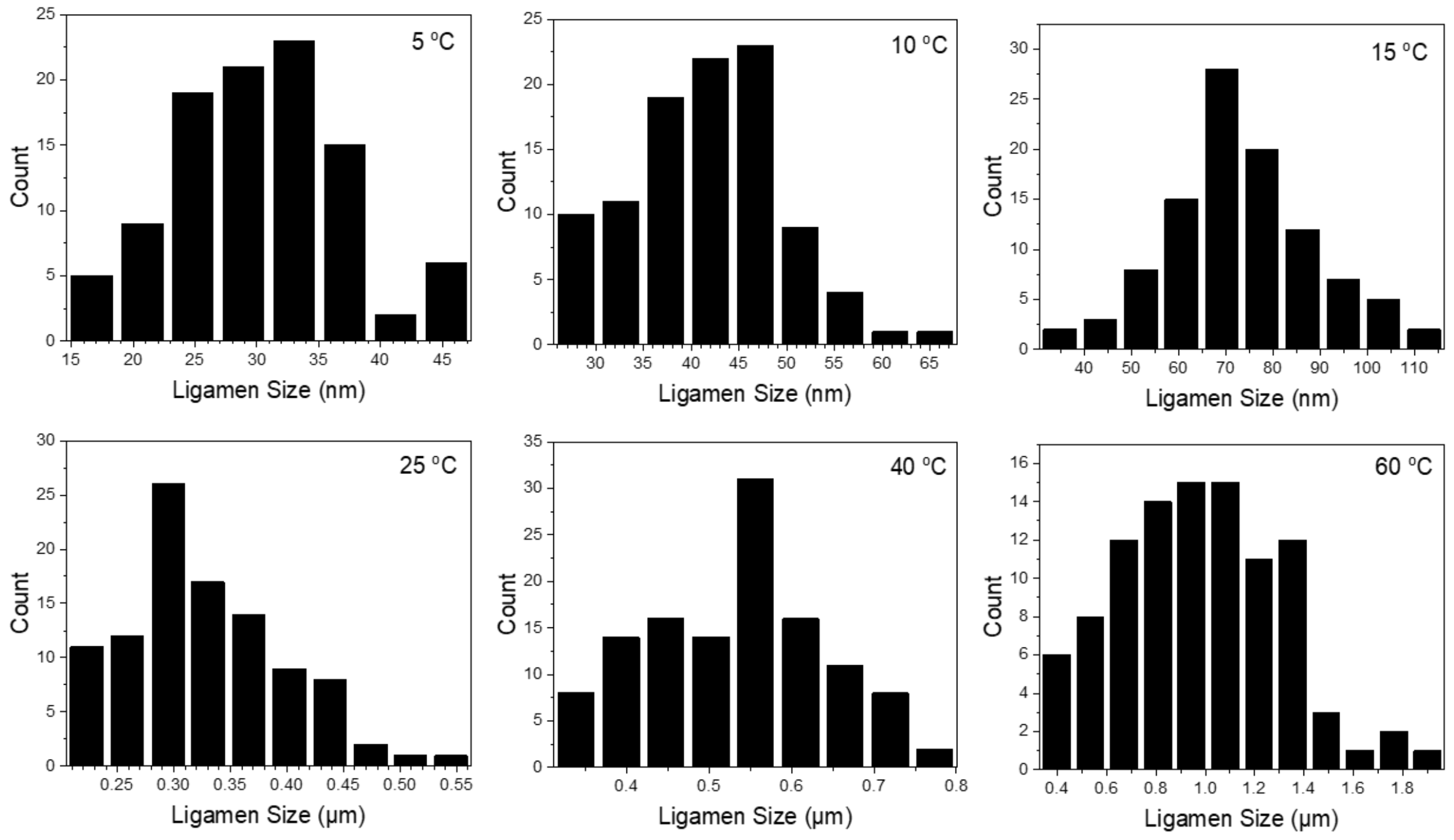

**Figure S10.** Bar charts showing the ligament sizes distribution of all the NPC samples.

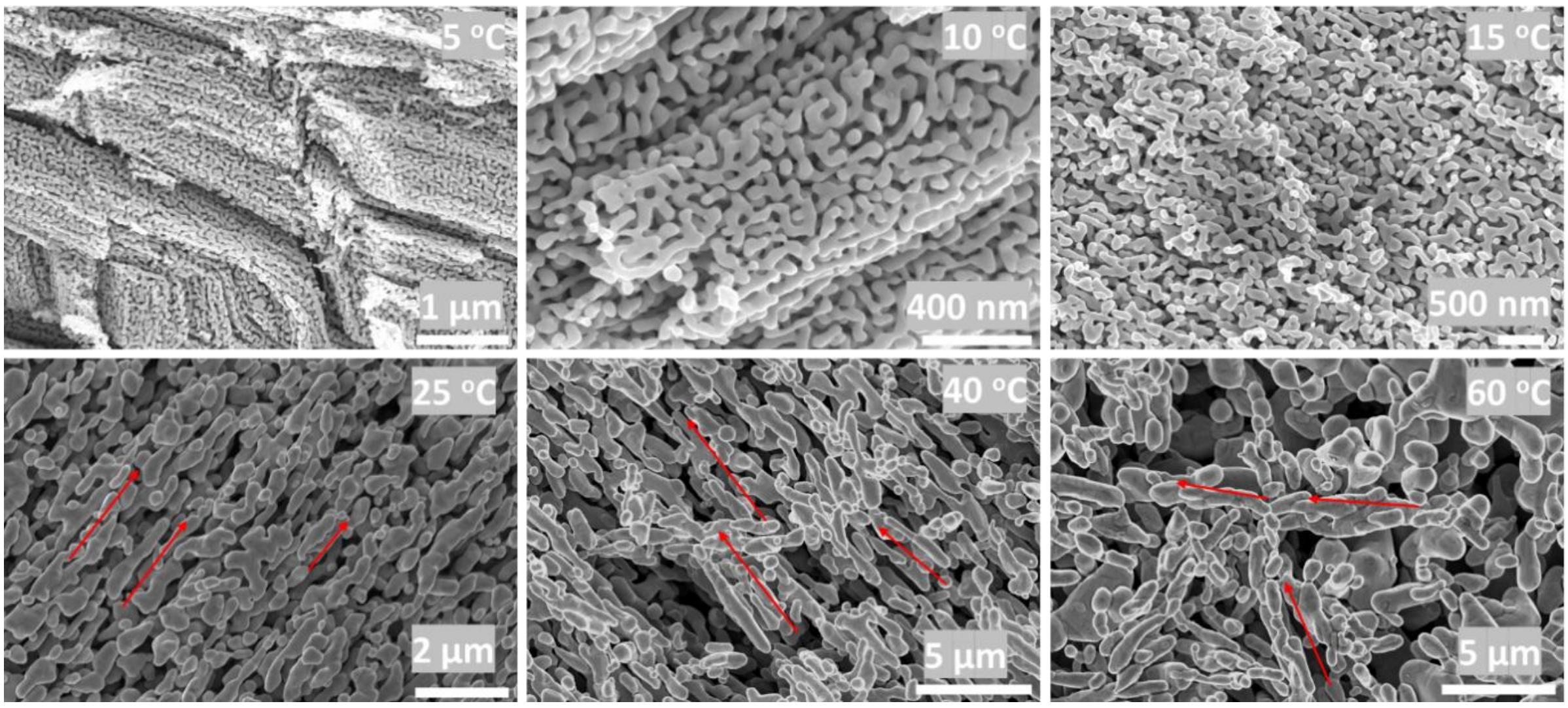

**Figure S11.** SEM images of the NPC dealloyed in 5 M $H_3PO_4$ for 24 hours under different temperatures, with arrows showing the elongation of the ligaments.

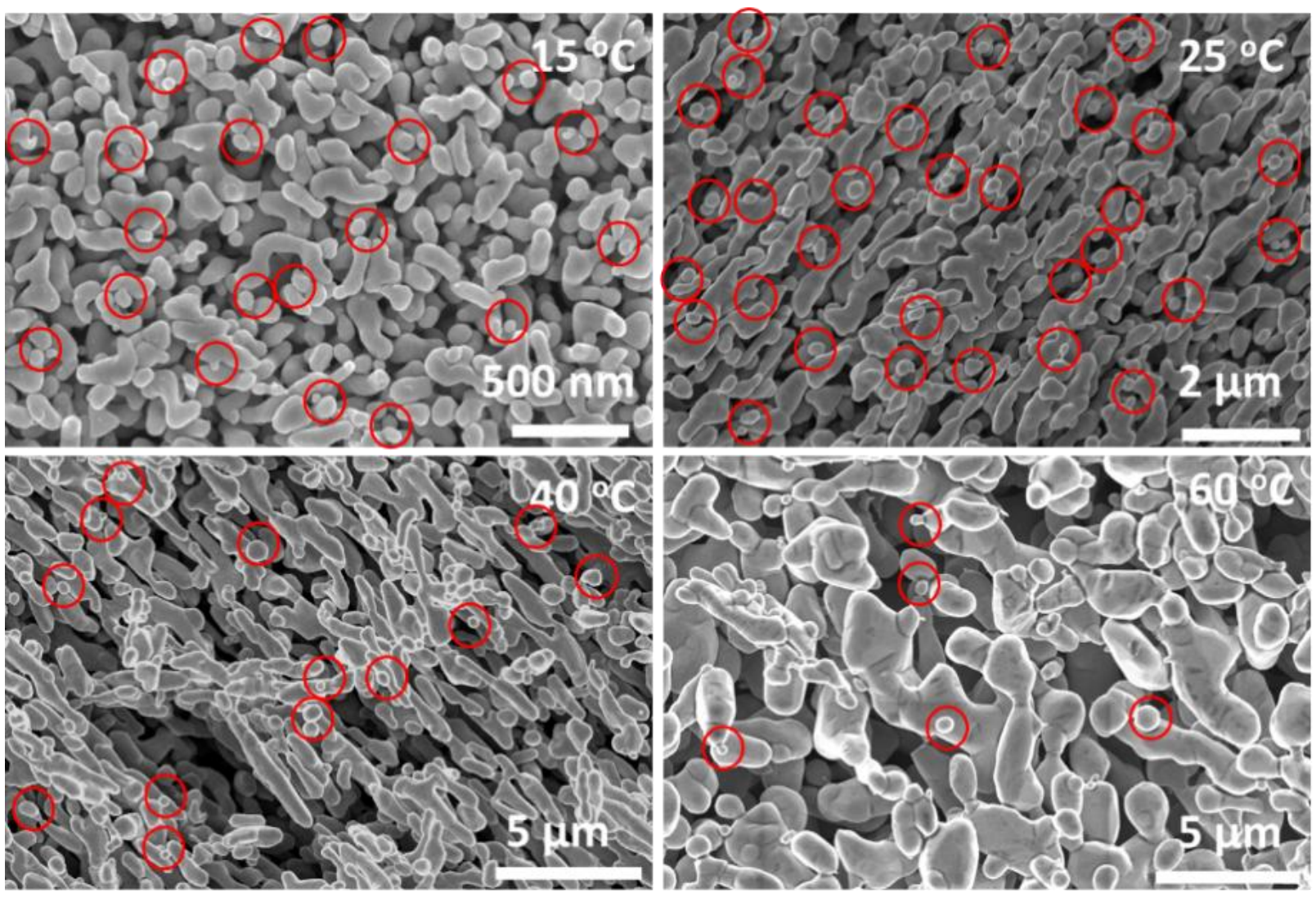

**Figure S12.** SEM images of the NPC dealloyed in 5 M $H_3PO_4$ for 24 hours under different temperatures. The small nanoscale spherical ligaments are highlighted by the red circles.

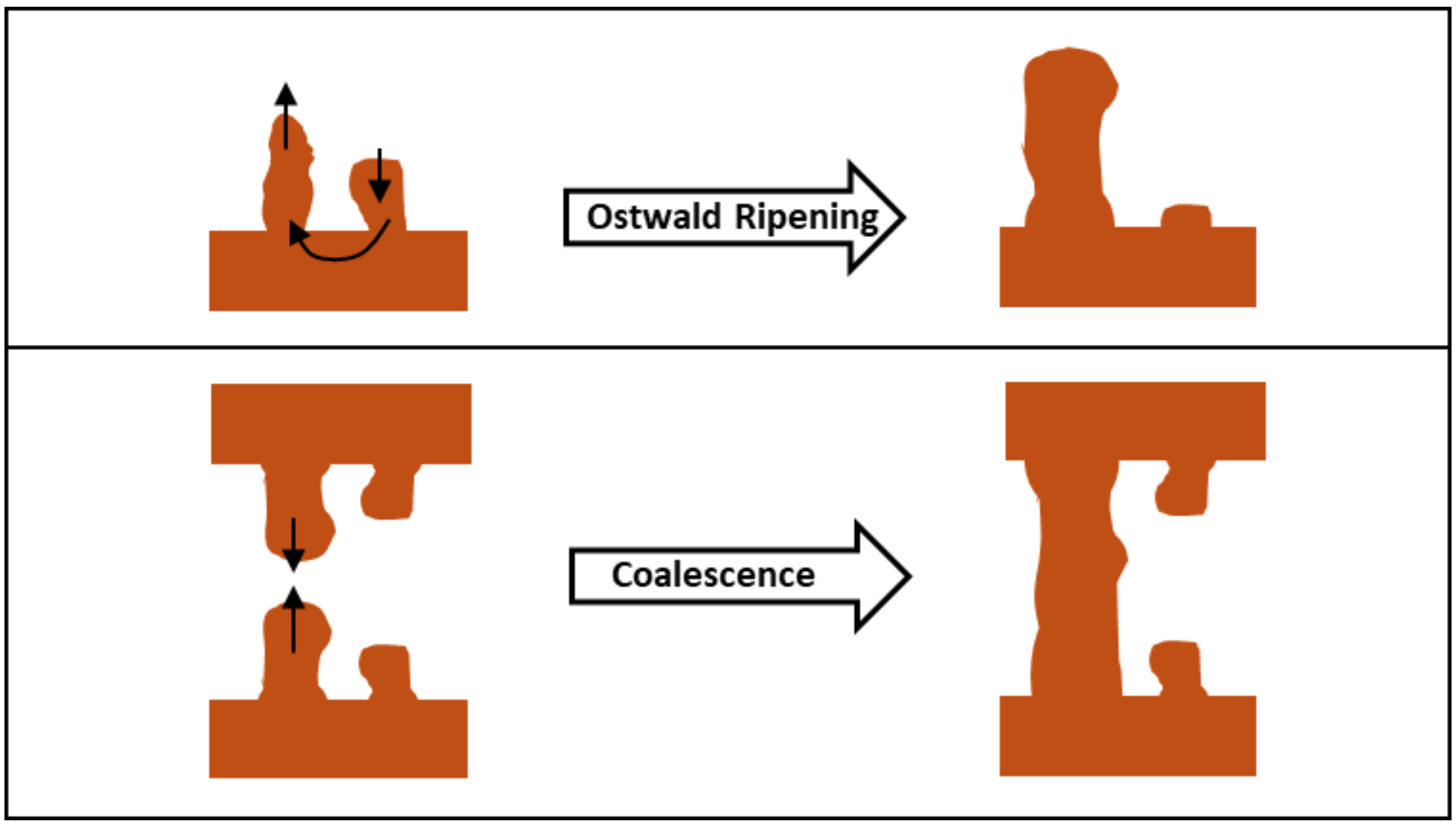

**Figure S13.** Schematic illustration of Ostwald ripening and coalescence-controlled coarsening.

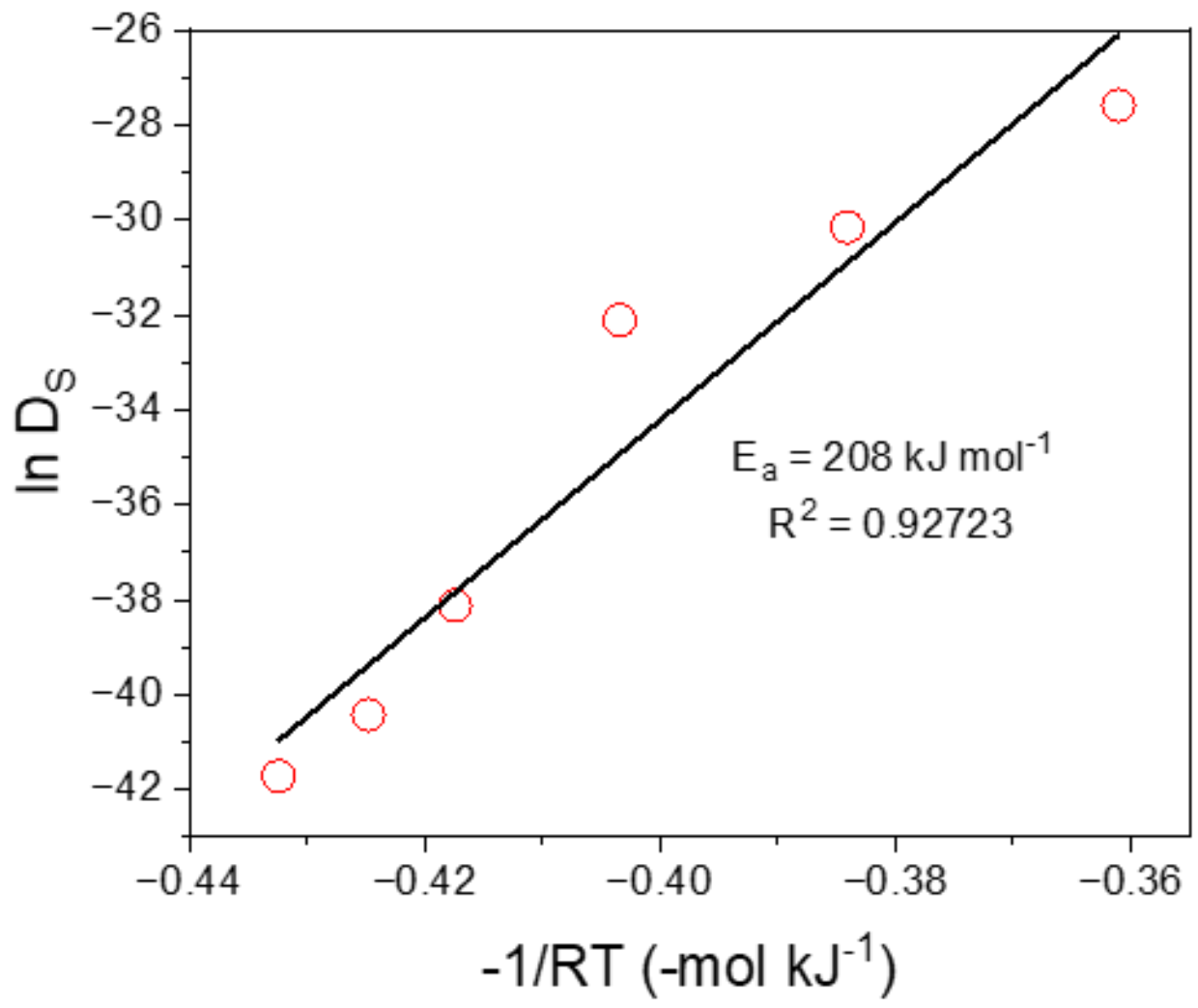

**Figure S14.** The Arrhenius plot for the electrochemical surface diffusion of Cu atoms in 5 M $H_3PO_4$, where the slope represents its activation energy ($E_a$) in kJ mol$^{-1}$. The y axis represents the natural logarithm of the surface diffusion coefficient ($D_s$).

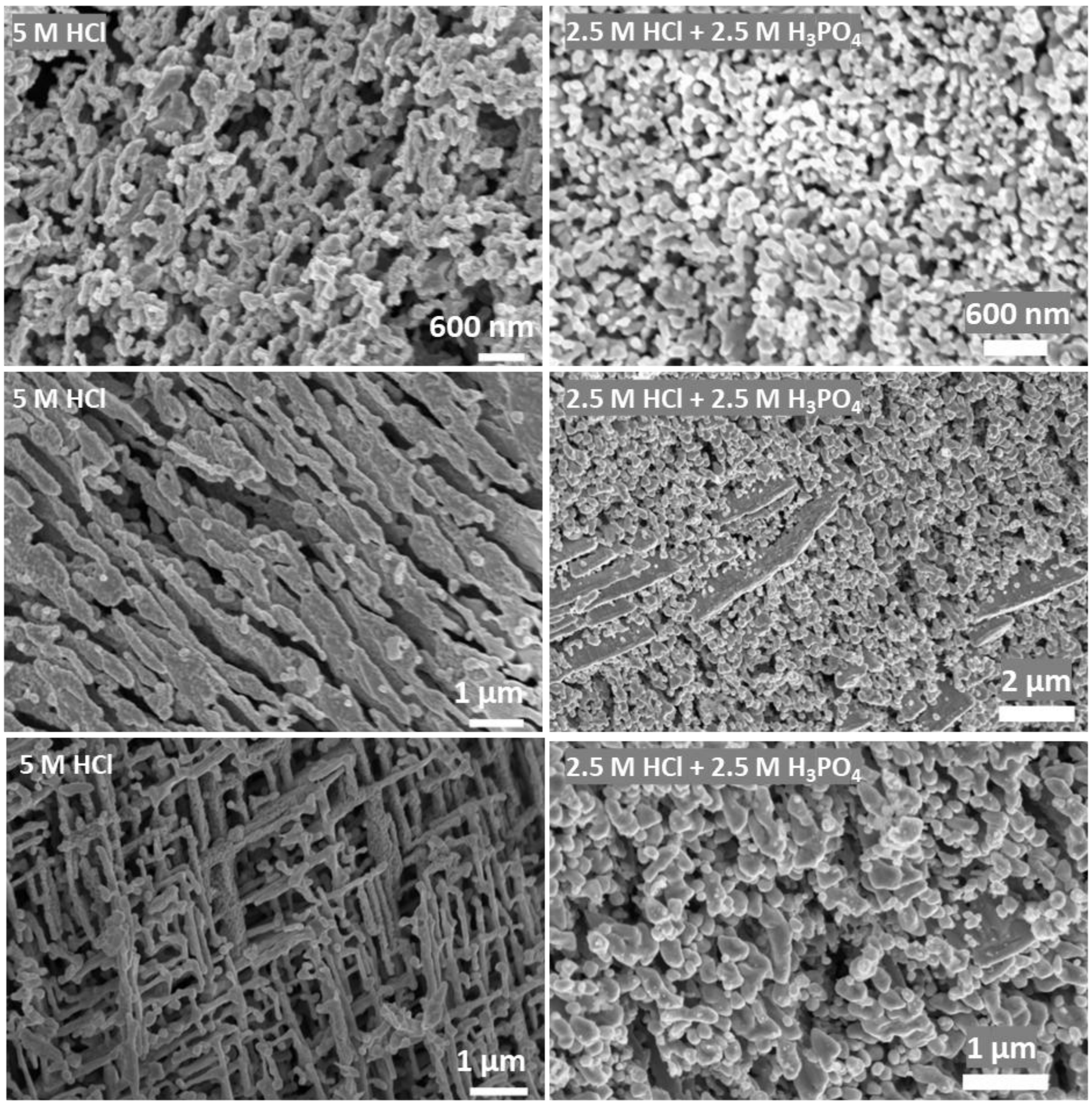

**Figure S15.** SEM images of the NPC dealloyed in HCl, and the mixture of HCl and $H_3PO_4$ for 24 h under room temperature. Less homogeneous structures were obtained compared with 5 M $H_3PO_4$.

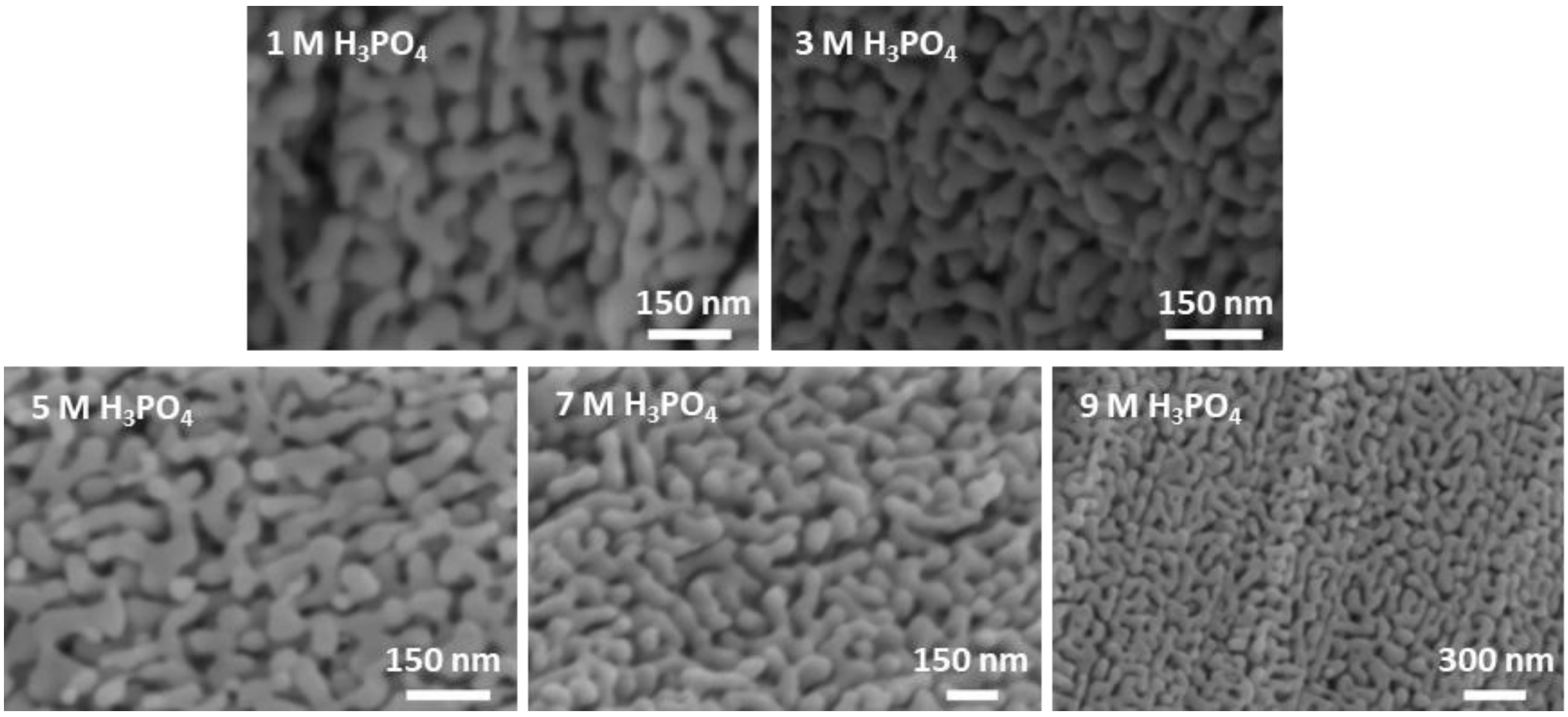

**Figure S16.** SEM images of the NPC dealloyed in different concentrations of $H_3PO_4$ for 24 h under room temperature.

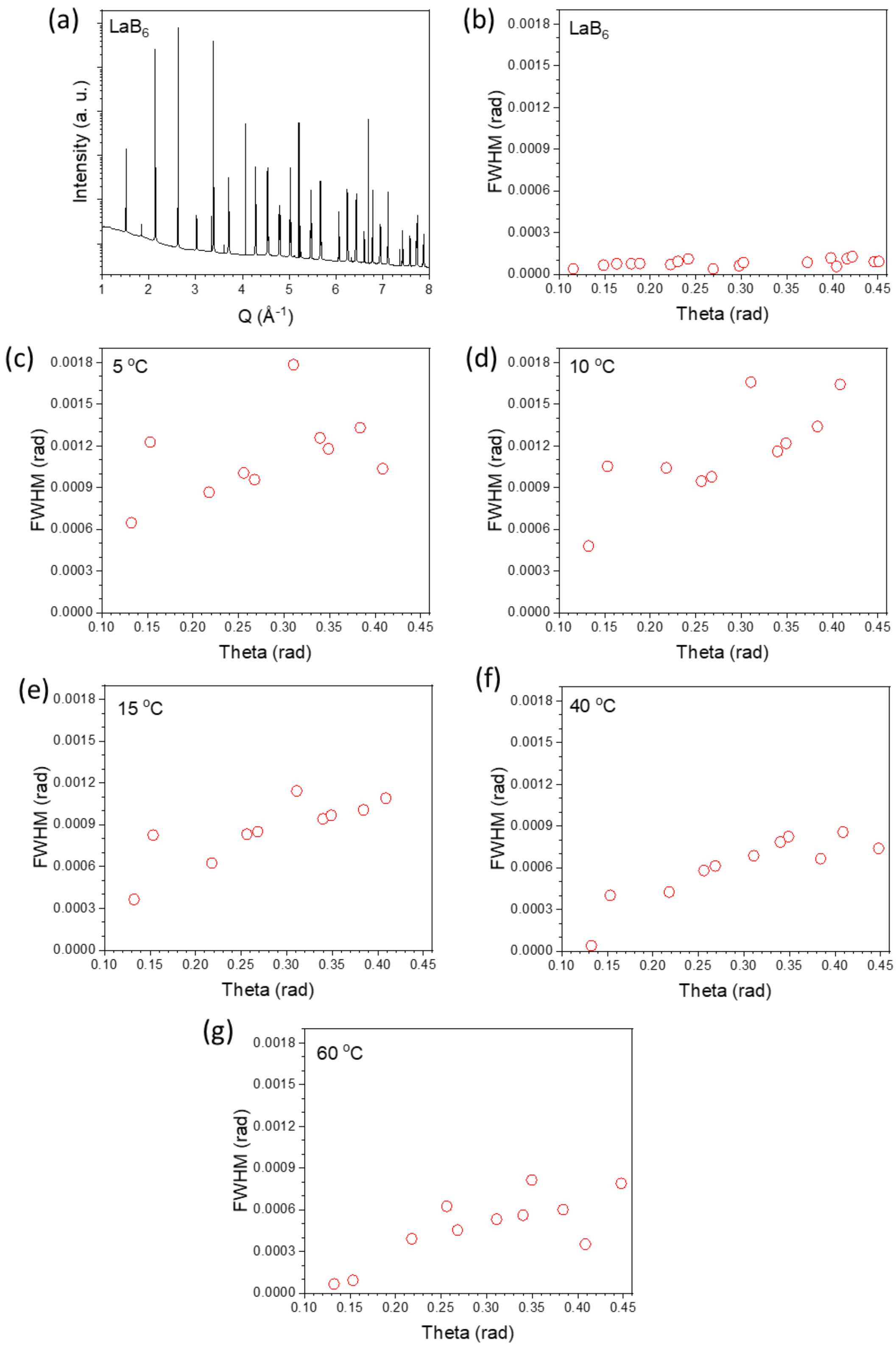

**Figure S17.** (a) XRD pattern of the $LaB_6$ standard measured under the same conditions as the NPC samples. (b) – (g) Plots of the peak broadening (FWHM) against the diffraction peak positions. The scattering vector, Q ($Å^{-1}$) was converted to X-ray diffraction angle, Theta (rad).

The y-axis range for (b) to (g) was kept same for a more conspicuous comparison, showing the peak broadening caused by the instrument error is negligible.

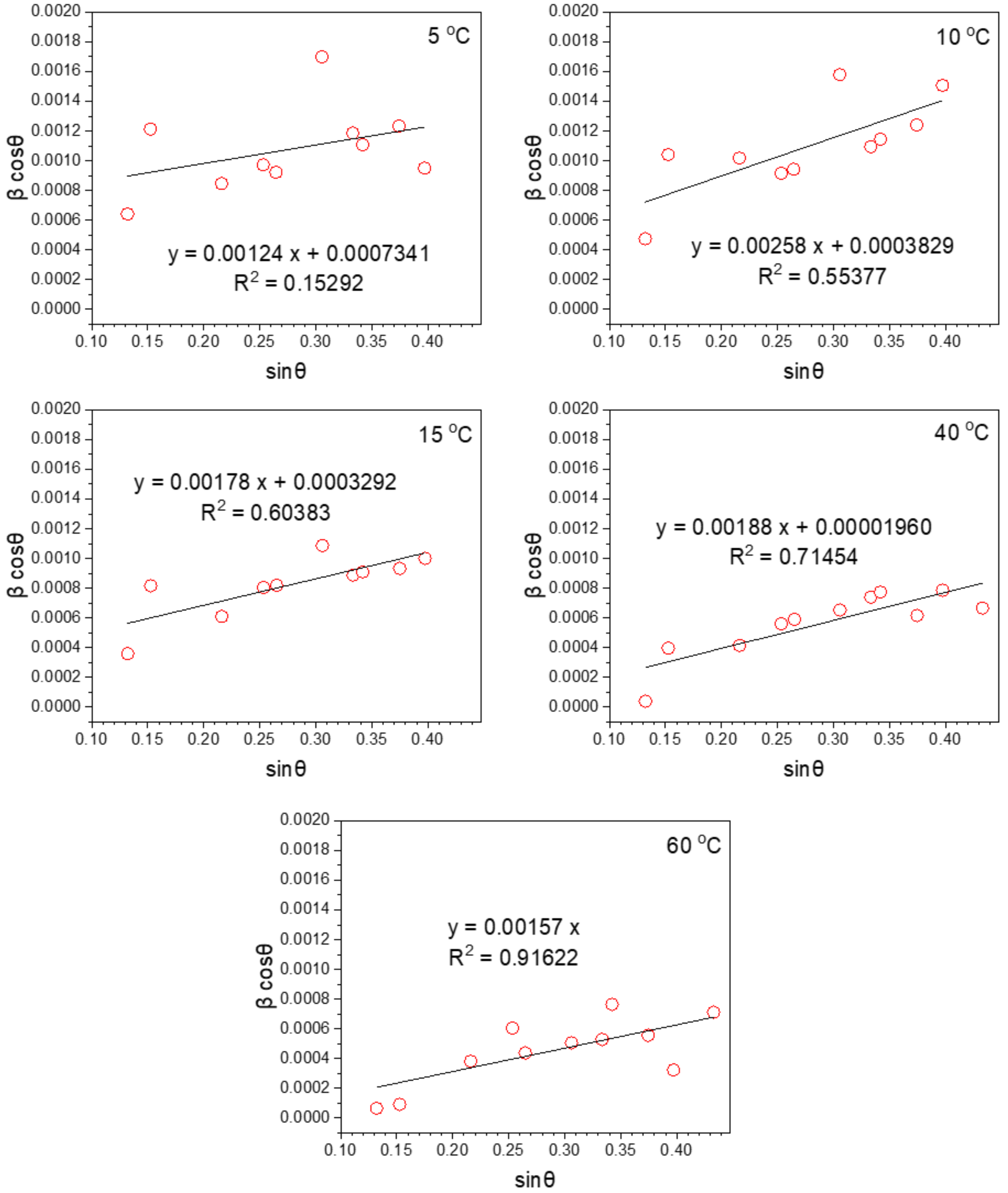

**Figure S18.** Williamson-Hall plots for the NPC samples. Poor linearity was obtained.

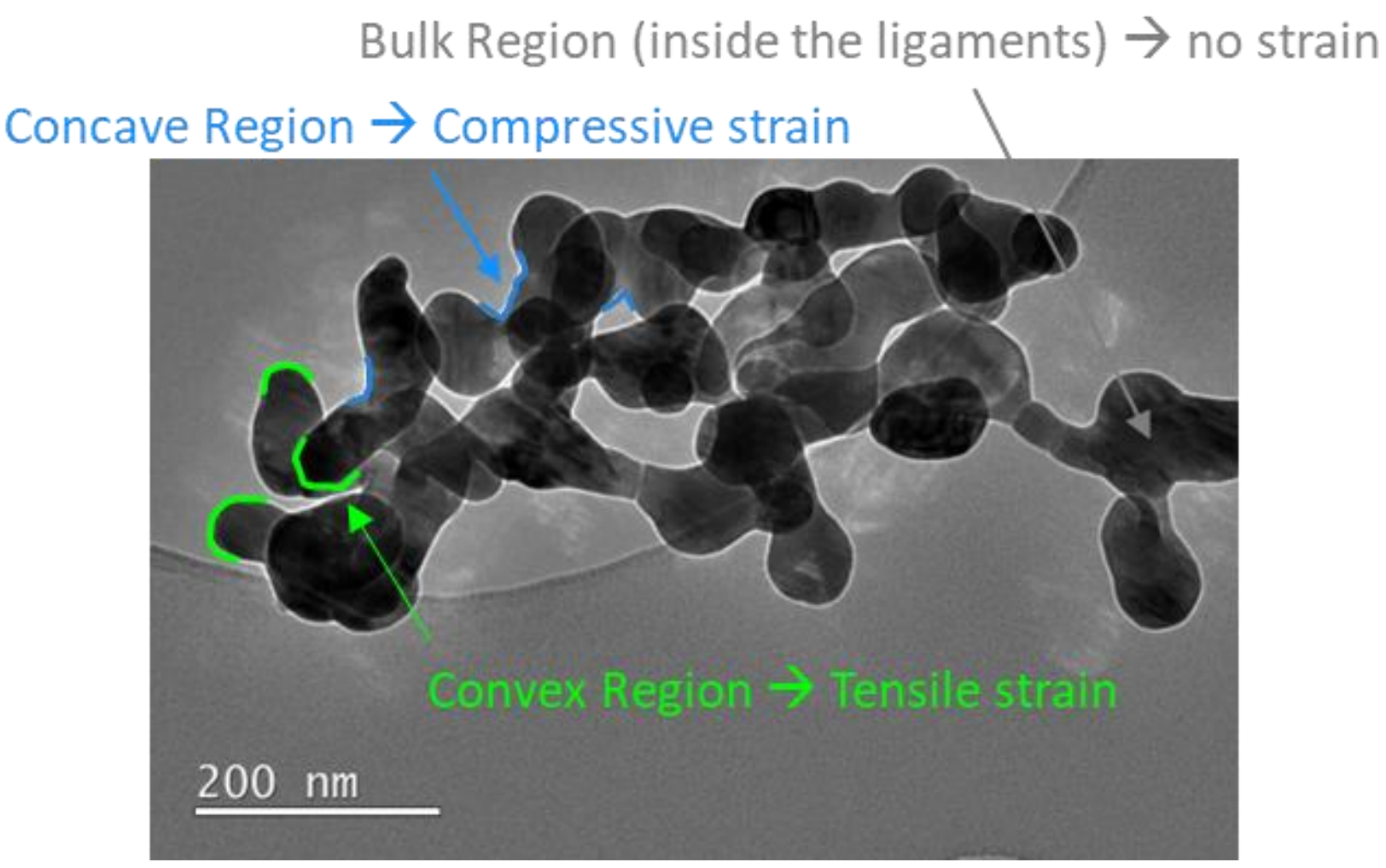

**Figure S19.** TEM image of the NPC dealloyed at 5 °C, with schematic illustration of the origins of the strain.

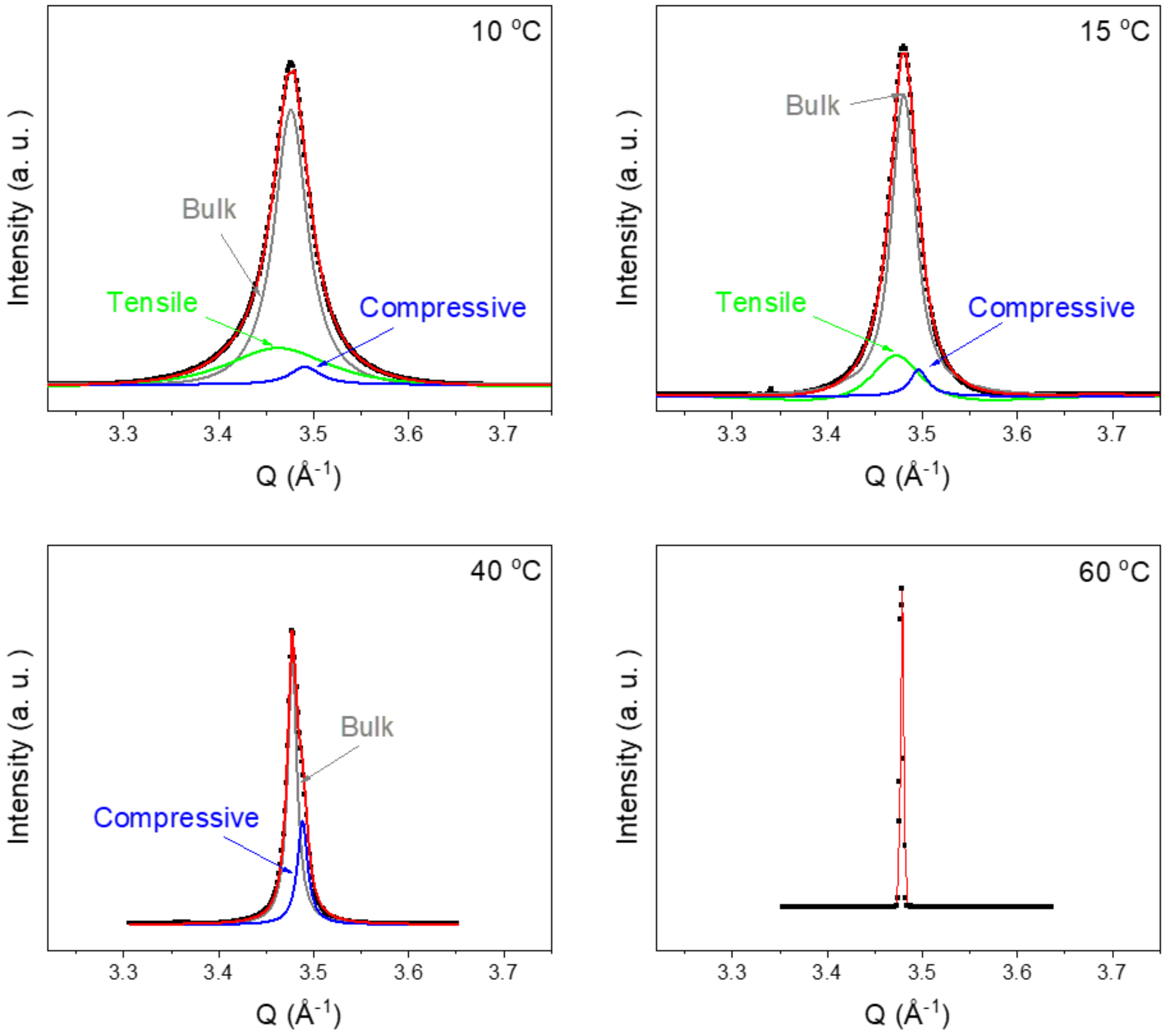

**Figure S20.** Fitting of the (200) diffraction peak of the NPC dealloyed at different temperatures. Black dotted line is the diffraction pattern after the baseline subtraction by a polynomial

equation. Grey, green, and blue solid lines are the pseudo-Voigt fittings to the data, and red solid line is the summary of all 3 fitting components, representing the quality of fittings.

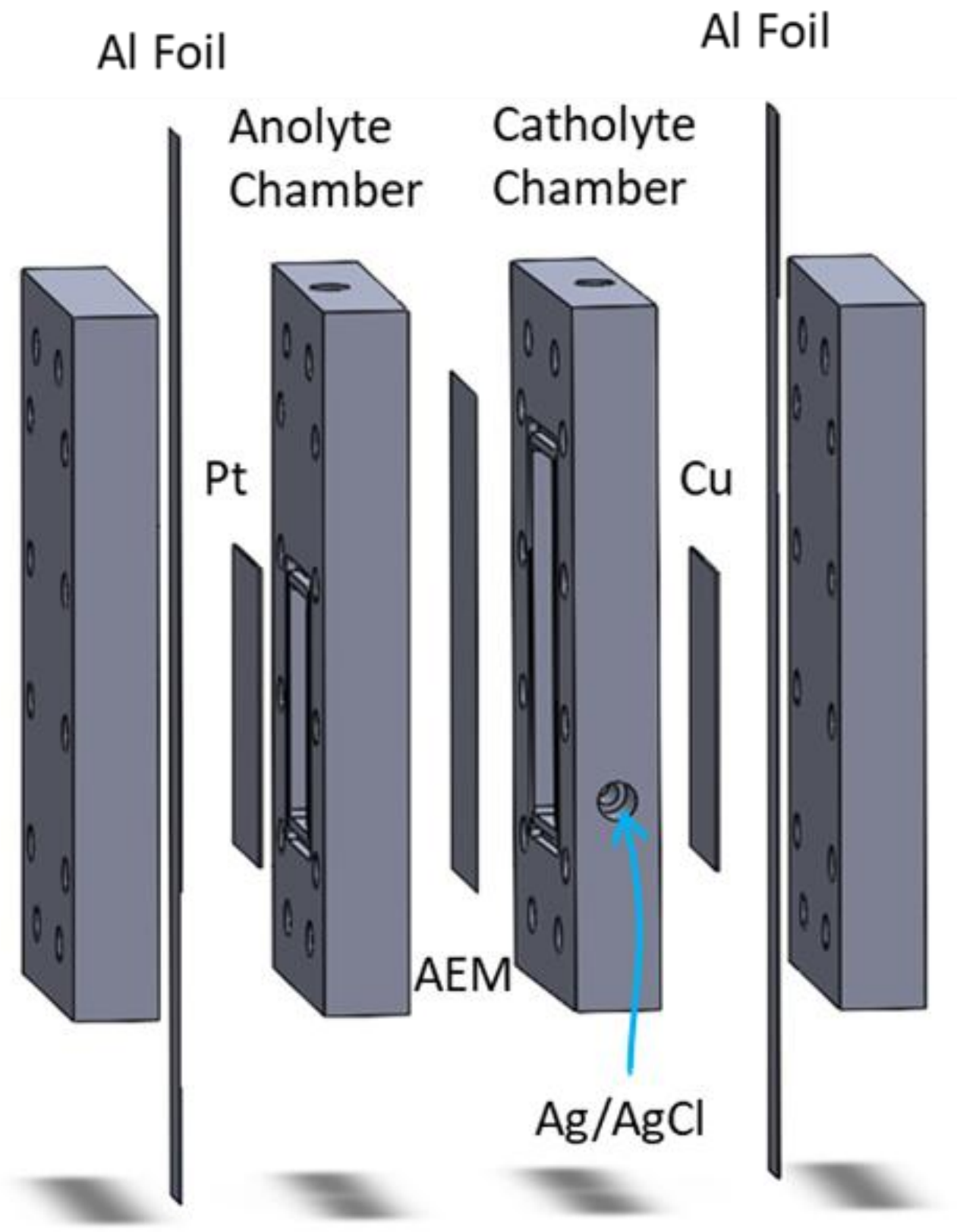

**Figure S21.** Schematic illustration of the H-cell setup used for $CO_2$RR test in this work.[1]

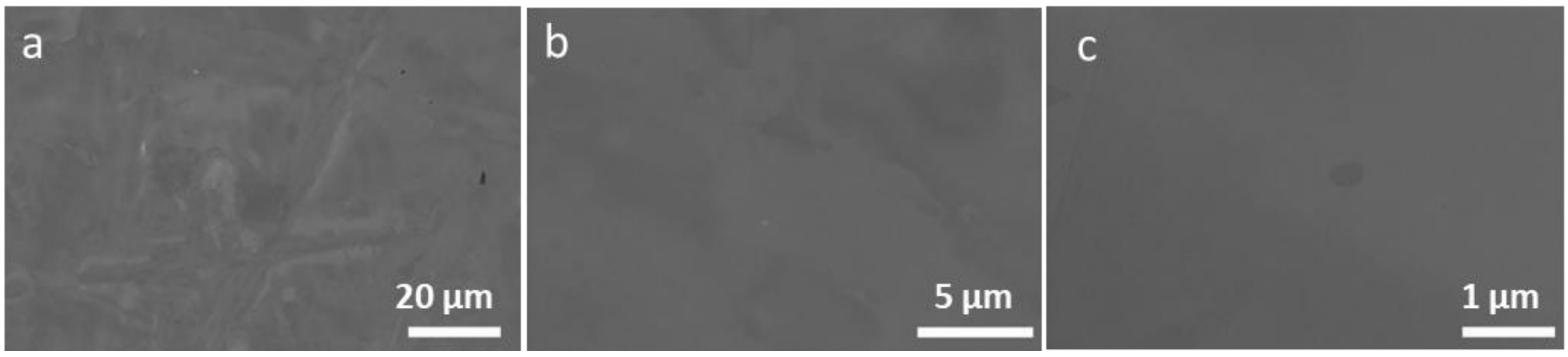

**Figure S22.** SEM images of the electropolished polycrystalline Cu used for benchmarking for both $CO_2$RR and CORR test.

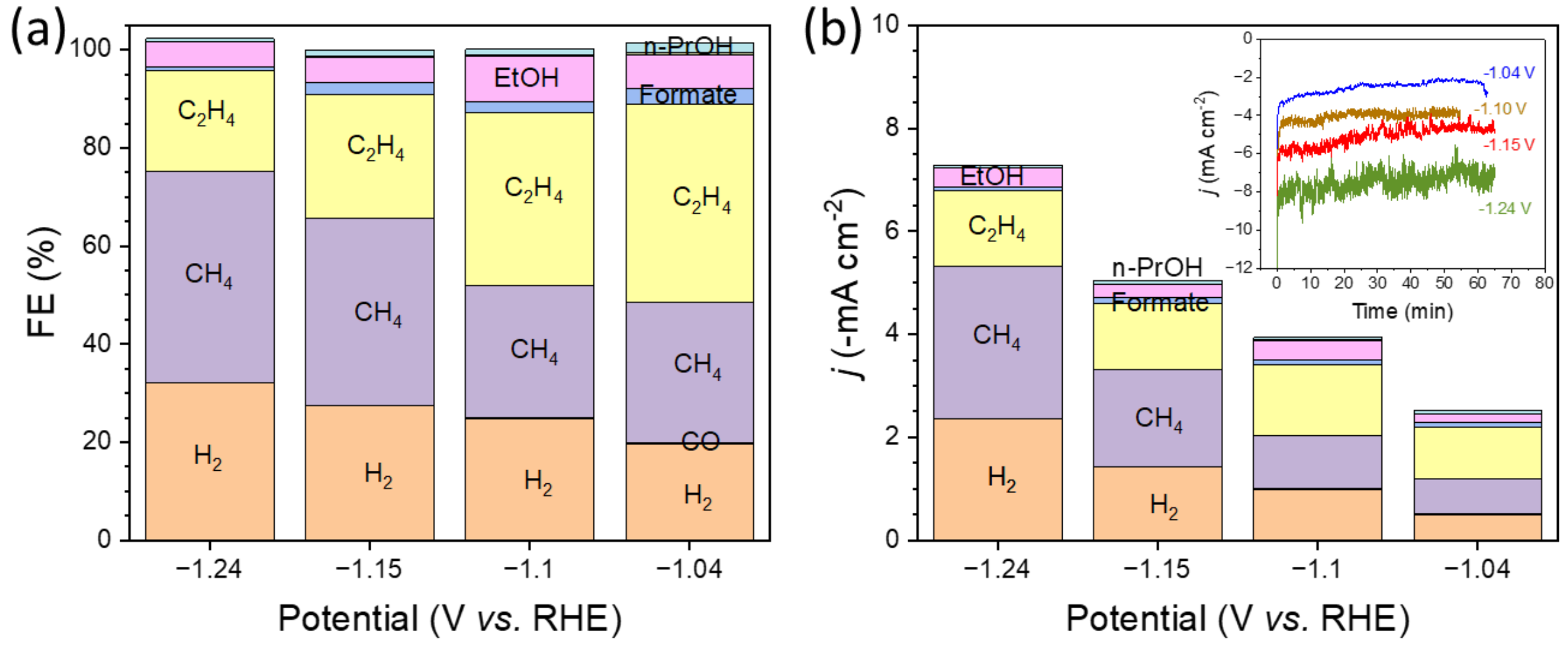

**Figure S23.** Plots of (a) FEs and (b) partial current densities of the $CO_2$RR products detected on an electropolished polycrystalline Cu, in $CO_2$ saturated 0.1 M $KHCO_3$, at room temperature, in the H-cell, against the working electrode potentials in RHE scale. Insert: The transient current curve for each chronoamperometry measurement. Each chronoamperometry measurement was conducted for 1 h.

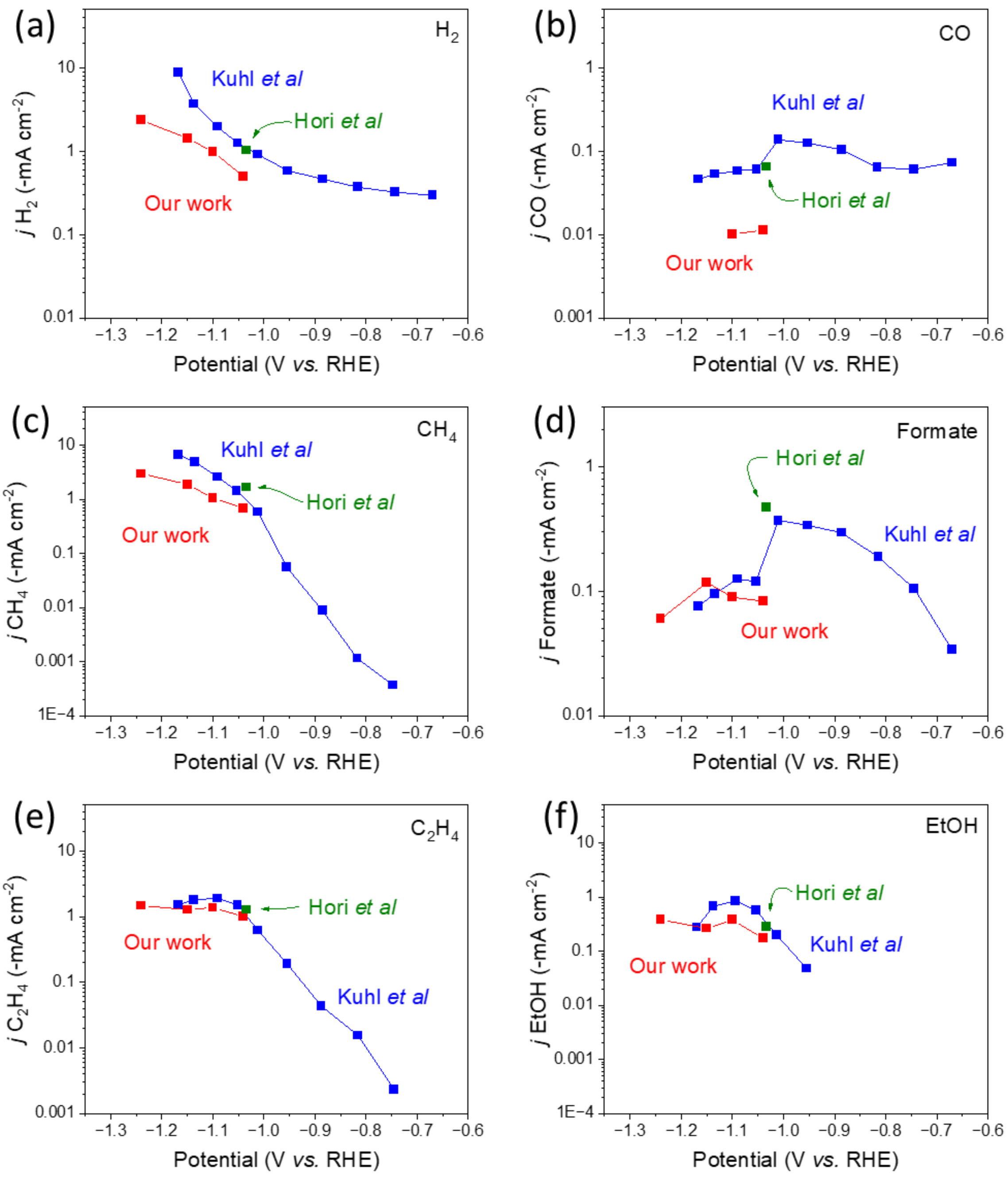

**Figure S24.** Comparison of partial current densities (normalised by the geometric surface area) of major $CO_2$RR with literature values measured under the similar conditions. Each data point represents a chronoamperometry measurement conducted for 1 h, in $CO_2$ saturated 0.1 M $KHCO_3$, at room temperature, and in the H-cell. Literature data was obtained from Hori *et al*,[21] and Kuhl *et al*.[1]

**Figure S24** compared our benchmarking results on the electropolished polycrystalline Cu (**Figure S22**) with the work from Hori *et al* [21] and Kuhl *et al*.[1] Kuhl's work was carried out in the identical 'sandwich' H-cell as ours. Same electrolyte ($CO_2$ saturated 0.1 M $KHCO_3$) was also used for both Hori and Kuhl in **Figure S24**. The similar Tafel slopes for all products were obtained compared with Kuhl's work. However, our partial current densities for $H_2$, CO, and $CH_4$ are slightly smaller, while the activities towards formate, ethylene, and ethanol are similar. This difference could be caused by the changed exposed crystalline facets of Cu, temperature, and systematic errors. Note that numerous papers have shown that different CVs can be obtained on different batched of polycrystalline Cu foils, as explained in our previous paper.[22] Considering this active nature of Cu, such difference we observed in the partial current densities of $H_2$, CO, and $CH_4$ should be acceptable.

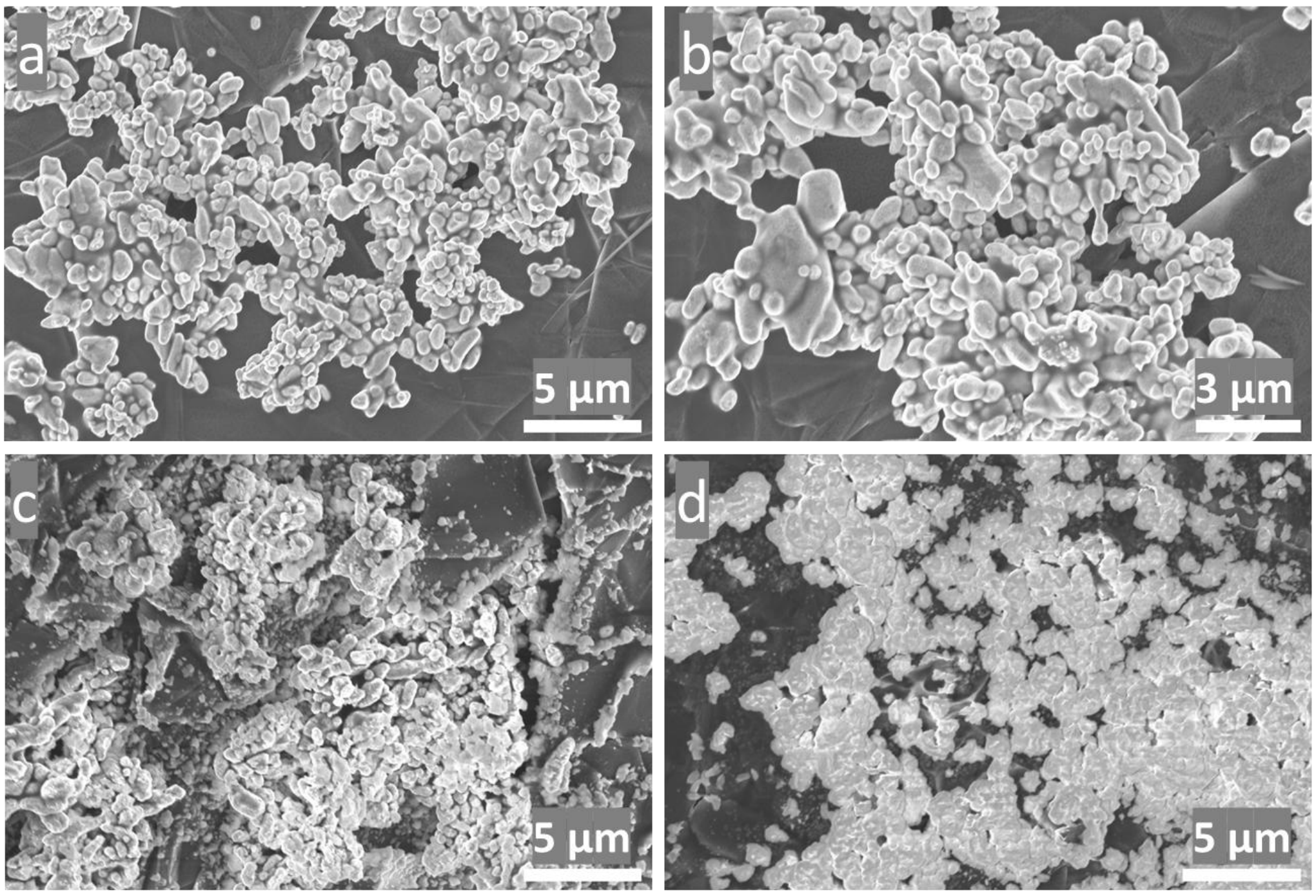

**Figure S25.** SEM images of (a) and (b) pristine graphite foil drop-casted with 150 µg cm$^{-2}$ of 40 $^{o}$C dealloyed NPC; (c) after 25 min of cathodic condition at -1.17 V *vs.* RHE in the $CO_2$ saturated 0.1 M $KHCO_3$; and (d) after 1 h of the same cathodic conditions.

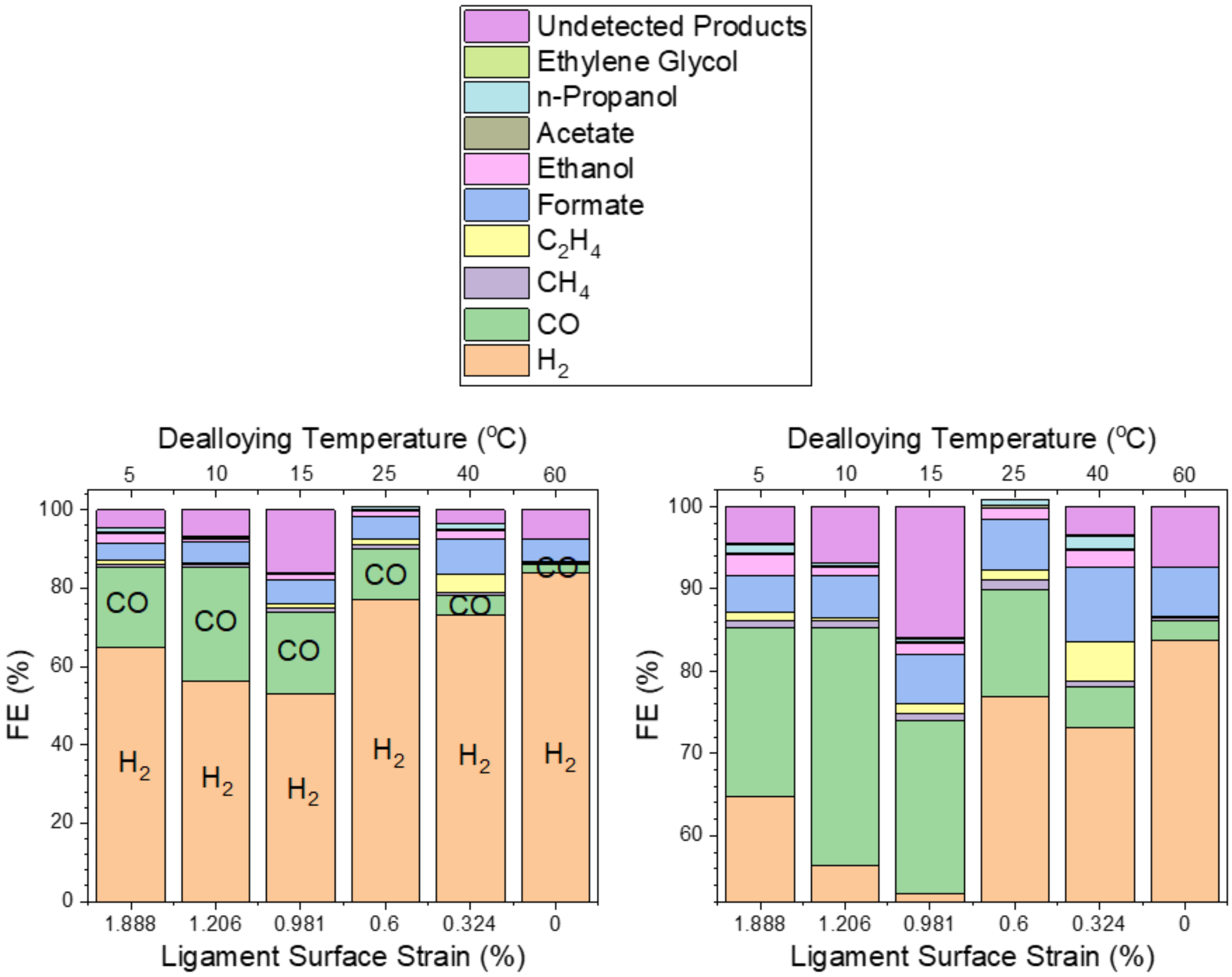

**Figure S26.** Plots of the product distributions (FE) on all the dealloyed NPC samples in $CO_2$ saturated 0.1 M $KHCO_3$ solution at -1.17 V *vs.* RHE.

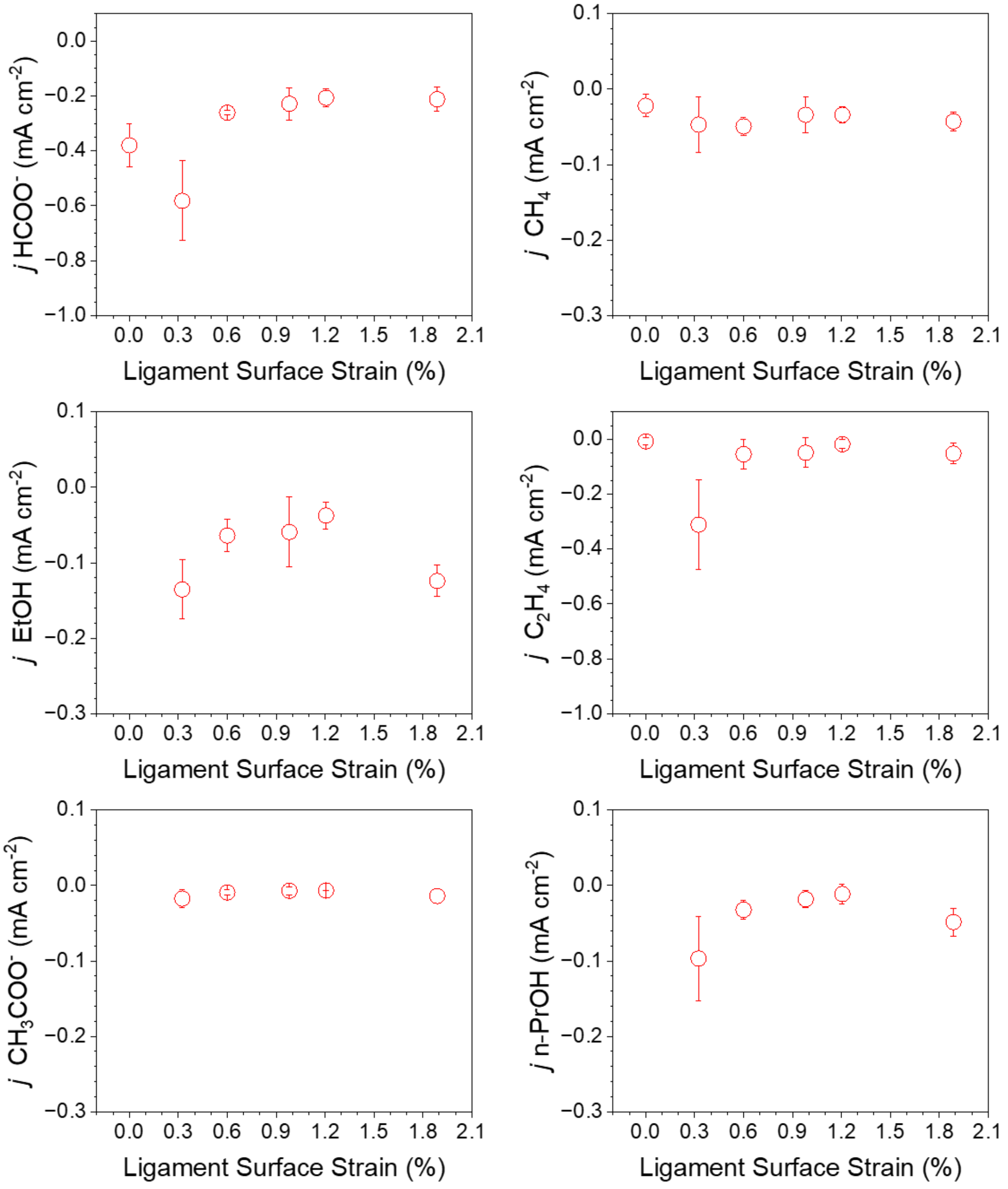

**Figure S27.** Plots of the individual $CO_2$RR product geometric surface area normalised current densities against the ligament surface strain of NPC samples. The error bar on the *y*-axis was obtained by repeating the $CO_2$RR tests on differently drop-casted electrodes for at least 3 times.

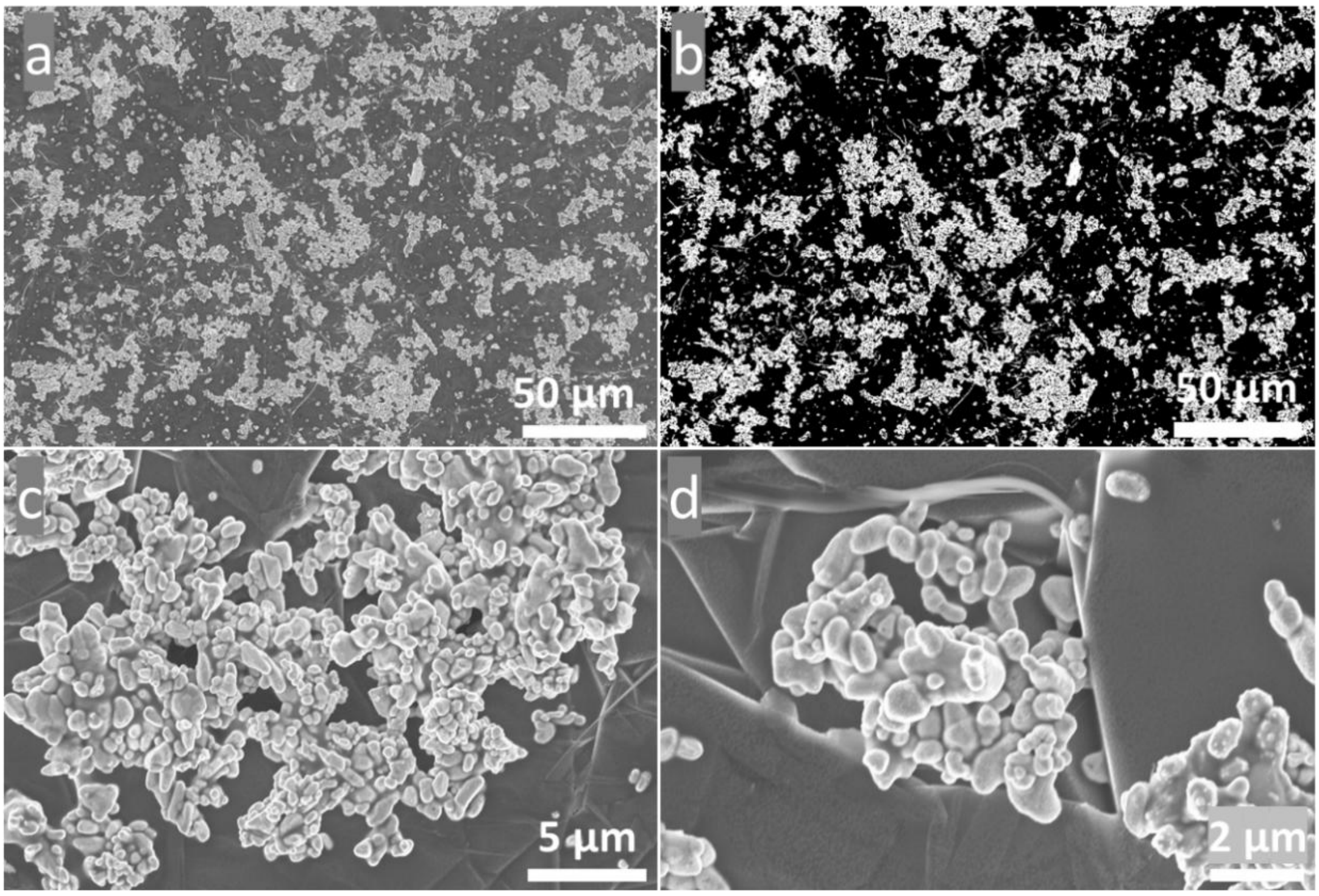

**Figure S28.** The SEM images of 150 µg $cm^{-2}$ of 40 $^{o}C$ dealloyed NPC drop-casted onto the graphite foil. (b) highlights the NPC covered area in white pixels, used for the surface coverage calculations in ImagJ. The surface coverage was calculated to be around 20%.

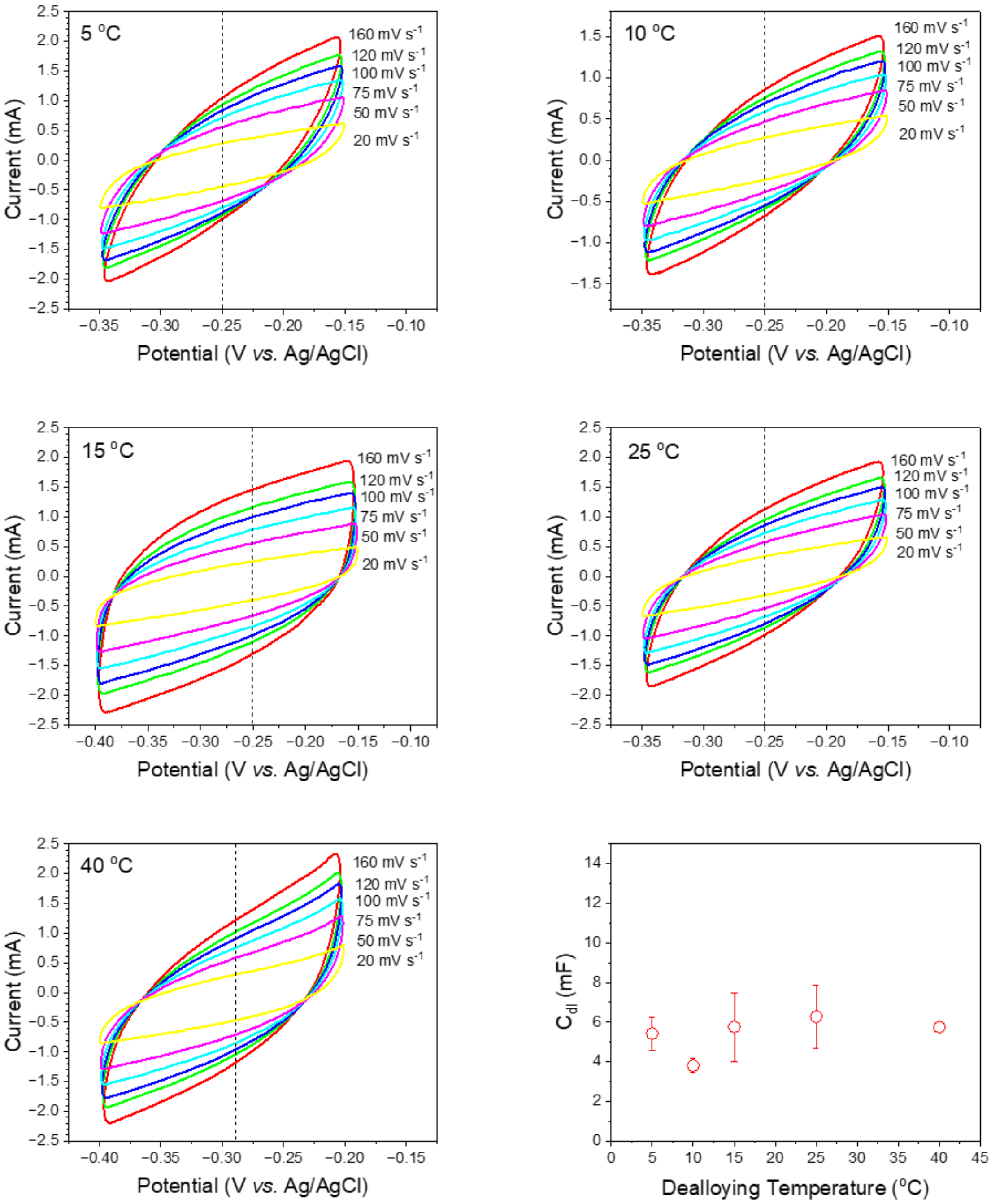

**Figure S29.** Double layer capacitance measurements of the dealloyed NPC samples with the mass loading of 150 μg cm$^{-2}$, drop-casted onto graphite foil (for $CO_2$RR test), in 0.1 M $HClO_4$. Dashed lines represent the potential where the current value was taken for the calculation of the double layer capacitance ($C_{dl}$) value plotted in the final plot.

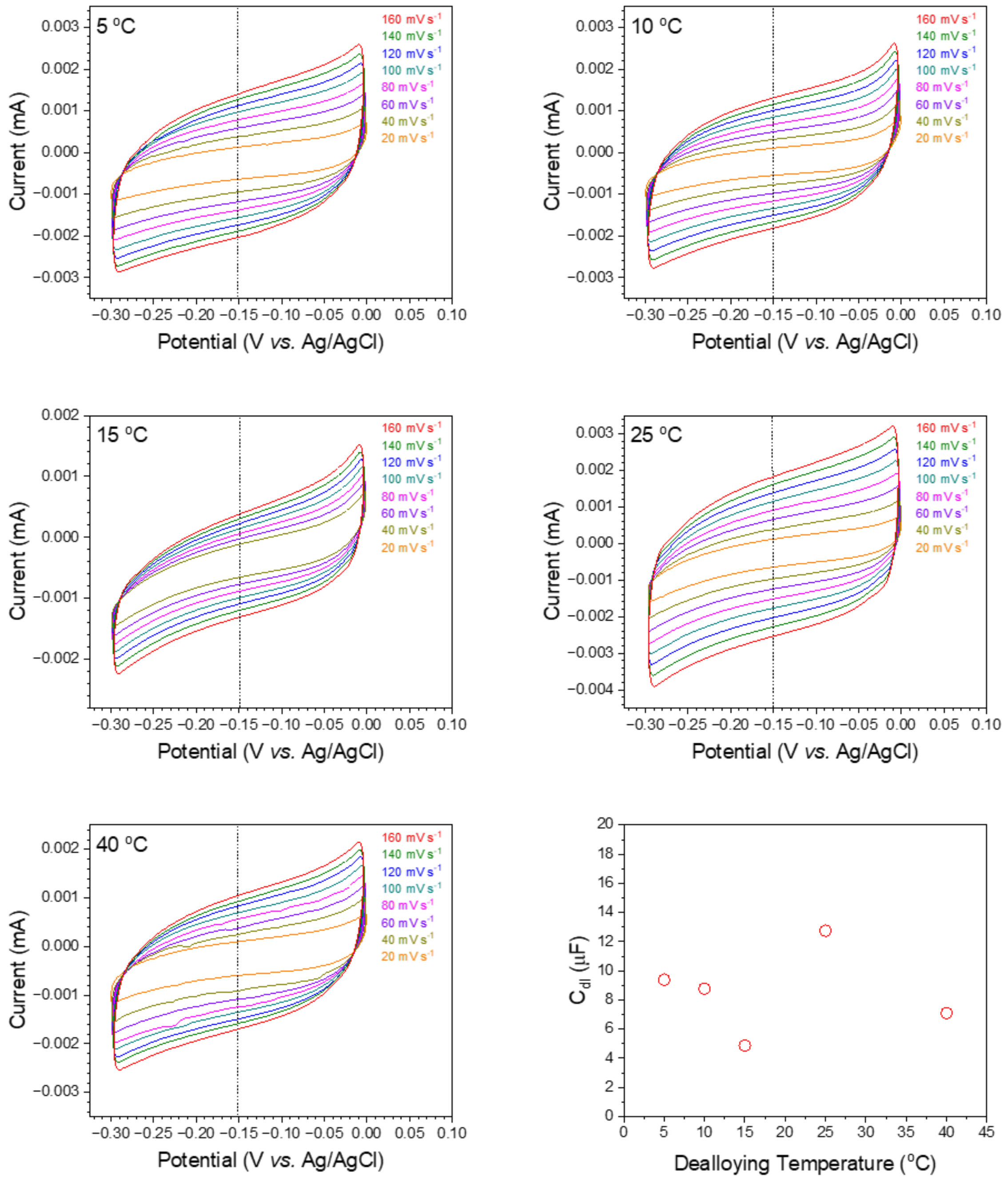

**Figure S30.** Double layer capacitance measurements of the dealloyed NPC samples with the mass loading of 4.3 µg cm$^{-2}$, drop-casted onto glassy carbon (for CORR test), in 0.1 M $HClO_4$. Dashed lines represent the potential where the current value was taken for the calculation of the double layer capacitance ($C_{dl}$) value plotted in the final plot.

The low mass loading (**Figure S28, S35**) also caused a similar ECSA between different dealloyed NPC samples, confirmed by the double layer capacitance measurement (**Figure S29-S30**), conducted in the same H-cell, and EC-MS setup used for $CO_2RR$ and CORR measurements, respectively. 20 mL min$^{-1}$ of Ar for H-cell and 1 mL min$^{-1}$ of He for EC-MS setup were purged into the cells for at least 30 min before the measurement for deaeration. Error bar in **Figure S29** was obtained by repeating the experiment on differently drop-casted electrodes for at least 3 times. Therefore, in this work, the intrinsic activity can be represented by the geometric surface area normalised partial current density while making comparison between different NPC samples.

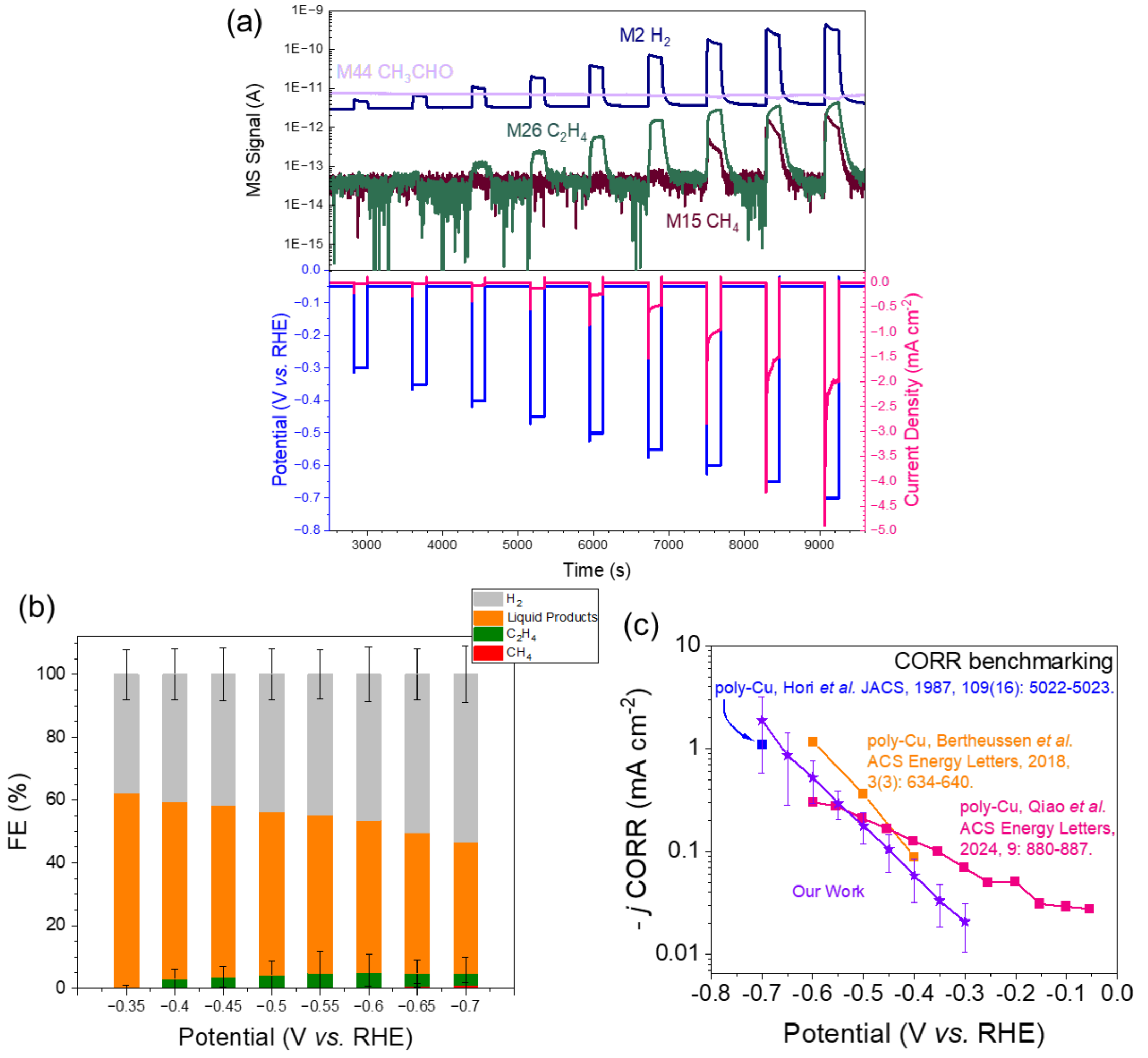

**Figure S31.** (a) EC-MS plot for the CO reduction on the electropolished polycrystalline Cu on 0.1 M KOH. No mass signal increasing for mass 44 fragment was detected, indicating M15 should come from methane only, as reaching the detection limits for alcohols (around 1% shown in **Figure S34**) is not possible. (b) CO reduction products distribution on the electropolished polycrystalline Cu. The FE of liquid products was calculated by $100\% - FE_{H_2} - FE_{CH_4} - FE_{C_2H_4}$. (c) Comparison of geometric surface area normalized partial current

density of CORR on electropolished polycrystalline Cu in 0.1 M KOH between this setup and literature benchmarking experiments.[23-25]

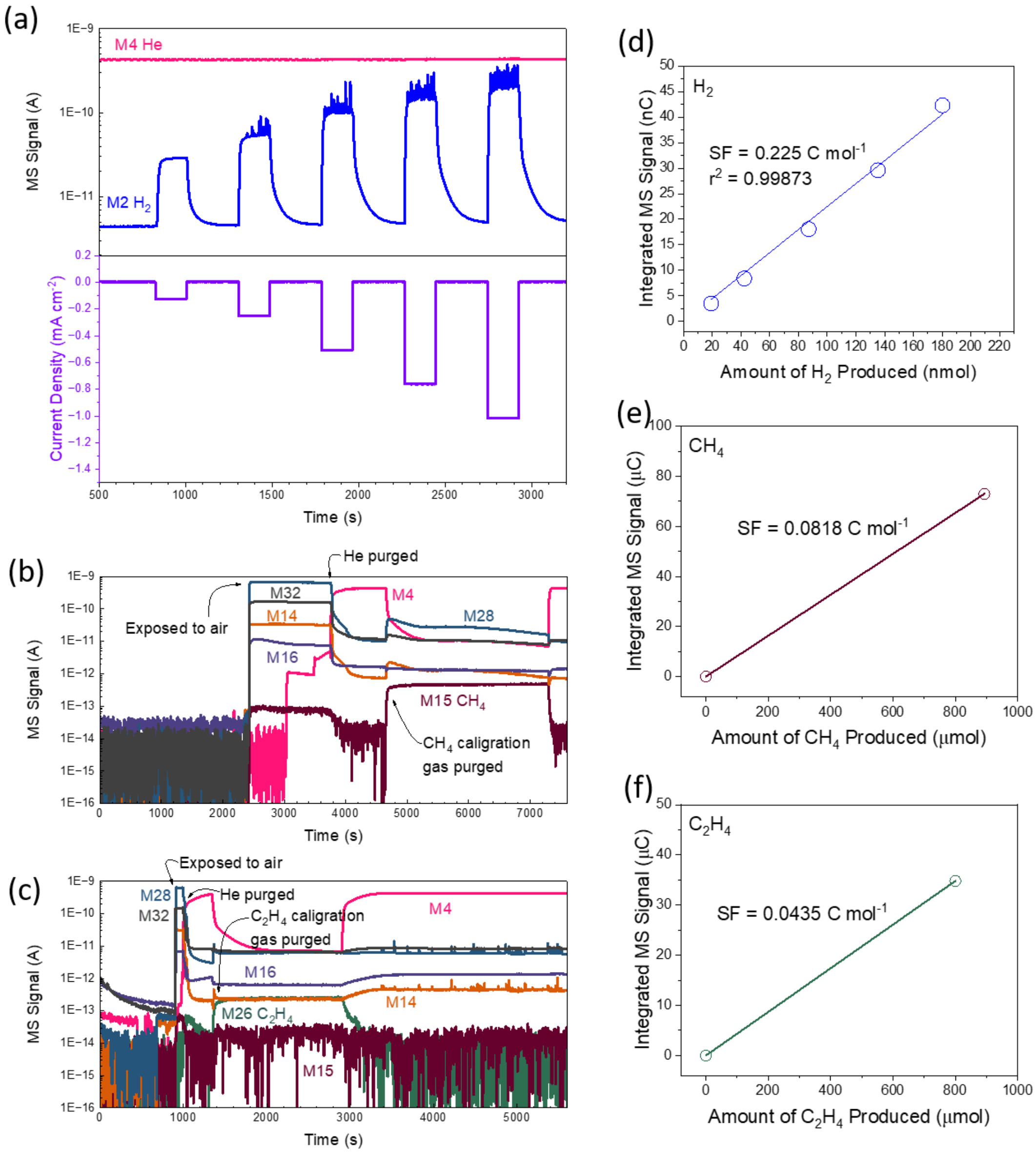

**Figure S32.** (a) – (c) EC-MS plots for $H_2$, $CH_4$, and $C_2H_4$ calibration. The mass 27 signal was also recorded for (c), the $C_2H_4$ calibration, and as it completely overlaps with mass 26, it is not presented here. (d) – (f) Calibration curves for $H_2$, $CH_4$, and $C_2H_4$, respectively.

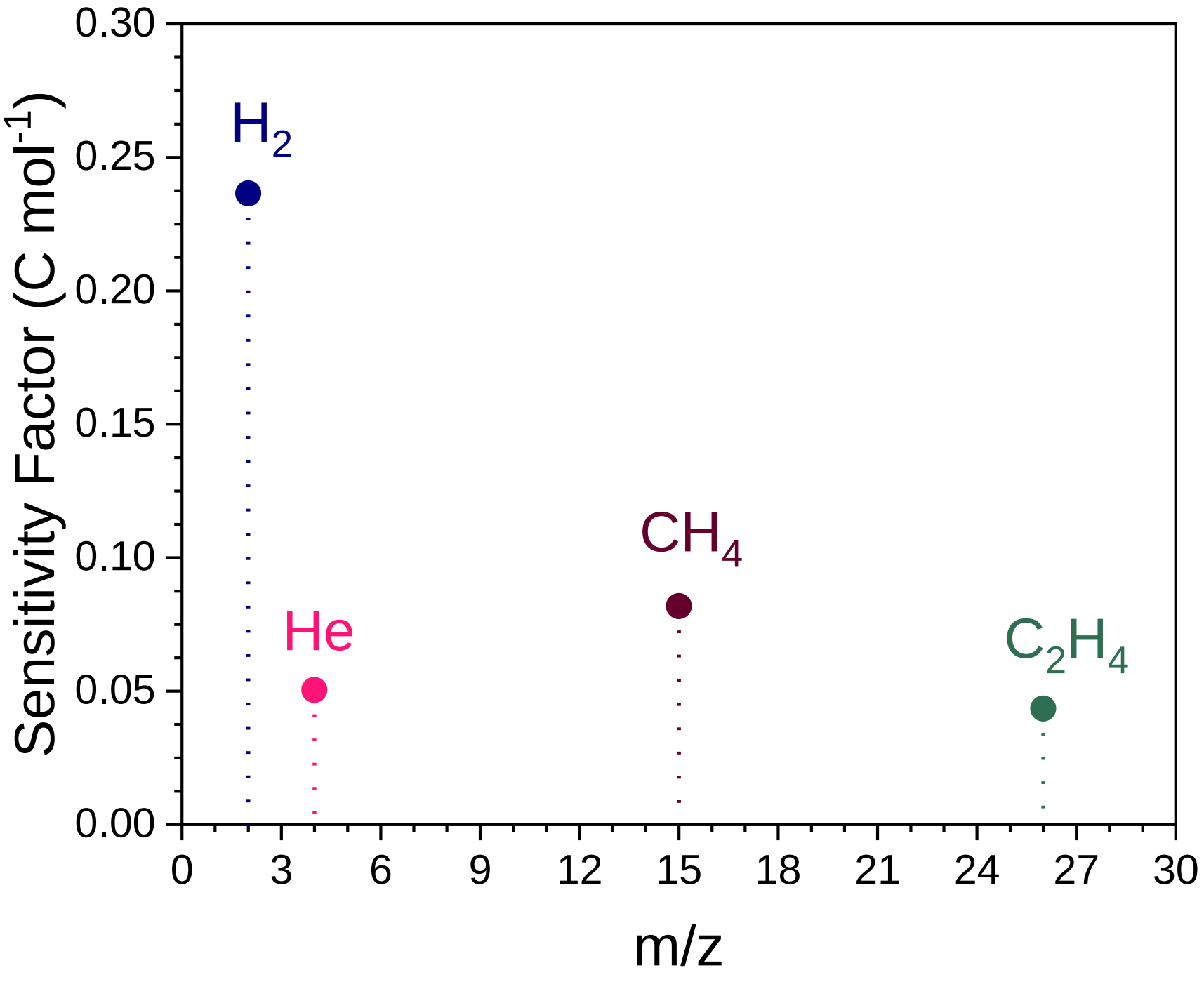

**Figure S33.** Summary of the sensitivity factors for EC-MS calibration

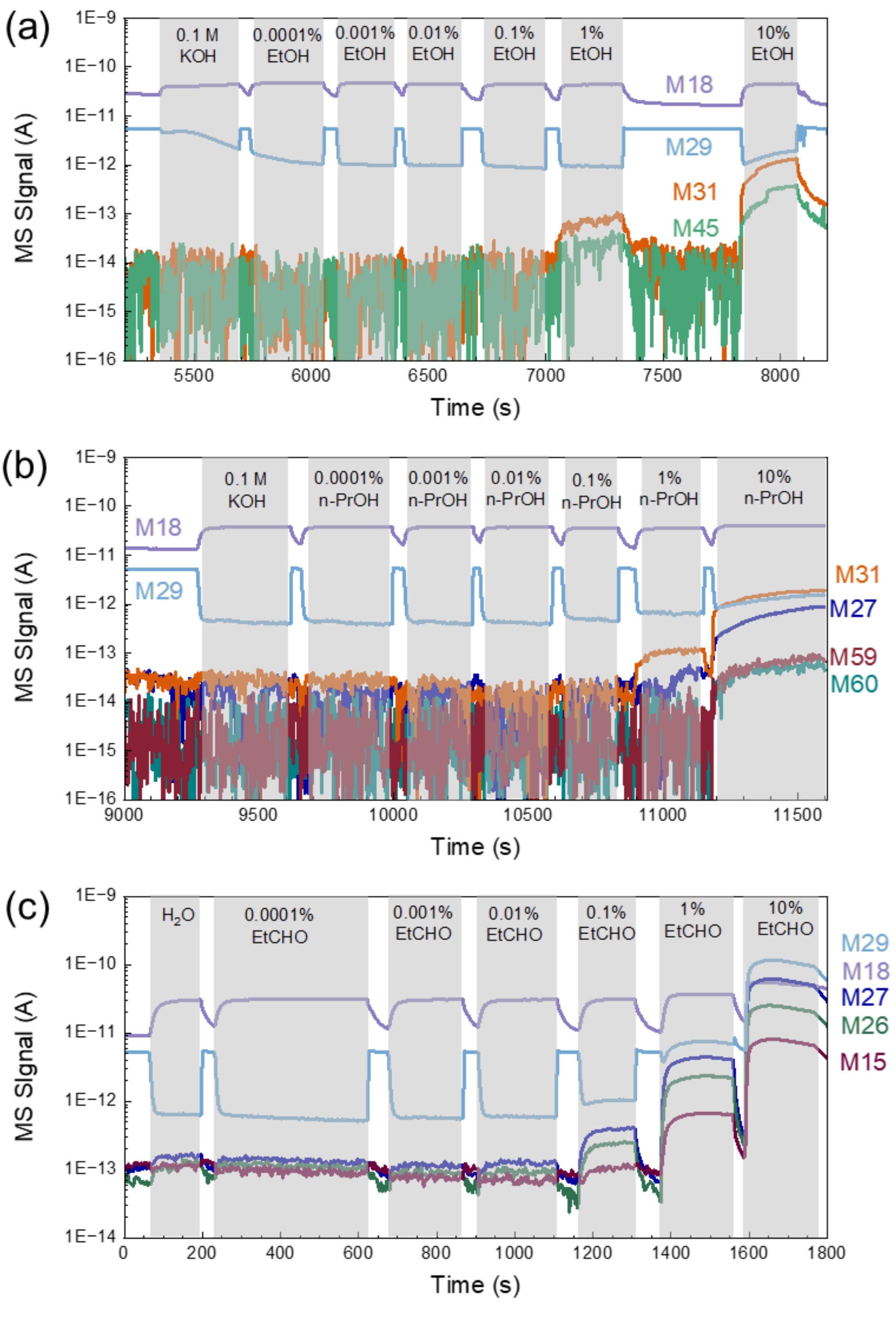

**Figure S34.** EC-MS detection limits test for the possible CO reduction liquid products, (a) ethanol, (b) n-propanol, and (c) propionaldehyde. Due to the similar Henry's law constants, methanol, and acetaldehyde should have similar detection limits with ethanol, and propionaldehyde, respectively.[26]

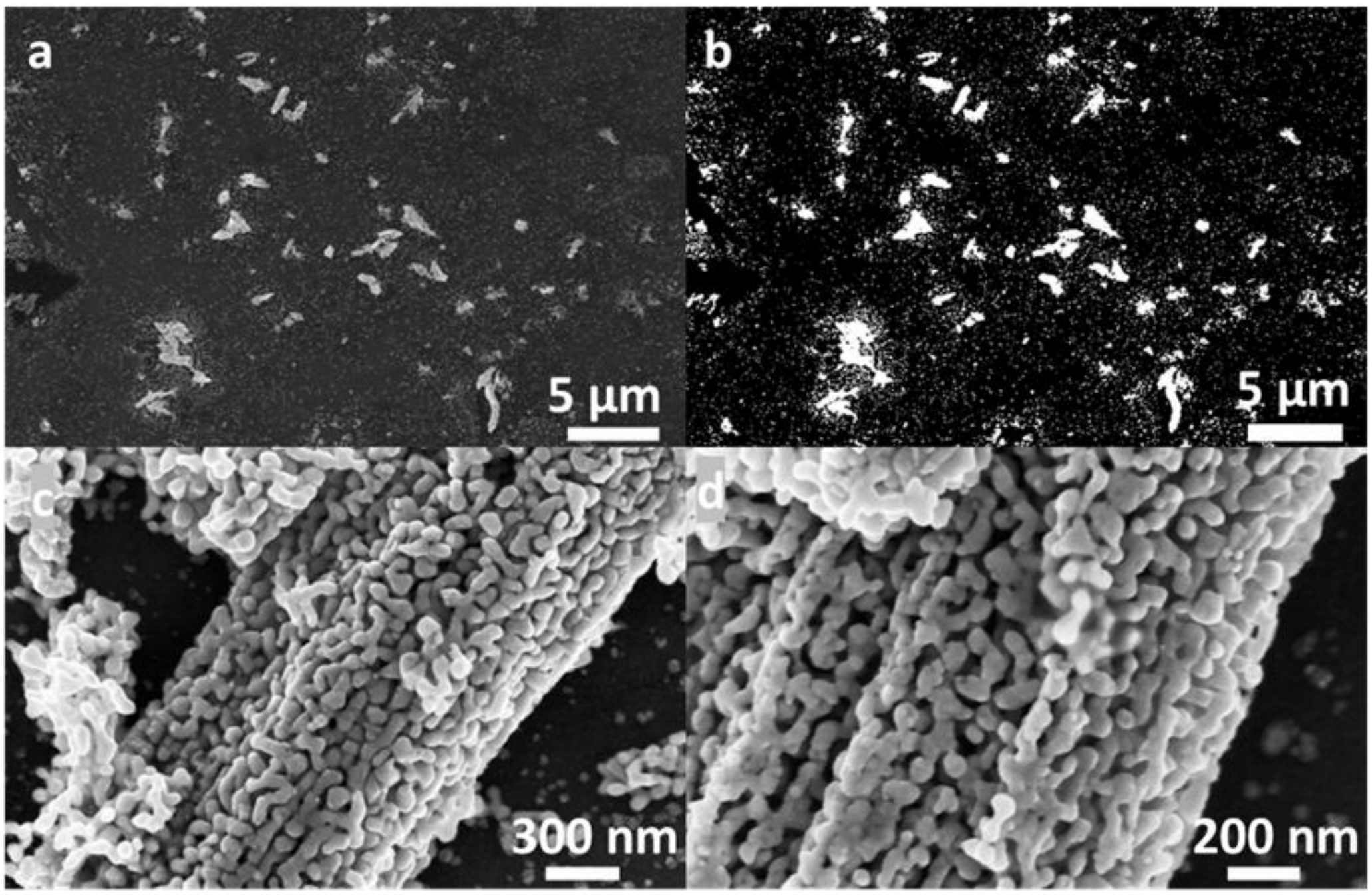

**Figure S35.** The SEM images of the glassy carbon RDE insert drop-casted with 4.3 µg $cm^{-2}$ of 15 $^{o}$C dealloyed NPC after the CO reduction measurement. (b) highlights the NPC covered area in white pixels while the uncovered area is in black pixels, used for the surface coverage calculations in ImageJ. The surface coverage was calculated to be round 8.9%.

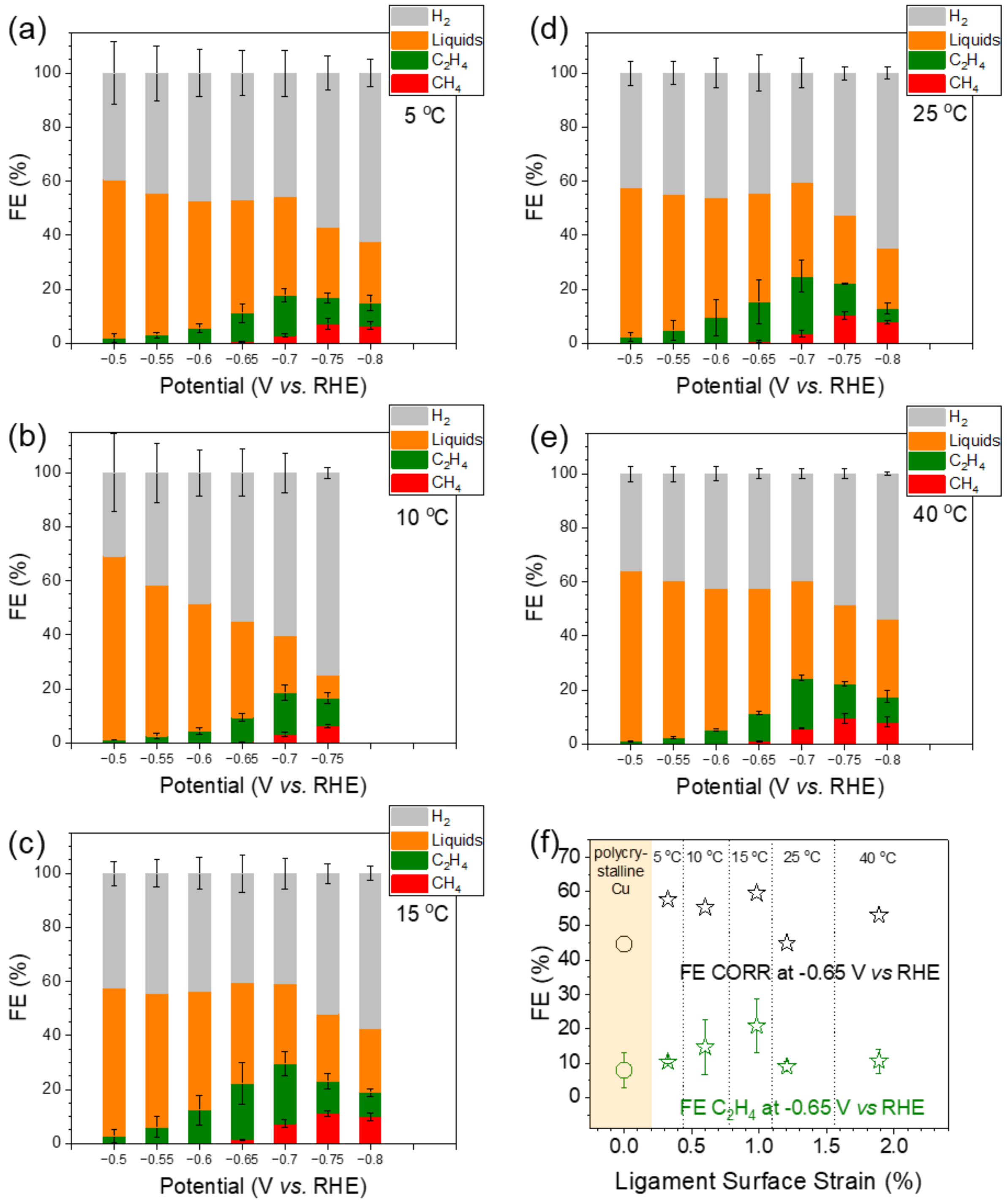

**Figure S36.** (a) – (e) Plots of the EC-MS CO reduction product FEs on the dealloyed NPC samples in 0.1 M KOH solution. Sample names are labelled in each figure. (f) Comparison of the FE towards CORR and $C_2H_4$ at -0.65 V *vs.* RHE for NPC samples and polycrystalline Cu.

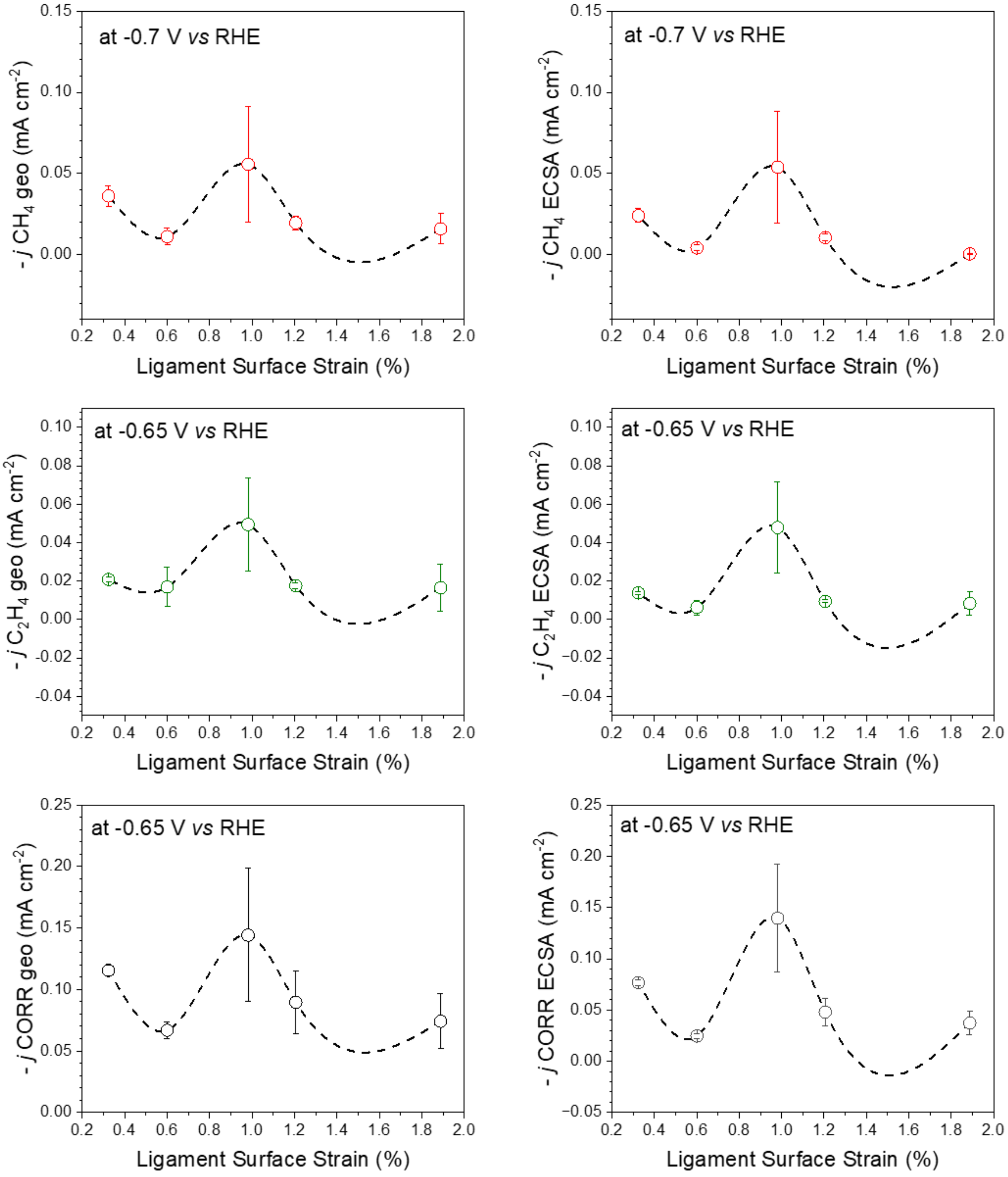

**Figure S37.** Plots of the partial current densities normalized by the geometric surface area (left) and ECSA (right) of $CH_4$, $C_2H_4$, and CORR for the NPC materials against the ligament surface strain.

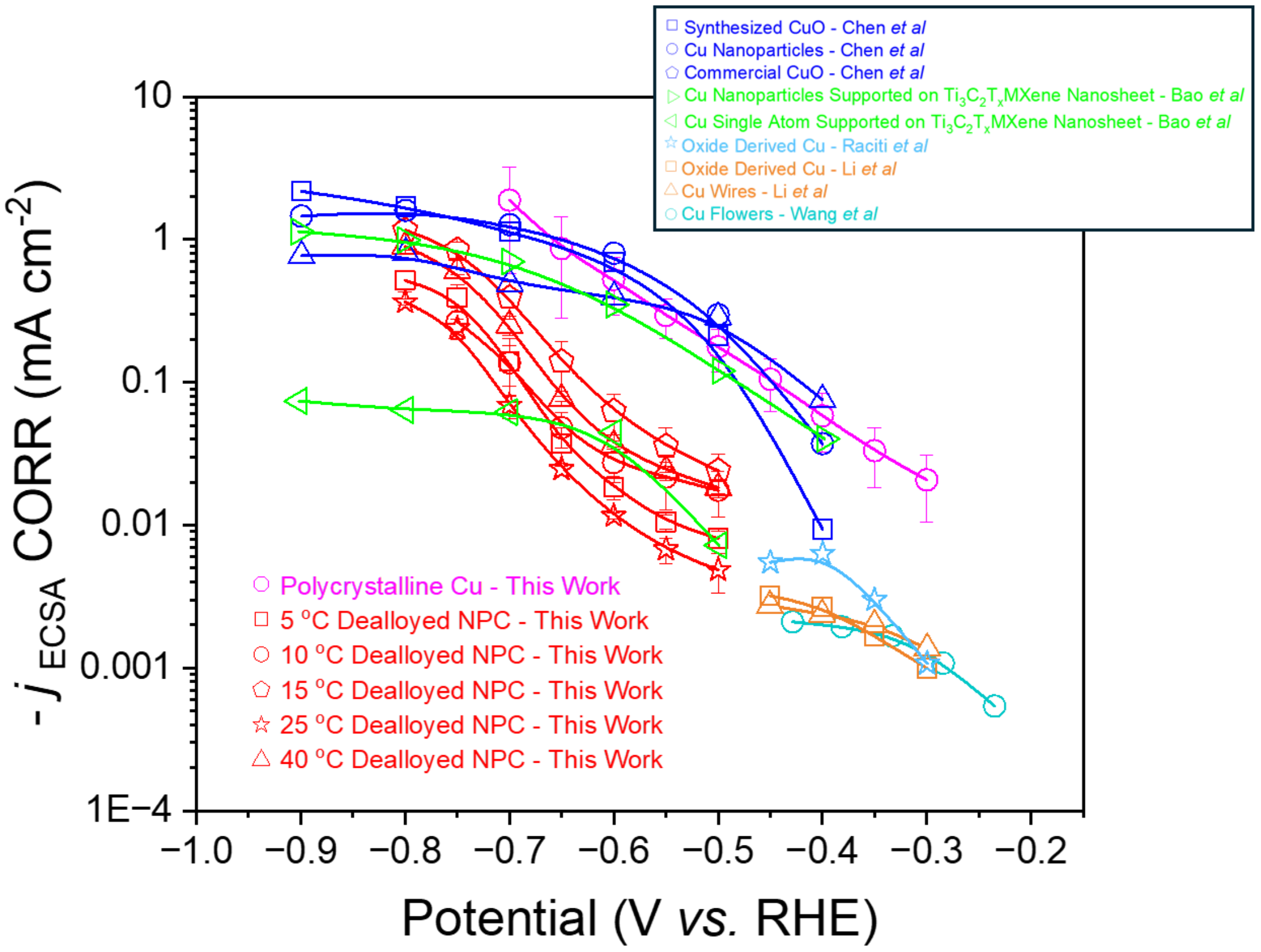

**Figure S38.** Comparison of ECSA normalized partial current density of CORR on NPC samples in 0.1 M KOH between this setup and selected high-performance literature values.[27-31]

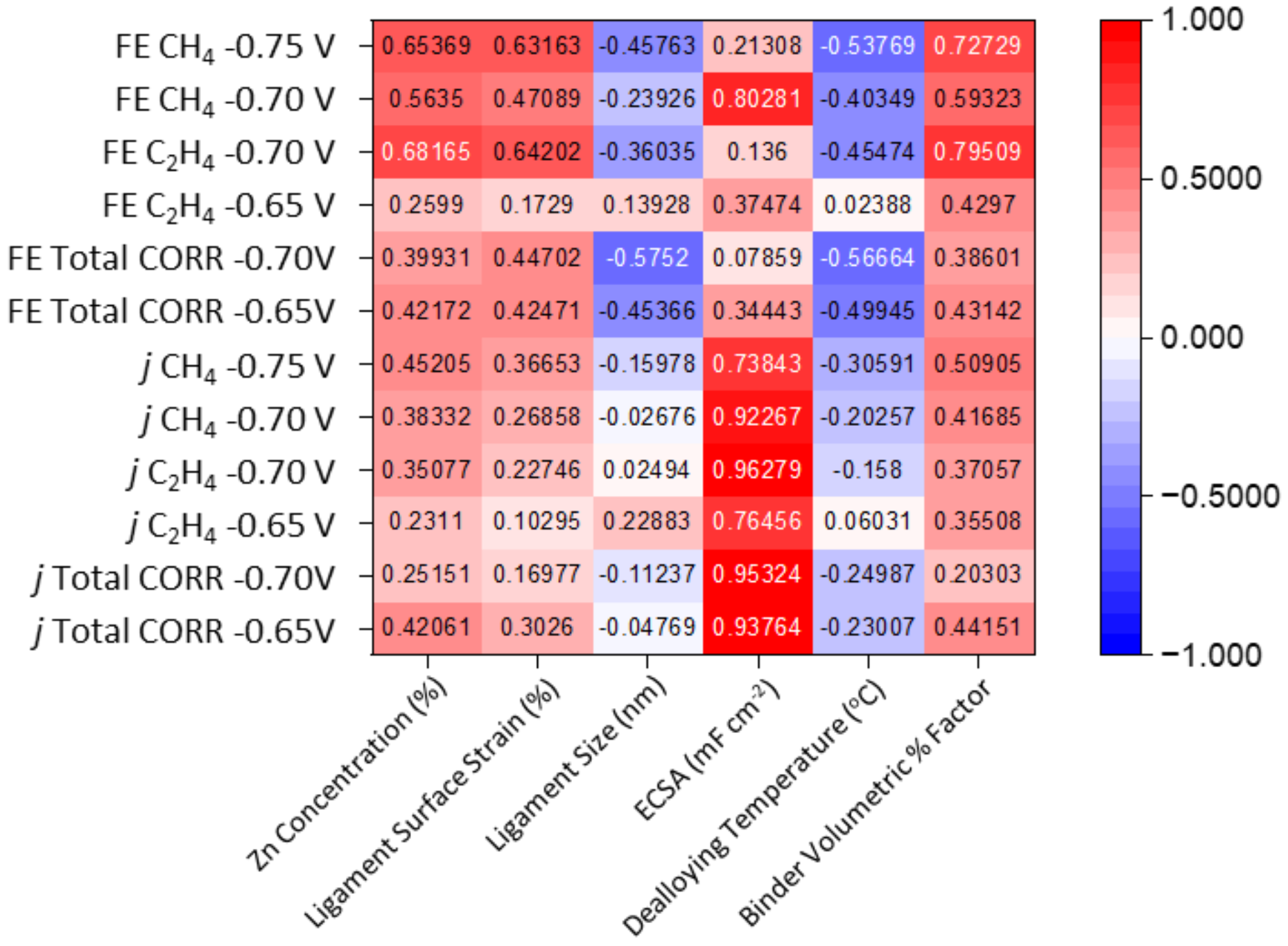

**Figure S39.** Heatmap of Pearson correlation between the physical characteristic analysis of NPC samples (x-axis) and their geometric surface area normalized current densities and Faradaic Efficiencies towards different CORR products at different potentials(all on RHE scale) (y-axis). Sustainion Solution Volume / (Ligament Size)[3] was used to calculate the Binder Volumetric % Factor.

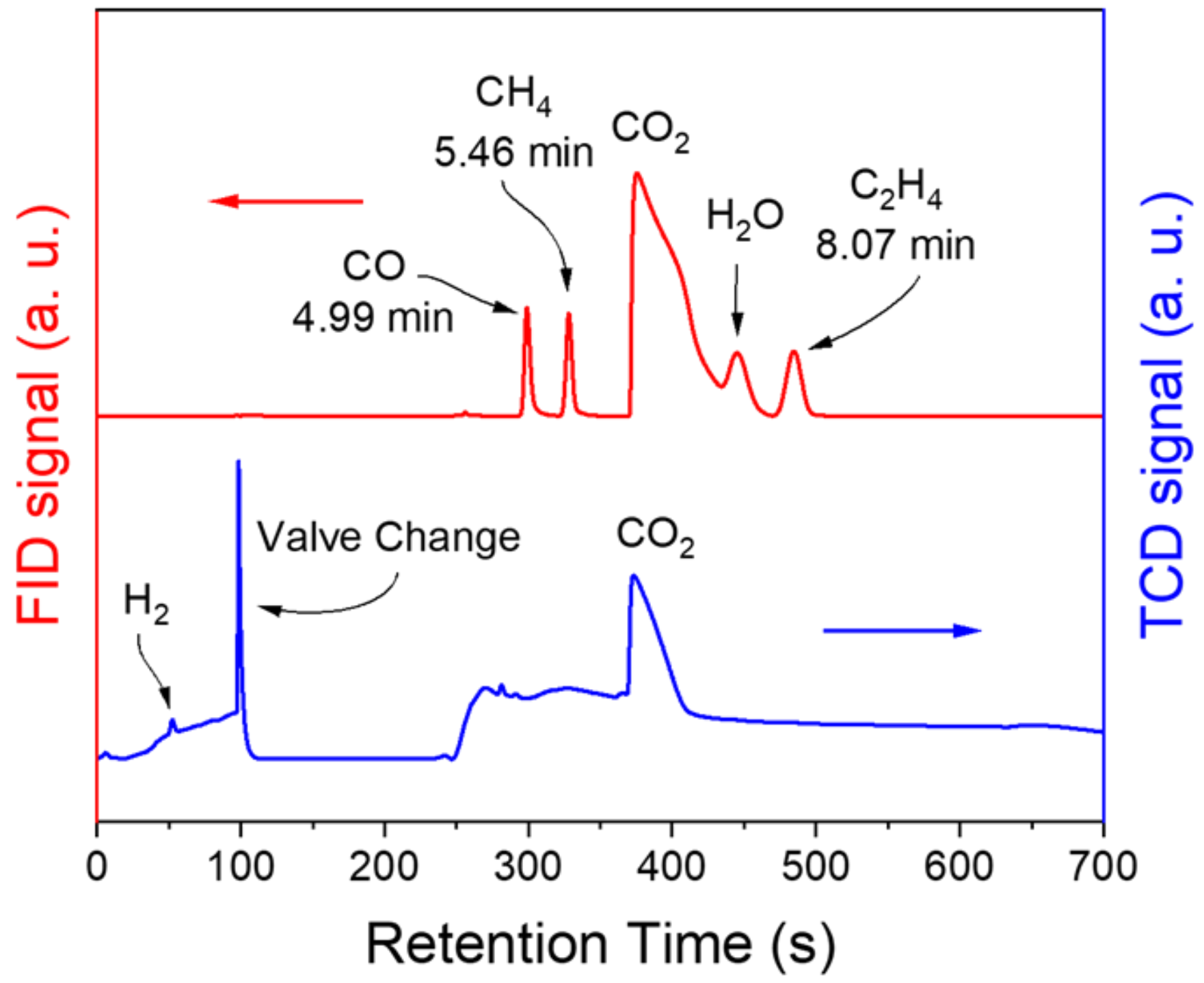

**Figure S40.** An example of the gas chromatogram measured with the mixture of 16 mL min$^{-1}$ Calibration gas and 4 mL min$^{-1}$ pure $CO_2$.

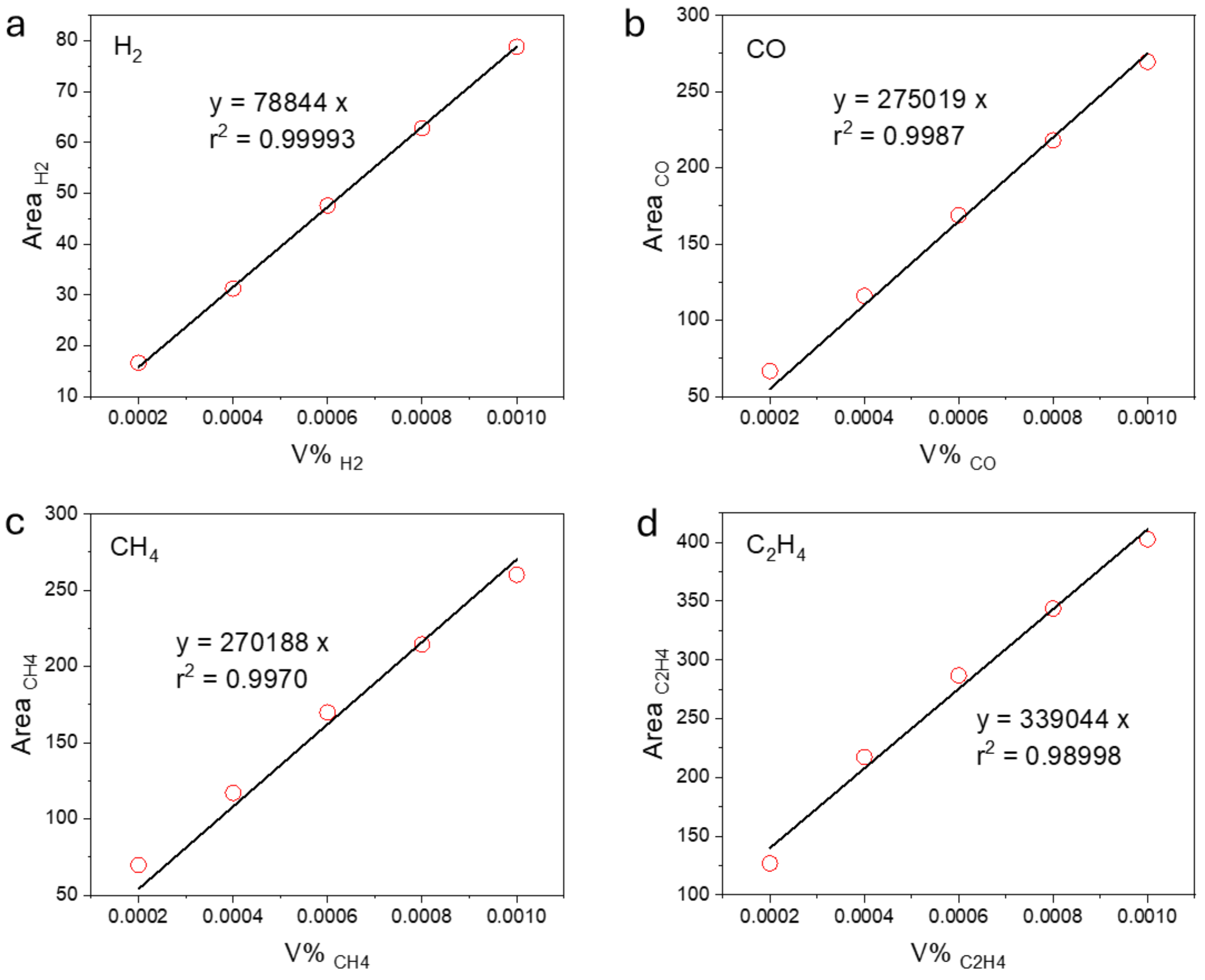

**Figure S41.** The GC calibration curves of $H_2$, CO, $CH_4$, and $C_2H_4$. The calibration curves are linear fits of the experimental data points.

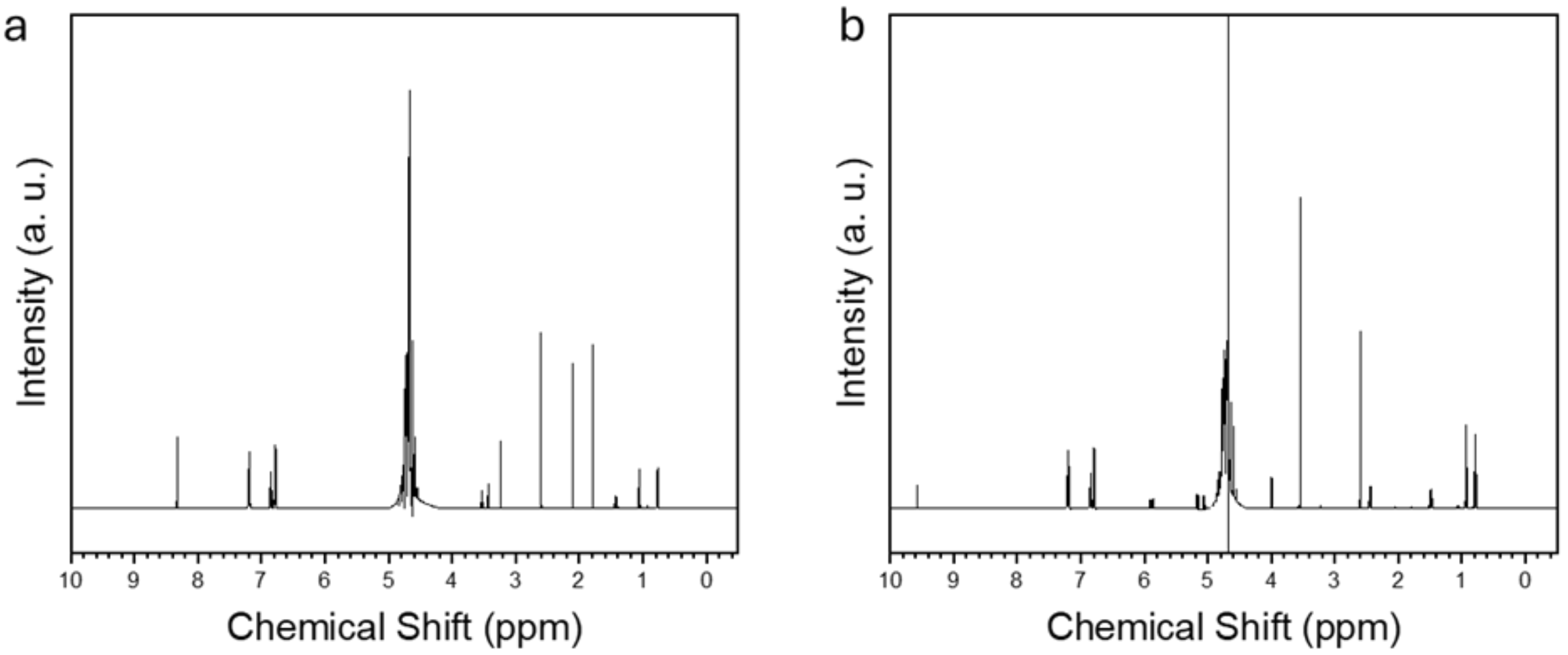

**Figure S42.** Examples of the NMR spectrum obtained from calibration measurements. Each sample contains 250 μL of the standard solution, 350 μL of $D_2O$, and 12.3 μL of the internal standards (10 mM phenol and 50 mM DMSO in $D_2O$). The standard solution for (a) contains 0.5 Mm of EtOH, 0.5 mM of n-PrOH, 0.5 mM of acetone, 0.5 mM of MeOH, 0.91 mM of potassium acetate, and 1.15 mM of potassium formate. The standard solution for (b) contains 1.25 mM of allyl alcohol, 1.25 mM of ethylene glycol, 2.50 mM of propionaldehyde.

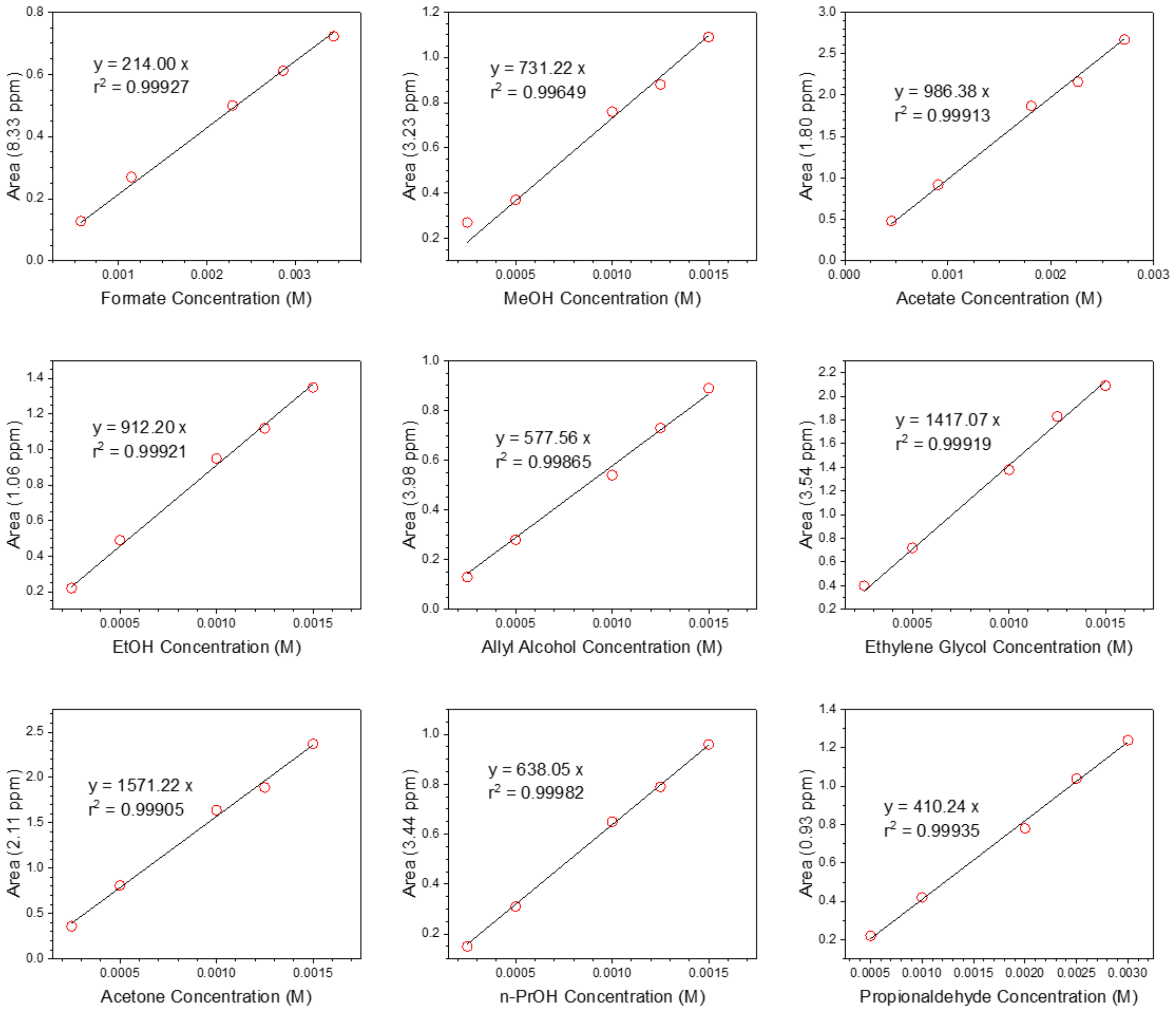

**Figure S43.** The $^{1}$H-NMR calibration curves of the liquid products. The calibration curves are linear fits of the experimental data points.